\documentstyle[12pt,subeqnarray,psfig]{article}
\begin{document}
%
%                      --------------
%                      | versione G |
%                      -------------- 
%
%\lhead[\thepage]{R. Caimmi and C. Marmo: Fitting simulated density
%       profiles in dark matter haloes}
%\rhead[Astron. Nachr./AN~{\bf XXX} (200X) X]{\thepage}
%\headnote{Astron. Nachr./AN {\bf 32X} (200X) X, XXX-XXX}
%
\newenvironment{lefteqnarray}{\arraycolsep=0pt\begin{eqnarray}}
{\end{eqnarray}\protect\aftergroup\ignorespaces}
\newenvironment{lefteqnarray*}{\arraycolsep=0pt\begin{eqnarray*}}
{\end{eqnarray*}\protect\aftergroup\ignorespaces}
\newenvironment{leftsubeqnarray}{\arraycolsep=0pt\begin{subeqnarray}}
{\end{subeqnarray}\protect\aftergroup\ignorespaces}
\newcommand{\appleq}{\stackrel{<}{\sim}}
\newcommand{\appgeq}{\stackrel{>}{\sim}}
\newcommand{\arcsinh}{\mathop{\rm arcsinh}\nolimits}
\newcommand{\arctg}{\mathop{\rm arctg}\nolimits}
\newcommand{\diff}{{\rm\,d}}
\newcommand{\displayfrac}[2]{\frac{\displaystyle #1}{\displaystyle #2}}
\newcommand{\Erfc}{\mathop{\rm Erfc}\nolimits}
\newcommand{\Int}{\mathop{\rm Int}\nolimits}
\newcommand{\Nint}{\mathop{\rm Nint}\nolimits}
\newcommand{\pprime}{{\prime\prime}}

\title{Dark matter haloes: an additional criterion % \\
       for the choice of fitting density profiles}
\author{{R. Caimmi and C. Marmo\footnote{
{\it Astronomy Department, Padua Univ., Vicolo Osservatorio 2,
I-35122 Padova, Italy} 
email: caimmi@pd.astro.it%, marmo@pd.astro.it
}
\phantom{agga}}}

%\institute{

%\date{Received {\it...................................}$\quad$ 
%Accepted {\it.................................}}

\maketitle
\begin{quotation}
\section*{}
\begin{Large}
\begin{center}

Abstract

\end{center}
\end{Large}
\begin{small}

%\abstract{

Simulated dark matter haloes are fitted
by self-similar, universal density profiles, where
the scaled parameters depend only on a scaled
(truncation) radius, $\Xi=R/r_0$, which, in turn,
is supposed to be independent on the mass and the
formation redshift.   The further assumption
of a lognormal distribution (for a selected mass
bin) of the scaled radius, or concentration,
in agreement with the data from a large statistical
sample of simulated haloes (Bullock et al. 2001),
allows (at least to a first
extent) a normal or lognormal distribution for
other scaled parameters, via the same procedure
which leads to the propagation of the errors.
A criterion for the choice of the best fitting
density profile is proposed, with regard to a set
of high-resolution simulations, where 
some averaging procedure on scaled density
profiles has been performed, in connection with
a number of fitting density profiles.   To this
aim, a minimum value of the ratio, $\vert x_
{\overline{\eta}}\vert/\sigma_{s\,\overline{\eta}}=
\vert\overline{\eta}-\eta^\ast\vert/
\sigma_{s\,\overline{\eta}}$, is required to
yield the best fit, where $\overline{\eta}$ is the
arithmetic mean over the whole set; $\eta^\ast$ is
its counterpart related to the fitting density profile;
$\sigma_{s\,\overline{\eta}}$ is the standard deviation
from the mean; and $\eta$ is a selected, scaled i.e.
dimensionless parameter.   The above criterion is
applied to a pair of sets each made of a dozen of
high-resolution simulations, FM01 (Fukushige \& Makino
2001) and KLA01 (Klypin et al. 2001), in connection
with two currently used fitting density profiles,
NFW (e.g., Navarro et al. 1997) and MOA (e.g.,
Moore et al. 1999), where the dependence of the scaled
radius on the mass and the formation redshift, may
be neglected to a first extent.
%, and in any case for asufficiently narrow mass range.
With regard to FM01 and KLA01 samples, the best fits
turn out to be MOA and NFW, respectively.
In addition, the above results
also hold in dealing with rms errors
derived via the propagation of the errors, with
regard to the distributions of scaled parameters.
The sensitivity error of simulations is also estimated
and shown to be less than the related, standard deviation,
that is a necessary condition for detectability of
accidental errors.   Some features of the early
evolution of dark matter haloes represented by fitting
density profiles, are discussed in the limit of the
spherical top-hat model.   Though the related matter
distributions appear to be poorly representative of simulated
haloes, unless the (mean) peak height is an increasing function
of the mass, the results are shown to be consistent,
provided considerable acquisition
of angular momentum takes place during the expansion phase. 

\noindent
{\it keywords - cosmology: dark matter.}
%}%                end of abstract

%\correspondence{caimmi@pd.astro.it}

%\maketitle

\end{small}
\end{quotation}

%
%________________________________________________________________

\section{Introduction}\label{intro}
According to a wide number of both analytical and numerical
studies (e.g., Cole \& Lacey 1996; Syer \& White 1998; 
Navarro et al. 1995, 1996, 1997, hereafter quoted as NFW97; 
Moore et al. 1998, 1999, hereafter
quoted as MOA99; Fukushige \& Makino 2001, hereafter quoted
as FM01; Klypin et al. 2001, hereafter quoted as KLA01;
Fukushige \& Makino 2003, hereafter quoted as FM03), dark
matter haloes which virialize from hierarchical clustering 
show universal density profiles, $\rho=\rho(r; \rho_0, r_0)$,
where $\rho_0$ is a scaling density and $r_0$ is a 
scaling radius.    In this view, smaller haloes formed first from
initial density fluctuations and then merged with each 
other, or were tidally disrupted from previously formed 
mergers, to become larger haloes.

The density profile is (i) self-similar, in the sense 
that it has the same expression, independent of time
(e.g., FM01), and (ii) universal, in the sense that it
has the same expression, independent of halo mass,
initial density perturbation spectrum, or value of
cosmological parameters (e.g., NFW97; FM01; FM03).   A
satisfactory fit to the results of numerical simulations
is the family of density profiles (e.g., Hernquist 1990;
Zhao 1996):
\begin{equation}
\label{eq:runi}
\rho\left(\frac r{r_0}\right)=\frac{\rho_0}{(r/r_0)^\gamma
[1+(r/r_0)^\alpha]^\chi}~~;\quad\chi=\frac{\beta-\gamma}
\alpha~~;
\end{equation}
for a suitable choice of exponents, $\alpha$, $\beta$, 
and $\gamma$.

This family includes both cuspy profiles first proposed by
Navarro et al. (1995, 1996), NFW97, $(\alpha,\beta,\gamma)=
(1,3,1)$, hereafter quoted as NFW density profile, and the 
so called modified isothermal profile,
$(\alpha,\beta,\gamma)=(2,2,0)$, which is the most widely
used model for the halo density distribution in analyses
of observed rotation curves.   It also includes the perfect
ellipsoid (e.g., De Zeeuw 1985), $(\alpha,\beta,\gamma)=
(2,4,0)$, which is the sole (known) ellipsoidal density
profile where a test particle admits three global integrals
of motion.   Finally, it includes the Hernquist (1990)
density profile, $(\alpha,\beta,\gamma)=(1,4,1)$, which
closely approximates the de Vaucouleurs $r^{1/4}$ law
for elliptical galaxies.
In dealing with the formation of dark matter 
haloes from hierarchical clustering in both CDM and 
$\Lambda$CDM scenarios,
recent high-resolution simulations allow $(\alpha,\beta,\gamma)=
(3/2,3,3/2)$, hereafter quoted as MOA density profile, as a best 
fit (e.g., Ghigna et al. 2000; FM01; KLA01; FM03), as first 
advocated by Moore et al. (1998) and MOA99%
\footnote{More precisely, a slope $\alpha=1.4$ was derived
by Moore et al. (1998), while the value $\alpha=1.5$ was
established in MOA99.}.
But for a different point of view, concerning the trend near 
the centre of the system, see e.g. M\"ucket \& Hoeft (2003).

In addition, purely dark matter structures which fulfill
Jeans equation, exhibit $-3\le\gamma\le-1$ for density profiles
following an exact power-law, $\rho\propto r^{-\gamma}$,
and this constraint weakens only slightly for a more general
mass distribution where the density-power slope, $\gamma(r)$,
is a function of the radius; if otherwise, the system cannot
be considered as in equilibrium and/or the effects of baryionic
component have to be investigated (Hansen 2004).

Leaving aside peculiar situations such as the occurrence
of major mergers, the average evolution may be approximated
as self-similar to a good extent.   Accordingly, 
a single halo may be characterized by
two parameters: the (fiducial) total mass, $M$, and a
dimensionless quantity, $\delta$, related to the amplitude
of the density perturbation at the collapse (NFW97, FM01).
The scaling density, $\rho_0$, and the scaling radius, $r_0$, 
may also be expressed in terms of $M$ and $\delta$ (NFW97,
FM01).

Though Eq.\,(\ref{eq:runi}) implies null density at infinite
radius, profile fits are necessarily performed within the
virialized region of a halo, bounded by a truncation radius.
The presence of
neighbouring systems makes the tidal radius as an upper limit.
On the other hand, isolated objects cannot extend outside the
Hubble sphere of equal mass.   The region enclosed within the
truncation boundary has to be intended as representative of
the quasi static halo interior, leaving aside the surrounding
material which is still infalling.   Numerical simulations
show that the quasi static halo interior is defined by a
mean density, $\bar{\rho}_{200}\approx200\rho_{crit}$ (e.g.,
Cole \& Lacey 1996; NFW97; FM01), where $\rho_{crit}$ is
the critical density of the universe.    An alternative
definition is found in KLA01, where the quasi static halo
interior has same mean density as predicted by the top-hat
model.

Given a set of simulated, dark matter haloes, the choice
of a fitting density profile, expressed by Eq.\,(\ref
{eq:runi}), implies the following steps (e.g., Dubinski
\& Carlberg 1991; FM01; KLA01; FM03).
\begin{description}%                          Beccari,  p. 26
\item[\rm{(a)}] Select a choice of exponents $(\alpha,
\beta,\gamma)$, for defining the universal density
profile.
\item[\rm{(b)}] Use a nonlinear
least-squares method to determine the best fit
for the scaling density, $\rho_0$, and the
scaling radius, $r_0$, with regard to each
simulation.
\item[\rm{(c)}] Determine the scaled,
averaged density profile, and related
values of the scaling parameters,
($\overline{r}_0$, $\overline{\rho}_0$).
\item[\rm{(d)}] Particularize the
fitting formula, expressed by Eq.\,(\ref
{eq:runi}), to $(r_0,\rho_0)=(\overline{r}_0,
\overline{\rho}_0)$, and calculate the
parameters of interest, including
the residuals related to the scaled, averaged
and fitting density profile.
\end{description}

At present, no general consensus exists on the details
of the above mentioned procedure.   The scaling density
and the scaling radius may be constrained to yield
$M(R)=M_{trn}$, where $M(R)$ is the mass within the
truncation radius, $R$, related to the fitting density
profile, and $M_{trn}$ is the mass within the virialized
region (e.g., FM03).   An average all over the simulations
may be performed with regard to the scaling density and
the scaling radius, or any other two equivalent parameters
(e.g., FM01).   The minimization procedure may be applied
on the sum of either the squares of logarithmic residuals,
$[\log(\rho_{sim}/\rho_0)-\log(\rho_{uni}/\rho_0)]^2$ 
(e.g., Dubinski \& Carlberg 1991; FM03), or the absolute
values of logarithmic residuals, $\vert\log(\rho_{sim}/\rho_0)
-\log(\rho_{uni}/\rho_0)\vert$ (e.g., KLA01), where $\rho_
{sim}$ and $\rho_{uni}$ denote simulated and universal density
profile, respectively, related to an assigned radius, $r$.
For further details on the fitting procedures see e.g.,
Fukushige et al. (2004), Tasitsiomi et al. (2004).

Though the above mentioned method allows a selection
between different, universal density profiles (e.g.,
FM01, FM03), it does not seem to hold in general (e.g.,
KLA01).   In other words a simulated density profile
may be fitted, to an acceptable extent, by universal
density profiles with several choices of exponents,
$(\alpha,\beta,\gamma)$, appearing in Eq.\,(\ref{eq:runi}).
Further investigation on additional criterions, in
fitting universal to simulated density profiles,
could be useful in lowering the above mentioned
degeneracy.

Data from a statistical sample of about five thousands
of simulated, dark matter haloes (Bullock et al. 2001),
are consistent with a lognormal distribution of the
concentration, $\Xi_{trn}=R/r_0$, i.e. the ratio of the
truncation radius, $R$, to the scaling radius, $r_0$,
with regard to NFW density profiles.
The distribution is related to masses within a range
(0.5-1.0)$\times10^nh^{-1}{\rm M}_\odot$, where $11\le n\le14$
and $n$ is an integer.   The scatter is large, of about
$\sigma_{\log\Xi_{trn}}=0.18$ (Bullock et al. 2001).

The existence of a lognormal distribution is a necessary,
but not sufficient condition, for the validity of the
central limit theorem.   In this view, the concentration
is related to the final properties of a simulated halo,
which are connected with the initial conditions, $\alpha
_1$, $\alpha_2$, ..., $\alpha_n$, by a transformation,
$\Xi_{trn}=\alpha_1\alpha_2...\alpha_n$, as in dealing
with the process of star formation, where the stellar mass
follows a lognormal distribution (for further details, see
Adams \& Fatuzzo 1996; Padoan et al. 1997). 

If, for a selected mass range, the density profile is
assumed to be universal, then the scaled physical 
parameters, related to e.g., mass, moment of inertia, 
and potential energy, depend on the scaled radius
only (e.g., Caimmi \& Marmo 2003).   Accordingly,
the distribution
of a scaled physical parameter, to a first extent, 
is expected to be normal or lognormal.

There is a well known analogon of the above procedure
in the theory of errors.   Let a physical quantity,
$\Xi$, be directly measured, and then follow a normal
distribution, characterized by an expected value,
$\Xi^\ast$, and a rms error, $\sigma_\Xi$.   Let one
other physical quantity, $\Psi$, depend on the former 
one, $\Psi=\phi(\Xi)$, and then be indirectly measured.
As a result of the theory of errors, the physical 
quantity, $\Psi$, at least to a first extent, also
follows a normal distribution, characterized by an 
expected value, $\Psi^\ast=\phi(\Xi^\ast)$, and a rms 
error, $\sigma_\Psi=\vert(\partial\phi/\partial\Xi)_{\Xi^
\ast}\vert\sigma_\Xi$.   

The current investigation is aimed to provide an
additional criterion in fitting universal to
simulated density profiles.   The related
procedure starts from a set of high-resolution
simulations, where (i) both mass and radius of
the virialized region are known for each sample
halo; (2) scaled density profiles are averaged
over the whole sample; (3) scaling parameters,
$(r_0,\rho_0)$, are deduced with regard to
different choices of exponents, $(\alpha,\beta,
\gamma)$, which appear in the fitting formula,
expressed by Eq.\,(\ref{eq:runi}).

A fitting, scaled density profile, is defined
by a choice of exponents, $(\alpha,\beta,
\gamma)$, and a scaled truncation radius,
$\Xi$, which allows the calculation of the
remaining scaled parameters.   Under the
assumption of invariant, scaled truncation
radius, a fitting, scaled density profile
represents a family of infinite density
profiles, hereafter quoted as {\it fitting
scaled halo},
each related to a particular choice of
scaling parameters, $(r_0,\rho_0)$.
Accordingly, the generic member of the
above mentioned family is defined by three
exponents, $(\alpha,\beta,\gamma)$, a
scaled truncation radius, $\Xi$, and two
scaling parameters, $(r_0,\rho_0)$.

A similar situation occurs for polytropes
(e.g., Caimmi 1980), where a scaled density
profile represents a family of infinite
density profiles.   It depends on one
exponent, $n$ (the polytropic index), and
a scaled radius, $\Xi$ (where the density
falls to zero).   The generic member of
the family depends, in addition, on two
scaling parameters, $(R,\lambda)$, which
represent the radius and the central density,
respectively.

Given a sample of simulated, dark matter
haloes, let us define a mean, scaled
density profile, as the result of some
averaging procedure over the whole sample.
Then a mean, scaled density profile
represents its parent set of simulations
and, in the following, it shall be quoted
as {\it mean scaled halo}.

With regard to a selected, fitting,
scaled density profile, $(\alpha,\beta,
\gamma)$, the dependence of the scaling
parameters, $(r_0,\rho_0)$, on a pair
of independent parameters, $(M,\delta)$,
can be deduced
from the mean, scaled halo (e.g., FM01).
Then a scaled mass, $M/M_0$, can be
explicitly expressed and compared with
its counterpart related to the fitting,
scaled density profile.   It, in turn,
allows the calculation of the scaled
truncation radius, $\Xi$, and other scaled
parameters, which define the fitting
scaled halo.

On the other hand, in dealing with a
generic, simulated halo, both mass and
radius of the virialized region, $M_
{trn}$ and $R_{trn}$, are known as
computer outputs.   Accordingly, a
scaled radius, $\xi_{trn}=r_{trn}/r_0$,
and a scaled mass, $M_{trn}/M_0$, can
be determined together with additional
scaled parameters, and the deviations
from their counterparts, related to the
fitting, scaled halo, may be analysed.
In particular, the arithmetic mean,
$\overline{\eta}$, the standard deviation
from the mean, $\sigma_{s\,\overline
{\eta}}$, and the standard deviation
from the standard deviation from the
mean, $\sigma_{s\,\overline{\mu}}$,
may be calculated for a selected scaled
parameter, $\eta$.

The fitting density profile to a fixed
simulated halo, is defined by three
exponents, $(\alpha,\beta,\gamma)$,
a scaled truncation radius, $\xi_
{trn}$, and two scaling parameters,
$(r_0,\rho_0)$.   It shall be hereafter
quoted as {\it fitting halo}.   Unlike the
fitting, scaled halo, fitting haloes
exhibit different scaled truncation
radii, related to different simulated
haloes, while the remaining parameters
are left unchanged.

At this stage, it is possible to see
to what extent different, fitting,
scaled density profiles, $(\alpha,\beta,
\gamma)$, make scaled parameters, $\eta$,
related to each simulated halo,
deviate from their counterparts,
$\eta^\ast$, related to the fitting,
scaled halo.   In other
words, one is able to recognize if the
inequality, $\overline{\eta}-\sigma_
{s\,\overline{\eta}}<\eta^\ast<\overline
{\eta}+\sigma_{s\,\overline{\eta}}$, is
fulfilled.   Finally, the best fitting
density profile among the ones under
consideration, is chosen as minimizing
the ratio, $\vert x_{\overline{\eta}}
\vert/\sigma_{s\,\overline{\eta}}=\vert
\overline{\eta}-\eta^\ast\vert/\sigma_
{s\,\overline{\eta}}$.

The current investigation shall be limited to samples 
of recent, high-resolution, virialized structures where
(a) the sample is homogeneous i.e. related to a fixed
cosmological model; (b) the sample is not extremely   
poor i.e. the number of objects exceeds ten; (c) the 
values of the scaling density, scaling radius, virial
mass, and virial radius, are reported or may be deduced
from the results.

The above conditions are satisfied by two samples, 
each made of a dozen of runs, namely FM01 and KLA01.
With regard to the latter, the twelve runs studied
therein are in fact three sets of simulations of
only four dark matter haloes with resolution varied
in each set.   They cannot be treated as twelve
independent runs but, on the other hand, they can
be conceived as measures of a same physical quantity,
but using different methods.

The two sets of runs 
%FM01 and KLA01,
differ in many respects, namely: 1) cosmological 
model; 2) criterium in making subsets of runs; 3) 
criterium in ending simulations; 4) mass range;
5) definition of the virial radius; 6) scaling 
between NFW and MOA density profiles; 7) choice of 
the pair of independent parameters, i.e. $(M,\delta)$ 
or $(\rho_0, r_0)$.   For further details, see
FM01 and KLA01.   

Both NFW and MOA density profiles are fitted to
simulated density profiles for the samples under
consideration (FM01, KLA01).   Accordingly, the
above mentioned criterion shall be used in the
present paper, to see what is the best fit, among
NFW and MOA, to each set of simulations.

The main limit of the current approach lies in the
assumption of scaled density profiles,
related to an invariant, scaled (truncation) radius,
$\Xi$, and other scaled parameters depending only 
on $\Xi$.    In general, the scaled radius, which
has the same formal definition as the concentration
(e.g., NFW97), depends on both the mass and the
redshift.   More precisely, the concentration is
lowered for increasing mass (constant redshift)
and redshift (constant mass), with a milder/steeper
dependence for CDM/$\Lambda$CDM cosmological models
(Bullok et al. 2001).   An investigation based on
a large statistical sample (Bullok et al. 2001)
has shown that, within a $\Lambda$CDM scenario,
the intrinsic spread in concentration related to
a mass bin of distinct haloes, is comparable to
the systematic change in the mean value of
concentration related to the above mentioned
mass bin, across the entire mass range studied 
therein ($10^{11}<M/{\rm M}_\odot<10^{14}$).

It will be shown that, for both FM01 and KLA01
simulations, the intrinsic
spread in concentration is dominant over
the systematic change in the mean value of
concentration on a mass bin, across the 
entire mass range studied therein.   
Accordingly, the fitting density profile may be
considered, to an acceptable extent, as
related to an invariant scaled radius.
In general, it is the case for a sufficiently
narrow mass range.

The current paper is organized in the following way.
Useful formulae related to NFW 
and MOA density profiles are summarized in section 
\ref{NMdp}.   The fitting, scaled haloes, related
to FM01 and KLA01 set of simulations, with regard
to both NFW and MOA density profiles, are determined 
in section \ref{sim}.   The deviations of simulated
haloes from their fitting counterparts, in connection
with a number of scaled parameters, is also analysed
therein.   The following
section \ref{disc} is dedicated to a discussion, within
which some features of the early evolution of fitting
haloes are discussed, in the limit of the spherical
top-hat model.
Some concluding remarks are drawn in section \ref{core}. 
Further investigation on a few special arguments is
performed in the Appendix.

\section{NFW and MOA density profiles}\label{NMdp}

With regard to the family of density profiles,
expressed by Eq.\,(\ref{eq:runi}), let us define
a scaled density, $f$, and a scaled radius, $\xi$,
as:
\begin{eqnarray}
\label{eq:f}
&& f(\xi)=\frac\rho{\rho_0}=\frac{2^\chi}{\xi^\gamma
(1+\xi^\alpha)^\chi}~~;\qquad f(1)=1~~; \\
\label{eq:csi}
&& \xi=\frac r{r_0}~~;\qquad\Xi=\frac R{r_0}~~;
\end{eqnarray}
where the normalization, $f(1)=1$, makes $\rho_0$ 
and $r_0$ be the density and the radius (i.e. 
radial coordinate),
respectively, of a reference isopycnic surface,
and $\Xi$ corresponds to the truncation isopycnic
surface, or the truncation radius, $R$.   On the
other hand, the normalization currently used in
the literature takes $\rho^\prime_0=2^\chi\rho_0$,
in particular $(\rho^\prime_0)_{NFW}=4\rho_0$
and $(\rho^\prime_0)_{MOA}=2\rho_0$.
The choice of exponents, $(\alpha,\beta,\gamma)=
(1,3,1)$, $(3/2,3,3/2)$, selects NFW and MOA density
profiles, respectively, from Eq.\,(\ref{eq:runi}).

The explicit expression of a number of scaled
parameters, related to global or local properties
of the parent density profile, are listed
in Tab.\,\ref{t:proMN}.   Local properties depend
on the scaled radius, $\xi$, and global properties
depend on the scaled truncation radius, $\Xi$,
which has the same formal definition as the
concentration (e.g., NFW97).
\begin{table}
\begin{tabular}{cccc}
\hline
\hline
\multicolumn{1}{c}{function or} &
\multicolumn{1}{c}{definition}  &
\multicolumn{1}{c}{NFW}  &
\multicolumn{1}{c}{MOA} \\
\multicolumn{1}{c}{parameter} & & & \\
\hline
$f(\xi)$ & $\frac{\rho(\xi)}{\rho_0}$ & $\frac4{\xi
(1+\xi)^2}$ & $\frac2{\xi^{3/2}(1+\xi^{3/2})}$ \\
$P(\xi)$ & $2\int f(\xi)\xi\diff\xi$ & $\frac8{1+\xi}$
& $-2[\omega_1(\xi)+\omega_2(\xi)+\omega_3(\xi)]$ \\
$F(\xi)$ & $2\int_\xi^\Xi f(\xi)\xi\diff\xi$
& $P(\Xi)-P(\xi)$ & $P(\Xi)-P(\xi)$ \\
$\nu_M$ & $\frac M{M_0}$ & $12\left[\ln(1+\Xi)-\frac
\Xi{1+\Xi}\right]$ & $4\ln(1+\Xi^{3/2})$ \\
$\nu_{\bar{\rho}}$ & $\frac{\bar{\rho}}{\rho_0}$ 
& $\frac{12}{\Xi^3}\left[\ln(1+\Xi)-\frac
\Xi{1+\Xi}\right]$ & $\frac4{\Xi^3}\ln(1+\Xi^{3/2})$ \\
$\nu_{eq}$ & $\frac{v_{eq}(\Xi)}{(v_0)_{eq}}$
& $\left[\frac1\Xi\frac{\ln(1+\Xi)-\Xi/
(1+\Xi)}{\ln2-1/2}\right]^{1/2}$ & $\left[\frac1\Xi
\frac{\ln(1+\Xi^{3/2})}{\ln2}\right]^{1/2}$ \\
$\nu_I$ & $\frac I{2MR^2}$ & $\frac{6(1+\Xi)\ln
(1+\Xi)+\Xi^3-3\Xi^2-6\Xi}{9\Xi^2[(1+\Xi)\ln(1+\Xi)-\Xi
]}$ & $\frac{\Xi^2-4\Xi^{1/2}+\omega_1(\Xi)+\omega_2
(\Xi)-\omega_3(\Xi)-\omega_1(0)}{\nu_M\Xi^2}$ \\
$\nu_{sel}$ & $\frac{-E_{sel}R}{2GM^2}$ & $\frac\Xi4
\frac{\Xi(2+\Xi)-2
(1+\Xi)\ln(1+\Xi)}{[(1+\Xi)\ln(1+\Xi)-\Xi]^2}$ & $\frac
9{16}\frac\Xi{\nu_M^2}\int_0^\Xi F^2(\xi)\diff\xi$ \\
$\nu_J$ & $\frac1{\nu_M\Xi}\int_0^\Xi f(\xi)\xi^3\diff\xi$ 
& $\frac4{\nu_M\Xi}\left[\frac{\Xi(2+\Xi)}{1+\Xi}-2\ln(1+
\Xi)\right]$ & $\frac{2\Xi-\omega_1(\Xi)+\omega_2(\Xi)-
\omega_3(\Xi)+\omega_1(0)}{\nu_M\Xi}$ \\
% & & & \\
\hline
\multicolumn{4}{l}{$\omega_1(\xi)=\frac4{\sqrt{3}}\arctg\frac
{2\xi^{1/2}-1}{\sqrt{3}}~;~~\omega_2(\xi)=\frac43\ln(1+\xi^
{1/2})~;~~\omega_3(\xi)=\frac23\ln(1-\xi^{1/2}+\xi)~;~~
\omega_1(0)=\frac{-2\pi}{3\sqrt{3}}~.$} \\
\hline\hline
 & & & \\
\end{tabular}
\caption{Comparison between functions (local
properties) and profile parameters (global
properties), related to NFW and MOA density
profiles, respectively.   The profile parameters
depend on a single unknown, i.e. the  
scaled radius, $\Xi$.   The profile parameter, 
$\nu_J$, is related to the special case of constant
rotational velocity on the equatorial plane.
Rigidly rotating configurations correspond to
$\nu_J=\nu_I$.   Caption of symbols:
$M$ - total mass within the truncation isopycnic 
surface; $M_0$ - mass of a homogeneous region 
with same density and boundary as the reference
isopycnic surface; $\bar{\rho}$ - mean density
within the truncation isopycnic surface;
$v_{eq}(\Xi)$, $(v_0)_{eq}$ - rotational velocity
with respect to the centre of mass, at a point
placed on the truncation and reference isopycnic
surface, respectively; $I$ - moment of inertia;
$R$ - radius; $E_{sel}$ - self potential-energy;
$G$ - constant of gravitation; $J$ - angular
momentum.}
\label{t:proMN}
\end{table}

\section{Mean and fitting dark matter haloes}
\label{sim}

It can be shown that both NFW and MOA density
profiles provide an acceptable fit to 
high-resolution simulations (e.g., FM01; 
KLA01), with the possible exception of
scales of the order of cluster of galaxies,
where MOA density profiles seem to be
preferred with respect to NFW density
profiles (e.g., FM03).

The explicit
expression of the scaling density, $\rho_0$, 
as a function of the independent parameters, 
$(M,\delta)$, prescribed in FM01, by
averaging over the whole set of simulations,
reads:
\begin{leftsubeqnarray}
\slabel{eq:rho0a}
&& \rho_0=C_\rho\delta\left(\frac M{{\rm M}_{10}}
\right)^{-1}\frac{{\rm M}_{10}}{{\rm kpc}^3}~~; 
\quad {\rm M}_{10}=10^{10}{\rm M}_\odot~~; \\
\slabel{eq:rho0b}
&& (C_\rho)_{NFW,FM}=\frac{7k_1^3}{40}~~;\quad
(C_\rho)_{MOA,FM}=\frac7{20}~~;
\label{seq:rho0}
\end{leftsubeqnarray}
and the analogon for the scaling 
radius, $r_0$, reads:
\begin{leftsubeqnarray}
\slabel{eq:r0a}
&& r_0=C_r\delta^{-1/3}\left(\frac M{{\rm M}_{10}}
\right)^{2/3}{\rm kpc}~~; \\
\slabel{eq:r0b}
&& (C_r)_{NFW,FM}=2~10^{-2/3}k_1^{-1}~~;\quad
(C_r)_{MOA,FM}=2~10^{-2/3}~~;
\label{seq:r0}
\end{leftsubeqnarray}
where the normalization constant, $k_1$,
provides a connection between NFW and MOA
density profiles, in fitting the results
of simulations.   For further details,
see FM01 and Caimmi \& Marmo (2003,
Appendix B).

With regard to a selected, scaled density
profile (NFW or MOA), scaled parameters
related to simulated haloes may be
calculated via the scaling parameters,
$(r_0,\rho_0)$, defined by Eqs.\,(\ref
{seq:rho0}) and (\ref{seq:r0}) which,
in turn, correspond to fixed choices
of independent parameters, $(M,\delta)$.
For sake of brevity, let us define any
parameter, related to a fitting,
halo, as {\it fitting parameter}
(e.g., the mass of a fitting
halo is referred to as fitting mass).

The value of the dimensionless
parameter, $\delta$, is considered to reflect
an amplitude of the density perturbation at
turnaround and, for this reason, it can be
thought of as constant during the evolution
of a halo (e.g., Cole \& Lacey 1996; NFW97;
FM01).   From the standpoint of top-hat,
spherical density perturbation, it is 
related to both the mass and the peak
height, as shown in Appendix A.
%\ref{rhov}.

The combination of Eqs.\,(\ref{seq:rho0}) and
(\ref{seq:r0}) yields:
\begin{equation}
\label{eq:MM0}
\nu_M=\frac M{M_0}=\frac3{4\pi}\frac1{C_\rho C_r^3}~~;
\end{equation}
where $M_0$ is the mass of a homogeneous region,
with same density and boundary as the reference 
isopycnic surface, $(r_0,\rho_0)$.
The shape factor, $\nu_M$, depends on the scaled
radius, $\Xi$, as shown in Tab.\,\ref{t:proMN}.
Then the last may be determined, with regard to
a selected, fitting density profile.

Let us define a dimensionless parameter, $\kappa$, as:
\begin{equation}
\label{eq:k1}
\kappa=\delta^{1/2}\left(\frac R{{\rm kpc}}\right)^{3/2}
\left(\frac M{{\rm M}_{10}}\right)^{-1}~~;
\end{equation}
where $R=\Xi r_0$ is the radius of the truncation
isopycnic surface.   The combination
of Eqs.\,(\ref{seq:r0}) and (\ref{eq:k1}) yields:
\begin{equation}
\label{eq:k2}
\kappa=C_r^{3/2}\Xi^{3/2}~~;
\end{equation}
which, in turn, is related to the whole set of
simulations, and depends on the fitting scaled
radius, $\Xi$.

The simulated parameters e.g., mass, mean density, 
radius, of the virialized configuration, and 
scaled radius, shall be labeled by the index, 
$trn$, where $trn=200, vir$, according if FM01 
or KLA01 runs are involved.   This is why 
different definitions of radius of the
virialized configuration, have been used in 
FM01 and KLA01.

The scaling density, $\rho_0$, and the scaling
radius, $r_0$, are taken as fundamental in 
KLA01, and then no counterpart to Eqs.\,(\ref
{seq:rho0}) and (\ref{seq:r0}) is provided
therein.   On the other hand, the results of
numerical simulations (e.g., NFW97; FM01; KLA01;
Bullock et al. 2001; FM03) provide additional 
support to the idea,
that density profiles of dark matter haloes in
hierarchically clustering universes have the
same shape, independent of the halo mass, the
initial density perturbation spectrum, and the
values of cosmological parameters.   Accordingly,
we suppose that Eqs.\,(\ref{seq:rho0}) and (\ref
{seq:r0}) hold even in averaging scaled density
profiles from KLA01
simulations, but different values must be
assigned to the coefficients, $C_\rho$ and $C_r$.
In doing this, the procedure will depend on
the density profile (NFW or MOA) under 
consideration.

With regard to NFW density profiles, the 
scaled radius, $\xi_{vir}=r_{vir}/r_0$,
is provided for each run in KLA01, and
the mean density of the virialized 
configuration, $\bar{\rho}_{vir}=3M_
{vir}/(4\pi r_{vir}^3)$, together with
the scaling radius, $r_0$, may be deduced
from KLA01 results.
% 
%the results of Table~\ref{t:SKL}.
%The above mentioned parameters are listed
%in Table~\ref{t:CKA}, together with two
%dimensionless parameters, $\delta$ and 
%$\kappa_{vir}$, which are deduced from 
%Eqs.\,(\ref{seq:r0}), (\ref{eq:k1}),
%and (\ref{eq:k2}).
%
%It is apparent that t
%
The twelve runs from KLA01
correspond to virial masses of the 
same order, which allows a comparison 
with four runs executed by FM01 i.e. $4M0$, 
$2M0$, $2M1$, $2M2$, where the virial
masses are also of the same order.

It can be seen that the related, averaged, 
scaling radius is:
\begin{equation}
\label{eq:r0ma}
(\bar{r}_0)_{NFW,KLA}=25.675~{\rm kpc}~~; 
\qquad(\bar{r}_0)_{NFW,FM}=12.87~{\rm kpc}~~;
\end{equation}
and the ratio equals two within the uncertainty
of the results. 
%
%listed in Tables~\ref{t:SKL} and \ref{t:CKA}.   
%
Then we assume that the
value of the constant, $C_r$, appearing in
Eqs.\,(\ref{seq:r0}), doubles its counterpart
related to FM01 simulations, that is:
\begin{equation}
\label{eq:Crrh}
(C_r)_{NFW,KLA}=4~10^{-2/3}k_1^{-1}~~;\qquad
(C_r)_{MOA,KLA}=4~10^{-2/3}~~.
\end{equation}

The parameters, $M_{vir}$, $r_{vir}$, and $\delta$,
are intrinsic to simulations (e.g., FM01) and for
this reason do not depend on the fitting density
profile.   Accordingly,
the combination of Eqs.\,(\ref{eq:r0a}) and (\ref
{eq:Crrh}) yields:
\begin{equation}
\label{eq:r0MN}
(r_0)_{MOA}=k_1(r_0)_{NFW}~~;
\end{equation}
where $k_1=2.275$ according to Caimmi \& Marmo
(2003, Appendix B).

The fitting scaled radius, $\Xi$, is determined by
averaging the results from KLA01 runs, as:
\begin{equation}
\label{eq:CSIKN}
\Xi=(\bar{\xi}_{vir})_{NFW}=13.50833~~;
\end{equation}
in the mass range $(0.68-2.10)\times10^{12}
h^{-1}{\rm M}_\odot$, which is consistent with
the expected value of the lognormal distribution
deduced from a statistical sample of about two
thousands of dark matter haloes in the mass
range $(0.5-1.0)\times10^{12}h^{-1}{\rm M}_\odot$
(Bullock et al. 2001).

Then the combination of Eqs.\,(\ref{eq:MM0}) 
and (\ref{eq:Crrh}) produces:
\begin{equation}
\label{eq:CrhoN}
(C_\rho)_{NFW,KLA}=\frac3{4\pi}\left[(\nu_M)_{NFW}
(C_r)^3_{NFW,KLA}\right]^{-1}~~;
\end{equation}
where $M=\nu_MM_0$ according to the results 
listed in Table\,\ref{t:proMN}.   The
combination of Eqs.\,(\ref{eq:CSIKN}) and
(\ref{eq:CrhoN}) yields $(C_\rho)_{NFW,KLA}=0.209911$.   
Replacing the truncation radius with the virial 
radius, the following relation is deduced from 
Table\,\ref{t:proMN}:
\begin{equation}
\label{eq:Mvir}
M_{vir}=12M_0\left[\ln(1+\xi_{vir})-\frac{\xi_
{vir}}{1+\xi_{vir}}\right]~~;
\end{equation}
which allows the calculation of the scaling
mass, $M_0$, the fitting mass, $M$, and then
the remaining parameters, for each run in
connection with NFW density profiles.

In dealing with MOA density profiles, the
constant, $C_r$, and the scaling radius,
$r_0$, are expressed by Eqs.\,(\ref{eq:Crrh}) 
and (\ref{eq:r0MN}), respectively, and the
constant, $C_\rho$, takes the expression:
\begin{equation}
\label{eq:CrhoM}
(C_\rho)_{MOA,KLA}=\frac2{k_1^3}(C_\rho)_
{NFW,KLA}=\frac{75}{32\pi}(\nu_M)_{NFW}^{-1}~~;
\end{equation}
where the profile parameter, $\nu_M$, may be 
calculated using the results listed in
Table\,\ref{t:proMN}
together with Eq.\,(\ref{eq:CSIKN}), yielding
$(C_\rho)_{MOA,KLA}=0.0356551$.   The combination of  
Eqs.\,(\ref{eq:MM0}), (\ref{eq:Crrh}), and 
(\ref{eq:CrhoM}) produces:
\begin{equation}
\label{eq:nuMN}
(\nu_M)_{MOA}=\frac12(\nu_M)_{NFW}~~;
\end{equation}
which implies $(M_0)_{MOA}=2(M_0)_{NFW}$
provided the fitting mass, $M$, is kept fixed
passing from MOA to NFW density profiles 
and vice versa.    To this respect, it has 
already been said that the parameters, 
$M_{vir}$, $r_{vir}$, and $\delta$, are 
also left unchanged.   Then the remaining
parameters may be calculated following
the same procedure used in connection
with FM01 simulations, as outlined in 
Appendix B.
%\ref{k1}.

\subsection{Fitting, scaled dark matter haloes
related to MOA and NFW density profiles}
\label{siNFW}
With regard to MOA density profiles, the 
combination of Eqs.\,(\ref{eq:rho0b}), 
(\ref{eq:r0b}) and (\ref{eq:MM0}) yields:
\begin{equation}
\label{eq:snuMM}
\nu_M=\frac{375}{14\pi}~~;
\end{equation}
and the comparison with the explicit
expression of the profile factor,
$\nu_M$, listed in Table\,\ref{t:proMN},
produces:
\begin{equation}
\label{eq:sCsiM}
\Xi=\left[\exp\left(\frac{375}{56\pi}\right)
-1\right]^{2/3}~~.
\end{equation}

With regard to NFW density profiles, similarly, 
we have:
\begin{equation}
\label{eq:snuMN}
\nu_M=\frac{375}{7\pi}~~;
\end{equation}
and the comparison with the explicit expression
of the profile factor, $\nu_M$, listed in
Table\,\ref{t:proMN}, yields:
\begin{equation}
\label{eq:sCsiN}
\frac1{1+\Xi}-\ln\frac1{1+\Xi}=1+\frac{125}{28\pi}~~.
\end{equation}

The knowledge of the scaled radius,
$\Xi$, via Eqs.\,(\ref{eq:sCsiM}) and (\ref
{eq:sCsiN}), allows the calculation of the 
profile parameters, $\nu_{\bar{\rho}}$, 
$\nu_{eq}$, $\nu_I$, $\nu_{sel}$, $\nu_J$,
according to the explicit expressions
listed in Table \ref{t:proMN}.   The
related physical parameters, $\bar
{\rho}$, $v_{eq}$, $I$, $E_{sel}$,
$J$, may, in turn, be
calculated, according to the definitions
listed in Table \ref{t:proMN} and via 
Eqs.\,(\ref{seq:rho0}), (\ref{seq:r0}),
provided the independent parameters,
$(M,\delta)$, are assigned.

The scaled radius, $\xi_{max}$, where either
NFW or MOA velocity profile, $v_{eq}(\xi)$,
related to
centrifugal support along a selected radial 
direction, attains its maximum value, may be
calculated by applying the standard methods
of analysis and then solving a transcendental
equation.

Numerical values of the scaled
radius, $\Xi$, the scaled radius where 
maximum centrifugal support along a selected,
radial direction occurs, $\xi_{max}$, and the
profile parameters, $\nu_\rho$, $\nu_{\bar
{\rho}}$, $\nu_M$, $\nu_{eq}$, $\nu_I$, $\nu_
{pot}$, $\nu_J$, $\nu_{rot}$, are listed in
Table\,\ref{t:pron} for fitting NFW and MOA
density profiles, to both FM01 and KLA01
simulations.
\begin{table}
\begin{tabular}{lllll}
\hline
\hline
\multicolumn{1}{c}{parameter} &
\multicolumn{2}{c}{FM01} & 
%\multicolumn{1}{c}{FM01} & 
\multicolumn{2}{c}{KLA01} \\
%\multicolumn{1}{c}{KLA01} \\
%\multicolumn{1}{c}{RC}  &
%\multicolumn{1}{c}{$l$} &
\multicolumn{1}{c}{} &
\multicolumn{1}{c}{NFW$\quad$}  &
\multicolumn{1}{c}{MOA$\quad$} &
\multicolumn{1}{c}{NFW$\quad$} &
\multicolumn{1}{c}{MOA$\quad$} \\
%\hline

\hline
$\Xi$              & $\phantom{1}$9.20678    & 3.80693 & 13.50833 & $\phantom{1}$5.43586\\
$\xi_{max}$        & $\phantom{1}$2.16258    & 1.24968 & $\phantom{1}$2.16258 & $\phantom{1}$1.24968 \\
$\nu_\rho$         & $\phantom{1}$0.00417037 & 0.0319486 & $\phantom{1}$0.00140677 & $\phantom{1}$0.0115409 \\
$\nu_{\bar{\rho}}$ & $\phantom{1}$0.0218504  & 0.154536 & $\phantom{1}$0.00848858 & $\phantom{1}$0.0651335\\
$\nu_M$            & 17.05231                & 8.52616 & 20.92379 & 10.46189\\
$\nu_{eq}$         & $\phantom{1}$0.893929   & 0.898766 & $\phantom{1}$0.647442 & $\phantom{1}$0.833159\\
$\nu_I$            & $\phantom{1}$0.0554130  & 0.0892287 & $\phantom{1}$0.0498080 & $\phantom{1}$0.0799089\\
$\nu_{sel}$        & $\phantom{1}$1.12890    & 0.561202 & $\phantom{1}$1.10549 & $\phantom{1}$0.632324\\
$\nu_J$            & $\phantom{1}$0.139180   & 0.146741 & $\phantom{1}$0.128641 & $\phantom{1}$0.135717 \\
$\nu_{rot}$        & $\phantom{1}$0.333333   & 0.333333 & $\phantom{1}$0.333333 & $\phantom{1}$0.333333 \\
$\kappa$           & $\phantom{1}$2.30269    & 2.10091 & 11.57498 & 10.13895 \\
\hline\hline
\end{tabular}
\caption{Values
of the scaled radius, $\Xi$, the 
scaled radius where maximum centrifugal support 
along a selected, radial direction occurs, 
$\xi_{max}$, the profile parameters, $\nu_\rho$, 
$\nu_{\bar{\rho}}$, $\nu_M$, $\nu_{eq}$, $\nu_I$, 
$\nu_{sel}$, $\nu_J$, $\nu_{rot}$, and the
dimensionless parameter, $\kappa$, related to
fitting NFW and MOA density profiles, to both
FM01 and KLA01 simulations.   The profile
parameters, $\nu_J$ and $\nu_{rot}$, are related to
the special case of constant rotational velocity on
the equatorial plane.   Rigidly rotating
configurations correspond to $\nu_J=\nu_I$, $\nu_
{rot}=\nu_I$. The
profile parameter, $\nu_{eq}$, attains values
which are very close each to the other, for FM01
simulations.}
\label{t:pron}
\end{table}
The profile parameter, $\nu_{eq}$, appears
to attain values which are very close each
to the other, for FM01 simulations.

\subsection{Deviation of simulated dark
matter haloes from their fitting counterparts}
\label{cse}

The above results allow 
the comparison between simulated and fitting,
dark matter haloes.   
Values of some relevant parameters related to 
simulations with high resolution, performed 
by FM01, are listed in Table\,\ref{t:SFM}.
%, and \ref{t:SFMM}
%, in connection with NFW and
%MOA density profiles, respectively.
\begin{table}
\begin{tabular}{rlllrllr}
\hline
\hline
\multicolumn{1}{c}{run} &
\multicolumn{1}{c}{$m/{\rm M}_{10}$} &
\multicolumn{1}{c}{$z_{start}$} &
\multicolumn{1}{c}{$z_{end}$} &
\multicolumn{1}{c}{$N_{200}$} & 
\multicolumn{1}{c}{$M_{200}/{\rm M}_{10}$}&
\multicolumn{1}{c}{$r_{200}/{\rm kpc}$} & 
\multicolumn{1}{c}{$\delta^{-1}$} \\
\hline
$16M0$ & $3.0~10^{-2}$ & 18.8 & 0.0 & 873170 &
$2.6~10^4$ & $1.7~10^3$ & 1.0 \\
$16M1$ & $6.0~10^{-2}$ & 18.5 & 0.0 & 1279383 &
$7.8~10^4$ & $2.4~10^3$ & 0.4 \\
$16M2$ & $6.1~10^{-2}$ & 20.4 & 0.0 & 1322351 &
$8.0~10^4$ & $2.4~10^3$ & 0.6 \\
$8M0$ & $3.7~10^{-3}$ & 22.3 & 0.58 & 745735 &
$2.8~10^3$ & $4.8~10^2$ & 2.5 \\
$8M1$ & $7.6~10^{-3}$ & 22.2 & 0.63 & 1186162 &
$9.0~10^3$ & $7.2~10^2$ & 1.0 \\
$8M2$ & $7.6~10^{-3}$ & 23.9 & 0.59 & 1015454 &
$7.7~10^3$ & $7.0~10^2$ & 3.0 \\
$4M0$ & $4.7~10^{-4}$ & 25.9 & 1.6 & 559563 &
$2.7~10^2$ & $1.3~10^2$ & 10.0 \\
$4M1$ & $9.5~10^{-4}$ & 25.9 & 1.6 & 846301 &
$8.0~10^2$ & $2.0~10^2$ & 3.0 \\
$4M2$ & $9.5~10^{-4}$ & 27.4 & 1.2 & 697504 &
$6.6~10^2$ & $2.2~10^2$ & 6.0 \\
$2M0$ & $5.9~10^{-5}$ & 29.7 & 2.1 & 643151 &
$6.6~10^1$ & $6.2~10^1$ & 35.0 \\
$2M1$ & $1.2~10^{-4}$ & 29.7 & 2.2 & 957365 &
$1.1~10^2$ & $8.5~10^1$ & 12.0 \\
$2M2$ & $1.2~10^{-4}$ & 30.9 & 1.8 & 923545 &
$1.0~10^2$ & $9.6~10^1$ & 30.0 \\
\hline\hline
\end{tabular}
\caption{Values of some parameters related to 
simulated, dark matter haloes, according to FM01,
for a standard CDM model with $H_0=50~$km s$
^{-1}~$Mpc$^{-1}$, $\Omega=1$, and $\sigma_8=0.7$.   
The particle masses are equal, and the total number
of particles for each simulation is $(2.0-2.1)
\times10^6$.
Captions: $m$ - mass of a single particle; $z$ -
redshift (at the start and end of simulation);
$N_{200}$ - total number of particles within the
sphere where $\bar{\rho}=200\rho_{crit}$ ($\rho_
{crit}$ is the critical density); $M_{200}$ -
mass enclosed within the above mentioned sphere;
$r_{200}$ - radius of the above mentioned sphere;
$\delta$ - dimensionless parameter related to the
amplitude of the density perturbation at the 
collapse.   The mass unit is ${\rm M}_{10}=10^
{10}{\rm M}_\odot$.
\label{t:SFM}}
\end{table}

\begin{table}
\begin{tabular}{rlllccrlr}
\hline
\hline
\multicolumn{1}{c}{run} &
\multicolumn{1}{c}{$\frac m{{\rm M}_{10}}$} &
\multicolumn{1}{c}{$z_{end}$} &
\multicolumn{1}{c}{$N_{vir}$} &
\multicolumn{1}{c}{$\left(\frac{N_{vir}m}{{\rm M}_{10}}
\right)^+$} &
\multicolumn{1}{c}{$\left(\frac{N_{vir}m}{{\rm M}_{10}}
\right)^-$} &
\multicolumn{1}{c}{$\frac{M_{vir}}{{\rm M}_{10}}$} & 
\multicolumn{1}{c}{$\frac{r_{vir}}{{\rm kpc}}$} &
\multicolumn{1}{c}{?}\\
\hline
$A1$ & $2.3~10^{-3}$ & 0 & $1.2~10^5$ & 294 & 
259 & 286 & 367 & Y \\
$A2$ & $1.9~10^{-2}$ & 0 & $1.5~10^4$ & 302 & 
268 & 300 & 373 & Y \\
$A3$ & $1.6~10^{-1}$ & 0 & $1.9~10^3$ & 322 & 
287 & 286 & 366 & N \\
$B1$ & $1.7~10^{-4}$ & 0 & $1.0~10^6$ & 184 & 
157 & 171 & 307 & Y \\
$B2$ & $1.7~10^{-4}$ & 0 & $1.5~10^4$ & $\phantom{2006}$2.71 & 
$\phantom{2006}$2.39 & 157 & 304 & N \\
$B3$ & $1.7~10^{-4}$ & 1 & $7.1~10^5$ & 125 & 
116 & 121 & 344 & Y \\
$C1$ & $1.7~10^{-4}$ & 0 & $1.1~10^6$ & 201 & 
173 & 186 & 321 & Y \\
$C2$ & $1.1~10^{-2}$ & 0 & $1.6~10^4$ & 190 & 
163 & 171 & 314 & Y \\
$C3$ & $1.7~10^{-4}$ & 1 & $5.0~10^5$ & $\phantom{2}$88 & 
$\phantom{2}$82 &  97 & 297 & N \\
$D1$ & $1.7~10^{-4}$ & 0 & $1.3~10^6$ & 236 & 
206 & 214 & 336 & Y \\
$D2$ & $1.1~10^{-2}$ & 0 & $2.0~10^4$ & 236 & 
205 & 214 & 334 & Y \\
$D3$ & $1.7~10^{-4}$ & 1 & $7.9~10^5$ & 139 & 
129 & 137 & 350 & Y \\
\hline\hline
\end{tabular}
\caption{Values of some parameters related to 
simulated, dark matter haloes, according to 
KLA01, for a $\Lambda$CDM model with $H_0=
70$ km s$^{-1}$ Mpc$^{-1}$, $\Omega_0=1-
\Lambda_0=0.3$, and $\sigma_8=0.9$.
All simulations were started at $z_{start}=
60$.   Captions: $m$ - mass of a single
particle; $z_{end}$ - redshift at the end of
simulation; $N_{vir}$ - total number of 
particles within the sphere where $\bar
{\rho}=\rho_{crit}\Omega_0\delta_{TH}$ 
($\rho_{crit}$ is the critical density and
$\delta_{TH}$ is the density excess predicted 
by the top-hat model); $M_{vir}$ - total mass
within the above mentioned sphere; $r_{vir}$
- radius of the above mentioned sphere; $(N_
{vir}m/{\rm M}_{10})^\mp$ - upper and lower value
of $N_{vir}m/{\rm M}_{10}$ deduced from the data.   
The mass unit is ${\rm M}_{10}=10^{10}~{\rm M}_\odot$.
A positive answer to the question mark means 
that the inequality, $M_{vir}^-\le M_{vir}\le
M_{vir}^+$ is satisfied.   The related,
explicit expression, is shown by Eqs.\,(\ref
{seq:disKl}).
\label{t:SKL}}
\end{table}

The framework is a standard CDM model with $H_0=
50$~km~s$^{-1}~$Mpc$^{-1}$, $\Omega=1$, and $\sigma_
8=0.7$.    The particle masses are equal, and the 
total number of particles for each simulation is 
$(2.0-2.1)\times10^6$.   For further details, see FM01.
The ending redshift, $z_{end}>0$, is determined
so that the truncation outside the sphere did not
influence the profile around $r_{200}$.   Then the
data listed in Table\,\ref{t:SFM} do not make a
homogeneous set, as simulations related to $z_{end}
=0$ satisfy a different condition in respect of
$z_{end}>0$.  To this aim, computations related to
$z_{end}=0$ would have been continued until the
above mentioned condition would be satisfied at
some $z_{end}<0$ in the future.

Values of some relevant parameters related to 
simulations with high resolution, performed
by KLA01, are listed in Table\,\ref{t:SKL}.
%in connection with NFW and MOA density 
%profiles, respectively.
The framework is a $\Lambda$CDM model with $H_0=
70$~km~s$^{-1}$~Mpc$^{-1}$, $\Omega_0=1-\Lambda_
0=0.3$, and $\sigma_8=0.9$.   All
simulations were started at $z_{start}=60$.
The twelve runs are, in fact, three sets
of simulations of only four haloes with
resolution varied in each set.   Thought
they cannot be considered as twelve 
independent runs, still they may be
conceived as measures of a same physical 
quantity, using different methods.
For further details, see KLA01.

It is worth of note that the virial radius in 
KLA01 is not defined as the radius, $r_{200}$,
within which the mean density is 200 times the
critical density, as done by e.g., NFW97 and 
FM01.   On the other hand, the virial radius 
is intended therein as the radius, $r_{vir}$, 
within which the mean density is equal to the
density predicted by the top-hat model, $\bar
{\rho}_{vir}=\delta_{TH}\Omega_0\rho_{crit}$,
where $\delta_{TH}$ is the density excess
predicted by the top-hat model and $\rho_
{crit}$ is the critical density.   In the
case of $\Omega_0=0.3$ cosmologies, it can
be seen that $r_{vir}\approx1.3r_{200}$ 
(KLA01).

In general, the scaled radius, $\Xi$ (or
concentration with regard to NFW density
profiles), is lowered for increasing
mass (constant redshift) and redshift (constant
mass).   The mass range is large $(6\cdot10^{11}
<M/{\rm M}_\odot\le8\cdot10^{14})$ for FM01 simulations,
but the use of a CDM cosmological model makes the
concentration mildly depend on the mass (at constant
redshift).   On the other hand, the concentration is 
decreased for low-mass haloes, which virialize earlier,
and increased for high-mass haloes, which virialize
later.   The net effect is an even milder dependence
of the concentration on the mass.

In any case, it can be said that the intrinsic spread
in concentration is dominant over the systematic
change in the mean value of the concentration on a
mass bin (e.g., Bullock et al. 2001), across the entire
mass ranges covered by FM01 and KLA01 simulations.
Accordingly, the fitting, scaled
density profile may be considered, to an acceptable
extent, as related to an invariant scaled radius,
$\Xi$.

The following inequalities may be useful for testing
the intrinsic spread of some data listed in 
Table\,\ref{t:SFM}:
\begin{leftsubeqnarray}
\slabel{eq:cricoa}
&& M_{200}^-\le M_{200}\le M_{200}^+~~;\quad \bar
{\rho}_{200}^-\le200\rho_{crit}(z_{end})\le\bar{\rho}_
{200}^+~~; \\
\slabel{eq:cricoab}
&& \bar{\rho}_{200}^-\le\bar{\rho}_{200}
\le\bar{\rho}_{200}^+~~;\quad\kappa_{200}^-\le\kappa_
{200}\le\kappa_{200}^+~~; \\
\slabel{eq:cricob}
&& M_{200}^\mp=(N_{200}\mp\Delta N_{200})(m\mp\Delta 
m)~~; \\
\slabel{eq:cricoc}
&& \bar{\rho}_{200}^\mp=\bar{\rho}_{200}\left[\frac
{1\mp\Delta M_{200}/M_{200}}{(1\pm\Delta r_{200}/r_
{200})^3}\right]~~; \\
\slabel{eq:cricod}
&& \kappa_{200}^\mp=(\delta\mp\Delta\delta)\frac
{r_{200}\mp\Delta r_{200}}{{\rm kpc}}\left(\frac{M_{200}^
\pm}{{\rm M}_{10}}\right)^{-1}~~; \\
\slabel{eq:cricoe}
&& \bar{\rho}_{200}=\frac3{4\pi}\frac{M_{200}}{r_{200}^3}~~; \\
\slabel{eq:cricof}
&& \rho_{crit}(z)=0.691785~10^{-8}(1+z)^3{\rm M}_{10}{\rm kpc}
^{-3}~~; \\
\slabel{eq:cricog}
&& \kappa_{200}=\delta^{1/2}\left(\frac{r_{200}}{{\rm kpc}}
\right)^{3/2}\left(\frac{M_{200}}{{\rm M}_{10}}\right)^{-1}~~;
\label{seq:crico}
\end{leftsubeqnarray}
where $z$ is the redshift, ${\rm M}_{10}=10^{10}{\rm M}_\odot$
and, in general, $\Delta\eta=5\cdot10^{-n-1}$ is the
uncertainty assumed for $\eta=u~10^{-n}$, $0\le u
<10$.   Upper and lower values are listed in
Table\,\ref{t:CFM}.
\begin{table}
\begin{tabular}{rrrrcllll}
\hline
\hline
\multicolumn{1}{c}{run} &
\multicolumn{1}{c}{$\frac{M_{200}}{{\rm M}_{10}}$} &
\multicolumn{1}{c}{$\left(\frac{N_{200}m}{{\rm M}_{10}}
\right)^+$} &
\multicolumn{1}{c}{$\left(\frac{N_{200}m}{{\rm M}_{10}}
\right)^-$} &
\multicolumn{1}{c}{?} &
\multicolumn{1}{c}{$\kappa_{200}~$} & 
\multicolumn{1}{c}{$\kappa_{200}^+$} &
\multicolumn{1}{c}{$\kappa_{200}^-$} &
\multicolumn{1}{c}{?} \\
\hline
$16M0$ & 26000$\quad~$ & 26632$\quad$ & 25758$\quad$ & Y & $\quad$2.70 & 2.92 & 2.46 & Y \\
$16M1$ & 78000$\quad~$ & 77403$\quad$ & 76123$\quad$ & N & $\quad$2.38 & 2.69 & 2.19 & Y \\
$16M2$ & 80000$\quad~$ & 81325$\quad$ & 80002$\quad$ & N & $\quad$1.90 & 2.04 & 1.74 & Y \\
$8M0$  & 2800 $\quad~$ & 2796 $\quad$ & 2722 $\quad$ & N & $\quad$2.37 & 2.51 & 2.32 & Y \\
$8M1$  & 9000 $\quad~$ & 9074 $\quad$ & 8955 $\quad$ & Y & $\quad$2.15 & 2.24 & 2.06 & Y \\
$8M2$  & 7700 $\quad~$ & 7768 $\quad$ & 7667 $\quad$ & Y & $\quad$1.39 & 1.42 & 1.37 & Y \\
$4M0$  & 270  $\quad~$ & 266  $\quad$ & 260  $\quad$ & N & $\quad$1.74 & 1.91 & 1.66 & Y \\
$4M1$  & 800  $\quad~$ & 808  $\quad$ & 800  $\quad$ & Y & $\quad$2.04 & 2.14 & 1.93 & Y \\
$4M2$  & 660  $\quad~$ & 666  $\quad$ & 659  $\quad$ & Y & $\quad$2.02 & 2.10 & 1.92 & Y \\
$2M0$  & 66   $\quad~$ & 38   $\quad$ & 38   $\quad$ & N & $\quad$1.25 & 2.20 & 2.14 & N \\
$2M1$  & 110  $\quad~$ & 120  $\quad$ & 110  $\quad$ & Y & $\quad$2.06 & 2.08 & 1.86 & Y \\
$2M2$  & 100  $\quad~$ & 115  $\quad$ & 106  $\quad$ & N & $\quad$1.72 & 1.63 & 1.48 & N \\
\hline\hline
\multicolumn{1}{c}{run} &
\multicolumn{1}{c}{$\frac{200\rho_{crit}(z_{end})}
{{\rm M}_{10}{\rm kpc}^{-3}}$} & 
\multicolumn{1}{c}{$\frac{\bar{\rho}_{200}^+}
{{\rm M}_{10}{\rm kpc}^{-3}}$} &
\multicolumn{1}{c}{$\frac{\bar{\rho}_{200}^-}
{{\rm M}_{10}{\rm kpc}^{-3}}$} &
\multicolumn{1}{c}{?} &
\multicolumn{1}{c}{$\frac{\bar{\rho}_{200}}
{{\rm M}_{10}{\rm kpc}^{-3}}$} &
\multicolumn{1}{c}{?} & & \\
\hline
$16M0$ & $1.38~10^{-6}\quad$ & $1.41~10^{-6}$ & $1.15~10^{-6}$ & Y & $1.26~10^{-6}$ & ~~Y & & \\
$16M1$ & $1.38~10^{-6}\quad$ & $1.42~10^{-6}$ & $1.24~10^{-6}$ & Y & $1.35~10^{-6}$ & ~~Y & & \\
$16M2$ & $1.38~10^{-6}\quad$ & $1.50~10^{-6}$ & $1.30~10^{-6}$ & Y & $1.38~10^{-6}$ & ~~Y & & \\
$ 8M0$ & $5.46~10^{-6}\quad$ & $6.23~10^{-6}$ & $5.70~10^{-6}$ & N & $6.04~10^{-6}$ & ~~Y & & \\
$ 8M1$ & $5.99~10^{-6}\quad$ & $5.93~10^{-6}$ & $5.61~10^{-6}$ & N & $5.76~10^{-6}$ & ~~Y & & \\
$ 8M2$ & $5.56~10^{-6}\quad$ & $5.52~10^{-6}$ & $5.22~10^{-6}$ & N & $5.36~10^{-6}$ & ~~Y & & \\
$ 4M0$ & $2.43~10^{-5}\quad$ & $3.25~10^{-5}$ & $2.52~10^{-5}$ & N & $2.93~10^{-5}$ & ~~Y & & \\
$ 4M1$ & $2.43~10^{-5}\quad$ & $2.60~10^{-5}$ & $2.22~10^{-5}$ & Y & $2.39~10^{-5}$ & ~~Y & & \\
$ 4M2$ & $1.47~10^{-5}\quad$ & $1.60~10^{-5}$ & $1.38~10^{-5}$ & Y & $1.48~10^{-5}$ & ~~Y & & \\
$ 2M0$ & $4.12~10^{-5}\quad$ & $3.90~10^{-5}$ & $3.72~10^{-5}$ & N & $6.61~10^{-5}$ & ~~N & & \\
$ 2M1$ & $4.53~10^{-5}\quad$ & $4.75~10^{-5}$ & $4.20~10^{-5}$ & Y & $4.28~10^{-5}$ & ~~Y & & \\
$ 2M2$ & $3.04~10^{-5}\quad$ & $3.15~10^{-5}$ & $2.82~10^{-5}$ & Y & $2.70~10^{-5}$ & ~~N & & \\
\hline\hline
\end{tabular}
\caption{Comparison between (i) the value of 
$M_{200}$ and the upper and lower value of 
$N_{200}m$, deduced from the data listed in
Table\,\ref{t:SFM}; (ii) the value of $\kappa
_{200}$ deduced 
from the data listed in Table\,\ref{t:SFM}
and the related upper and lower values, using 
Eqs.\,(\ref{eq:cricod}) and (\ref{eq:cricog}),
respectively; and (iii) the value of $200
\rho_{crit}(z_{end})$ calculated using 
Eq.\,(\ref{eq:cricof}), the upper and lower
value of $\bar{\rho}_{200}=3N_{200}m/(4\pi
r_{200}^3)$, and the value of $\bar{\rho}_
{200}=3M_{200}/(4\pi r_{200}^3)$, deduced 
from the data listed in Table\,\ref{t:SFM}.  
A positive answer to the question mark means 
that the related inequalities, among
Eqs.\,(\ref{eq:cricoa}) and (\ref{eq:cricob}),
are satisfied
for the values listed on the same lines of the
corresponding columns.
%$M_{200}^-\le M_{200}\le M_{200}^+$; 
%$\kappa_{200}^-\le\kappa_{200}\le\kappa_{200}^+$;
%$\bar{\rho}_{200}^-\le200\rho_
%{crit}(z_{end})\le\bar{\rho}_{200}^+$; 
%$\bar{\rho}_{200}^-\le\bar{\rho}_
%{200}\le\bar{\rho}_{200}^+$.
The related, explicit expressions, are shown by
Eqs.\,(\ref{eq:cricob})-(\ref{eq:cricog})
%, (\ref{eq:cricod}), (\ref{eq:cricof}), respectively
.}
\label{t:CFM}
\end{table}
A positive answer to a question mark therein,
with regard to the parameter on the
left of the column under consideration, means 
that the related inequalities, among (\ref{eq:cricoa})
and (\ref{eq:cricoab}), are satisfied, and
vice versa.

Similarly, for testing the intrinsic spread of masses 
listed in Table\,\ref{t:SKL}:
\begin{leftsubeqnarray}
\slabel{eq:disKla}
&& M_{vir}^-\le M_{vir}\le M_{vir}^+~~; \\
\slabel{eq:disKlb}
&& M_{vir}^\mp=(N_{vir}\mp\Delta N_{vir})(m\mp\Delta m)~~;
\label{seq:disKl}
\end{leftsubeqnarray}
where, in general, $\Delta\eta=5\cdot10^{-n-1}$ is the
uncertainty assumed for $\eta=u~10^{-n}$, $0\le u<10$.   
The upper and lower values are also listed in
Table\,\ref{t:SKL}.   A positive answer to the 
question mark therein means that the inequality
(\ref{eq:disKla}) is satisfied, and
vice versa.    

All the data produce an acceptable intrinsic
spread, with
the exception of run $2M0$, in FM01, and run $B2$,
in KLA01, which exhibit a
substantial inconsistency in any case.   The
larger discrepancies can be due to nothing but 
printing errors.   Accordingly, 
the value $M_{200}/{\rm M}_{10}=38$,
also deduced from FM01, is assigned instead
of 66, to run 2$M$0, and the value $m/{\rm M}_{10}
=1.1~10^{-2}$, also inferred from KLA01, is
assigned instead of $1.7~10^{-4}$, to run $B$2.

\begin{table}
\begin{tabular}{llllllll}
\hline
\hline
\multicolumn{1}{c}{run} &
\multicolumn{1}{c}{$(\xi_{vir})_{NFW}$} &
\multicolumn{1}{c}{$\frac{(r_0)_{NFW}}{\rm {\rm kpc}}$} &
\multicolumn{1}{c}{$\frac{\bar{\rho}_{vir}}{{\rm M}_{10}
{\rm kpc}^{-3}}$} &
%\multicolumn{1}{c}{$\frac{\zeta_r^3}\delta$} &
%\multicolumn{1}{c}{$\frac{\kappa_{vir}}{\zeta_r^{3/2}}$} &
\multicolumn{1}{c}{$\delta^{-1}$} & 
\multicolumn{1}{c}{$\kappa_{vir}$} \\
\hline
$A1$ & 17.4 & 21.1 & $1.38~10^{-6}$ 
%& 0.108 & 62.57 
& $\phantom{1}$2.68 & 16.9 \\
$A2$ & 16.0 & 23.3 & $1.38~10^{-6}$ 
%& 0.132 & 55.17 
& $\phantom{1}$3.05 & 14.9 \\
$A3$ & 16.6 & 22.0 & $1.39~10^{-6}$ 
%& 0.123 & 58.30 
& $\phantom{1}$2.94 & 15.8 \\
$B1$ & 15.6 & 19.7 & $1.41~10^{-6}$ 
%& 0.245 & 53.11 
& $\phantom{1}$5.51 & 14.4 \\
$B2$ & 16.5 & 18.4 & $1.33~10^{-6}$ 
%& 0.239 & 57.78 
& $\phantom{1}$5.66 & 15.6 \\
$B3$ & 12.3 & 28.0 & $7.10~10^{-7}$ 
%& 1.40  & 37.19 
& 25.0 & 10.1 \\
$C1$ & 11.2 & 28.7 & $1.33~10^{-6}$ 
%& 0.646 & 32.31 
& 10.3 & $\phantom{1}$8.74 \\
$C2$ & $\phantom{1}$9.8 & 32.0 & $1.32~10^{-6}$ 
%& 1.06  & 26.45 
& 14.7 & $\phantom{1}$7.15 \\
$C3$ & 11.9 & 25.0 & $8.84~10^{-7}$ 
%& 1.55  & 35.39 
& 26.7 & $\phantom{1}$9.57 \\
$D1$ & 11.9 & 28.2 & $1.35~10^{-6}$ 
%& 0.461 & 35.39 
& $\phantom{1}$7.95 & $\phantom{1}$9.57 \\
$D2$ & 13.4 & 24.9 & $1.37~10^{-6}$ 
%& 0.318 & 42.28 
& $\phantom{1}$6.17 & 11.4 \\
$D3$ & $\phantom{1}$9.5 & 36.8 & $7.64~10^{-7}$ 
%& 2.50  & 25.24 
& 33.7 & $\phantom{1}$6.83 \\
\hline\hline
\end{tabular}
\caption{The scaled, virialized radius  
related to NFW density profiles, $(\xi_
{vir})_{NFW}$, the scaling radius related 
to NFW density profiles, $(r_0)_{NFW}$,
the mean density inside the virialized
configuration, $\bar{\rho}_{vir}$, and
the dimensionless parameters, $\delta^
{-1}$ and $\kappa_{vir}$, taken from 
twelve runs in KLA01 or deduced from
Eqs.\,(\ref{seq:r0}), (\ref{eq:k1}),
and (\ref{eq:k2}).}
\label{t:CKA}
\end{table}

With regard to NFW density profiles, the 
scaled radius, $\xi_{vir}=r_{vir}/r_0$,
is provided for each run in KLA01, and
the mean density of the virialized 
configuration, $\bar{\rho}_{vir}=3M_
{vir}/(4\pi r_{vir}^3)$, together with
the scaling radius, $r_0$, may be deduced
from the results of Table~\ref{t:SKL}.
The above mentioned parameters are listed
in Table~\ref{t:CKA}, together with two
dimensionless parameters, $\delta$ and 
$\kappa_{vir}$, which are deduced from 
Eqs.\,(\ref{seq:r0}), (\ref{eq:k1}),
and (\ref{eq:k2}).     It is apparent that
the twelve runs correspond to virial masses
of the same order.

At this stage, the deviation of simulated,
dark matter haloes, from their fitting
counterparts, may be analysed along the
following steps.

\begin{description}%                          Beccari,  p. 26
\item[\rm{(i)}\hspace{3.5mm}] Select a fitting
halo among NFW and MOA density profiles.
\item[\rm{(ii)}~~] Select a simulated halo among
the twelve runs in Tables\,\ref{t:SFM} and \ref
{t:SKL}.   
\item[\rm{(iii)}~] Calculate the fitting mass, $M$,
using Eqs.\,(\ref{seq:rho0}), (\ref{seq:r0}), 
%(\ref{eq:sCsiM}), 
(\ref{eq:MM0}), and the related values of $M_
{trn}$ and $r_{trn}$, $trn=200, vir$, listed in 
Tables\,\ref{t:SFM} and \ref{t:SKL}.   For a formal 
derivation, see Appendix B.
%\ref{k1}.
\item[\rm{(iv)}\hspace{1.7mm}] Calculate the values 
of the scaling density, $\rho_0$, the scaling radius,
$r_0$, and the scaling mass, $M_0$.   It is worth
remembering the latter is the mass
of a homogeneous region, with same density and
boundary as the reference isopycnic surface,
$(r_0,\rho_0)$.
\item[\rm{(v)}~] Calculate the scaled mass, 
$M_{trn}/M_0$, the scaled radius, $r_{trn}/r_0$,
the scaled density, $\bar{\rho}_{trn}/\rho_0$,
and the dimensionless parameter, $\kappa_{trn}$,
$trn=200, vir$.
\item[\rm{(vi)}\hspace{1.0mm}] Return to (ii).
\item[\rm{(vii)}] Return to (i).
\end{description}

Simulated haloes may be characterized by four 
scaled parameters, $M_{trn}/M_0$, $r_{trn}/
r_0$, $\bar{\rho}_{trn}/\bar{\rho}_0$, and $\kappa_
{trn}$.   Their fitting counterparts are $\nu_M$,
$\Xi$, $\nu_{\bar{\rho}}$, and
$\kappa$, respectively, which have been listed in
Table\,\ref{t:pron}.   The fitting mass,
$M$, the scaling density, radius, mass, $\rho_0$,
$r_0$, $M_0$, and the ratios of three parameters
related to simulated haloes, to their fitting
counterparts, $\kappa_{trn}/\kappa$,
$M_{trn}/M$, $r_{trn}/R$, are listed in Tables \ref
{t:RFMN} and \ref{t:RKAN}, in connection with both 
NFW and MOA density profiles, for the twelwe runs 
from FM01 and KLA01, respectively.

\begin{table}
\begin{tabular}{rlllllll}
\hline
\hline
\multicolumn{1}{c}{run} &
\multicolumn{1}{c}{$\frac M{{\rm M}_{10}}$} &
\multicolumn{1}{c}{$\frac{\rho_0}{{\rm M}_{10}
{\rm kpc}^{-3}}$} &
\multicolumn{1}{c}{$\frac{r_0}{{\rm kpc}}$} &
\multicolumn{1}{c}{$\frac{M_0}{{\rm M}_{10}}$} & 
\multicolumn{1}{c}{$\frac{\kappa_{200}}
{\kappa}$} &
\multicolumn{1}{c}{$\frac{M_{200}}M$} &
\multicolumn{1}{c}{$\frac{r_{200}}R$} \\
\hline
 & & & & NFW & & & \\
\hline
$16M0$ & $2.37~10^4$ & $8.69~10^{-5}$ & $1.56~10^2$ 
& $1.39~10^3$ & 1.17 & 1.10 & 1.18 \\
$16M1$ & $7.64~10^4$ & $6.74~10^{-5}$ & $2.51~10^2$ 
& $4.48~10^3$ & 1.03 & 1.02 & 1.04 \\
$16M2$ & $9.09~10^4$ & $3.78~10^{-5}$ & $3.23~10^2$ 
& $5.33~10^3$ & 0.824 & 0.880 & 0.807 \\
$8M0$ & $2.75~10^3$ & $3.00~10^{-4}$ & $5.04~10^1$ 
& $1.61~10^2$ & 1.03 & 1.02 & 1.03 \\
$8M1$ & $9.41~10^3$ & $2.19~10^{-4}$ & $8.44~10^1$ 
& $5.52~10^2$ & 0.932 & 0.957 & 0.926 \\
$8M2$ & $1.24~10^4$ & $6.11~10^{-5}$ & $1.37~10^2$ 
& $6.59~10^2$ & 0.603 & 0.685 & 0.554 \\
$4M0$ & $3.27~10^2$ & $6.29~10^{-4}$ & $1.94~10^1$ 
& $1.92~10^1$ & 0.754 & 0.825 & 0.728 \\
$4M1$ & $8.64~10^2$ & $7.94~10^{-4}$ & $2.49~10^1$ 
& $5.07~10^1$ & 0.886 & 0.925 & 0.876 \\
$4M2$ & $7.19~10^2$ & $4.78~10^{-4}$ & $2.76~10^1$ 
& $4.21~10^1$ & 0.877 & 0.918 & 0.865 \\
$2M0$ & $3.94~10^1$ & $1.49~10^{-3}$ & $7.18~10^0$ 
& $2.31~10^0$ & 0.943 & 0.964 & 0.938 \\
$2M1$ & $1.18~10^2$ & $1.45~10^{-3}$ & $1.04~10^1$ 
& $6.94~10^0$ & 0.893 & 0.930 & 0.884 \\
$2M2$ & $1.22~10^2$ & $5.62~10^{-4}$ & $1.45~10^1$ 
& $7.17~10^0$ & 0.746 & 0.818 & 0.719 \\
\hline\hline
 & & & & MOA & & & \\
\hline
$16M0$ & $2.22~10^4$ & $1.58~10^{-5}$ & $3.40~10^2$ 
& $2.60~10^3$ & 1.28 & 1.17 & 1.31 \\
$16M1$ & $7.17~10^4$ & $1.22~10^{-5}$ & $5.48~10^2$ 
& $8.41~10^3$ & 1.13 & 1.09 & 1.15 \\
$16M2$ & $8.62~10^4$ & $6.76~10^{-6}$ & $7.09~10^2$ 
& $1.01~10^4$ & 0.903 & 0.928 & 0.889 \\
$8M0$ & $2.58~10^3$ & $5.43~10^{-5}$ & $1.10~10^2$ 
& $3.03~10^2$ & 1.13 & 1.09 & 1.15 \\
$8M1$ & $8.87~10^3$ & $3.55~10^{-5}$ & $1.85~10^2$ 
& $1.05~10^3$ & 1.02 & 1.01 & 1.02 \\
$8M2$ & $1.10~10^4$ & $1.06~10^{-5}$ & $3.08~10^2$ 
& $1.30~10^3$ & 0.661 & 0.698 & 0.597 \\
$4M0$ & $3.13~10^2$ & $1.12~10^{-4}$ & $4.28~10^1$ 
& $3.68~10^1$ & 0.826 & 0.863 & 0.798 \\
$4M1$ & $8.17~10^2$ & $1.43~10^{-4}$ & $5.43~10^1$ 
& $9.59~10^1$ & 0.972 & 0.980 & 0.968 \\
$4M2$ & $6.79~10^2$ & $8.59~10^{-5}$ & $6.05~10^1$ 
& $7.97~10^1$ & 0.961 & 0.972 & 0.955 \\
$2M0$ & $3.71~10^1$ & $8.60~10^{-6}$ & $1.57~10^1$ 
& $4.36~10^0$ & 1.03 & 1.02 & 1.04 \\
$2M1$ & $1.12~10^2$ & $2.61~10^{-4}$ & $2.29~10^1$ 
& $1.31~10^1$ & 0.979 & 0.985 & 0.976 \\
$2M2$ & $1.17~10^2$ & $9.98~10^{-5}$ & $3.20~10^1$ 
& $1.37~10^1$ & 0.817 & 0.856 & 0.788 \\
\hline\hline
\end{tabular}
\caption{The fitting mass,
$M$, the scaling density, radius, mass, $\rho_0$,
$r_0$, $M_0$, and the ratios of three parameters
related to simulated haloes to their fitting
counterparts, $\kappa_{200}/\kappa$,
$M_{200}/M$, $r_{200}/R$, in connection with 
NFW and MOA density profiles, and twelve
runs from FM01.}
\label{t:RFMN}
\end{table}

\begin{table}
\begin{tabular}{lllllll}
\hline\hline
\multicolumn{1}{c}{run} &
\multicolumn{1}{c}{$\frac M{{\rm M}_{10}}$} &
\multicolumn{1}{c}{$\frac{\rho_0}{{\rm M}_{10}
{\rm kpc}^{-3}}$} &
%\multicolumn{1}{c}{$\frac{r_0}{{\rm kpc}}$} &
\multicolumn{1}{c}{$\frac{M_0}{{\rm M}_{10}}$} &
\multicolumn{1}{c}{$\frac{\kappa_{vir}}
{\kappa}$} &
\multicolumn{1}{c}{$\frac{M_{vir}}M$} &
\multicolumn{1}{c}{$\frac{r_{vir}}R$} \\
\hline
\multicolumn{1}{c}{} &
\multicolumn{6}{c}{NFW} \\
% & & & NFW & & & & & & MOA & & & \\
\hline
$A1$ & 254 & $3.08~10^{-4}$ & 12.1 & 1.46 & 1.13 & 1.29 \\
$A2$ & 276 & $2.49~10^{-4}$ & 13.2 & 1.29 & 1.08 & 1.18 \\
$A3$ & 259 & $2.76~10^{-4}$ & 12.4 & 1.36 & 1.10 & 1.23 \\
$B1$ & 159 & $2.39~10^{-4}$ & 7.62 & 1.24 & 1.07 & 1.15 \\
$B2$ & 143 & $2.60~10^{-4}$ & 6.82 & 1.35 & 1.10 & 1.22 \\
$B3$ & 127 & $6.62~10^{-5}$ & 6.06 & 0.869 & 0.954 & 0.910 \\ 
$C1$ & 205 & $9.93~10^{-5}$ & 9.79 & 0.755 & 0.908 & 0.829 \\
$C2$ & 202 & $7.02~10^{-5}$ & 9.68 & 0.618 & 0.844 & 0.725 \\
$C3$ & 103 & $7.59~10^{-5}$ & 4.94 & 0.827 & 0.937 & 0.881 \\
$D1$ & 228 & $1.16~10^{-4}$ & 10.9 & 0.827 & 0.937 & 0.881 \\
$D2$ & 215 & $1.58~10^{-4}$ & 10.3 & 0.988 & 0.996 & 0.992 \\
$D3$ & 165 & $3.77~10^{-5}$ & 7.89 & 0.590 & 0.830 & 0.703 \\
\hline%\hline
\multicolumn{1}{c}{} & \multicolumn{6}{c}{MOA} \\
\hline
%\multicolumn{1}{c}{run} &
%\multicolumn{1}{c}{$\frac M{{\rm M}_{10}}$} &
%\multicolumn{1}{c}{$\frac{\rho_0}{{\rm M}_{10}
%{\rm kpc}^{-3}}$} &
%%\multicolumn{1}{c}{$\frac{r_0}{{\rm kpc}}$} &
%\multicolumn{1}{c}{$\frac{M_0}{{\rm M}_{10}}$} & 
%\multicolumn{1}{c}{$\frac{\kappa_{vir}}
%{\kappa}$} &
%\multicolumn{1}{c}{$\frac{M_{vir}}M$} &
%\multicolumn{1}{c}{$\frac{r_{vir}}R$}\\
%\hline
$A1$ & 236 & $5.63~10^{-5}$ & 22.6 & 1.48 & 1.21 & 1.48 \\
$A2$ & 258 & $4.54~10^{-5}$ & 24.6 & 1.36 & 1.16 & 1.36 \\
$A3$ & 241 & $5.02~10^{-5}$ & 23.1 & 1.41 & 1.18 & 1.41 \\
$B1$ & 149 & $4.35~10^{-5}$ & 14.2 & 1.32 & 1.15 & 1.32 \\
$B2$ & 133 & $4.74~10^{-5}$ & 12.7 & 1.40 & 1.18 & 1.40 \\
$B3$ & 118 & $1.20~10^{-5}$ & 11.3 & 1.04 & 1.02 & 1.04 \\
$C1$ & 191 & $1.81~10^{-5}$ & 18.3 & 0.950 & 0.972 & 0.948 \\
$C2$ & 190 & $1.28~10^{-5}$ & 18.1 & 0.837 & 0.901 & 0.829 \\
$C3$ & 96.6 & $1.38~10^{-5}$ & 9.23 & 1.01 & 1.00 & 1.01 \\
$D1$ & 213 & $2.10~10^{-5}$ & 20.4 & 1.01 & 1.00 & 1.01 \\
$D2$ & 200 & $2.88~10^{-5}$ & 19.2 & 1.13 & 1.07 & 1.13 \\
$D3$ & 155 & $6.83~10^{-6}$ & 14.8 & 0.812 & 0.884 & 0.802 \\
\hline\hline
\end{tabular}
\caption{The fitting mass,
$M$, the scaling density and mass, $\rho_0$,
and $M_0$, and the ratios of three parameters
related to simulated haloes, to their fitting
counterparts, $\kappa_{vir}/\kappa$,
$M_{vir}/M$, $r_{vir}/R$, in connection with 
NFW and MOA density profiles, and twelve
runs from KLA01.}
\label{t:RKAN}
\end{table}

\section{Discussion.}
\label{disc}
\begin{figure}
%\setlength{\unitlength}{1mm}
%\begin{center}
%\makebox[\linewidth]{
%\begin{picture}(120,140)
%\put(0,0){\special{insert masFM.ps}}
%   \centering
%   \resizebox{\hsize}{!}{\includegraphics{masFM.eps}} 
%\end{picture}
%}
%\end{center}
\centerline{\psfig{file=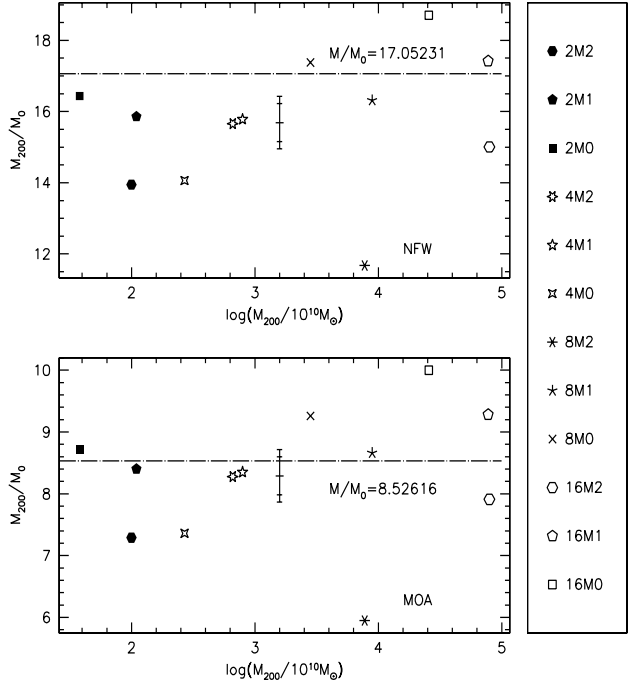,height=100mm,width=90mm}}
%\centerline{\psfig{file=masFM.ps,height=130mm,width=120mm}}
\caption{Deviation of simulated dark matter haloes
from their fitting counterparts, with regard to the
scaled, virial mass,
$M_{200}/M_0$, for both NFW (top) and MOA (bottom) 
density profiles.
%The computed mass, $M_{200}$, is taken as representative
%of its counterpart related to the whole (virialized)
%system, $M$.
Fitting haloes lie on the horizontal
line, while simulated haloes are represented by
different symbols.   The vertical bar is centred
on the mean value of plotted data, with respect to
the ordinates (and no connection with the abscissae), 
and is limited by the standard deviation from the 
mean, without (inner boundary) and with (outer 
boundary) addition of about twice the standard
deviation from the standard deviation from the mean,
deduced from Eqs.\,(\ref{eq:ma}), (\ref{eq:sd}), 
and (\ref{eq:ssd}), respectively.   Captions of
symbols on the right correspond to FM01 runs listed 
in Tables\,\ref{t:SFM}, \ref{t:CFM}, and \ref{t:RFMN}.}
% square-like -
%$iM0$; penthagon-like - $iM1$; exagon-like - 
%$iM2$.   Additional captions: open - $i=16$;
%skeleton - $i=8$; star - $i=4$; filled - $i=2$.}
\label{f:M200FM}    
\end{figure}

\begin{figure}
%   \centering
%   \resizebox{\hsize}{!}{\includegraphics{ragFM.eps}} 
\centerline{\psfig{file=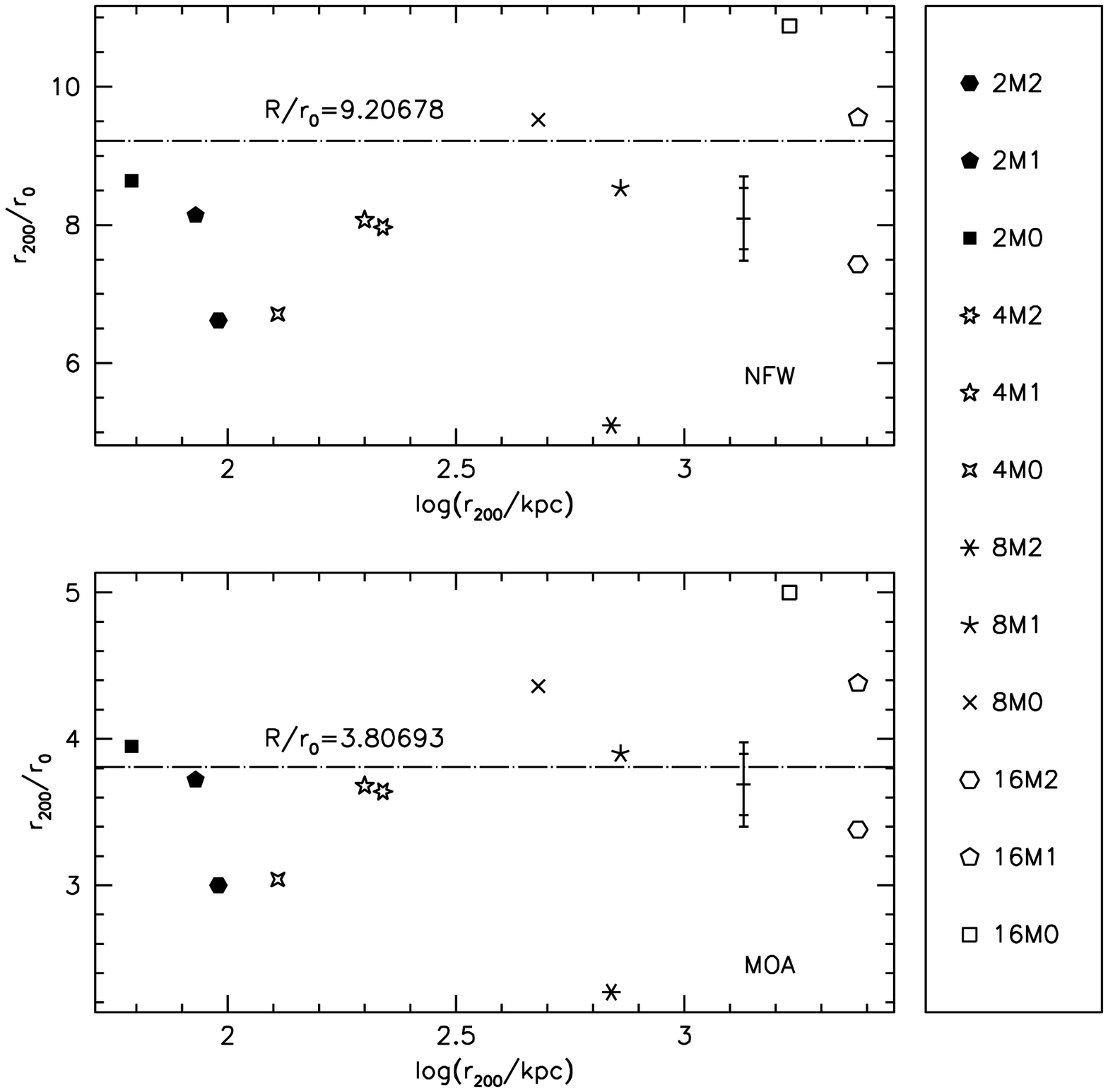,height=100mm,width=90mm}}
%\centerline{\psfig{file=ragFM.ps,height=130mm,width=120mm}}
\caption{Deviation of simulated dark matter haloes
from their fitting counterparts, with regard to the
scaled radius, $r_{200}
/r_0$, for both NFW (top) and MOA (bottom) 
self-similar, universal density profiles.
%The computed radius, $r_{200}$, is taken as representative
%of its counterpart related to the whole (virialized)
%system, $R$.
Other captions as in Fig.\,\ref
{f:M200FM}.}
\label{f:R200FM}    
\end{figure}

\begin{figure}
%   \centering
%   \resizebox{\hsize}{!}{\includegraphics{densFM.eps}} 
\centerline{\psfig{file=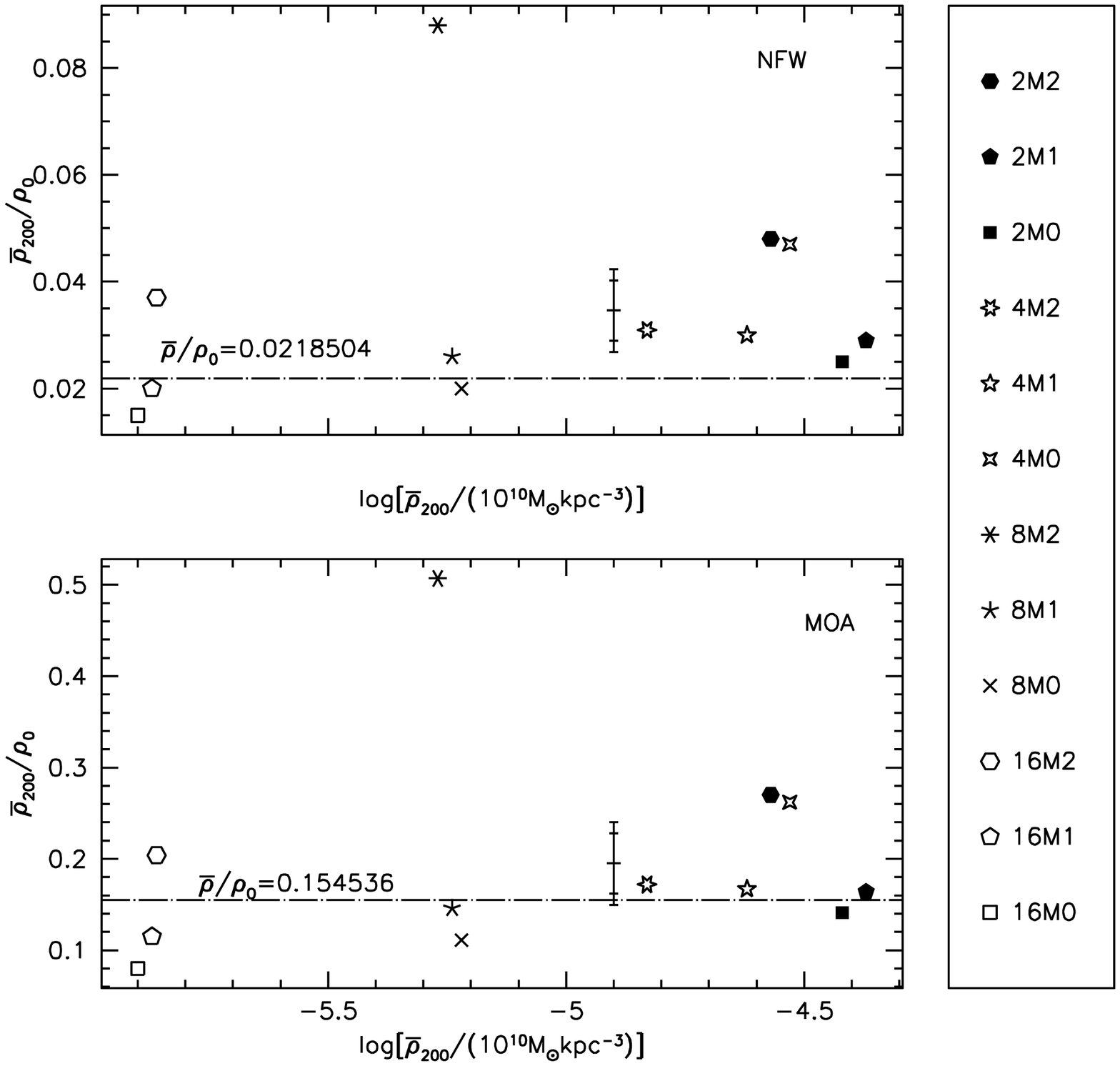,height=100mm,width=90mm}}
%\centerline{\psfig{file=densFM.ps,height=130mm,width=120mm}}
\caption{Deviation of simulated dark matter haloes
from their fitting counterparts, with regard to the
scaled density, $\bar
{\rho}_{200}/\rho_0$, for both NFW (top) and MOA 
(bottom) self-similar, universal density profiles.   
%The computed mean density, $\bar{\rho}_{200}$, is 
%taken as representative of its counterpart related 
%to the whole (virialized) system, $\bar{\rho}$.   
Other captions as in Fig.\,\ref{f:M200FM}.}
\label{f:D200FM}    
\end{figure}

\begin{figure}
%   \centering
%   \resizebox{\hsize}{!}{\includegraphics{kapFM.eps}} 
\centerline{\psfig{file=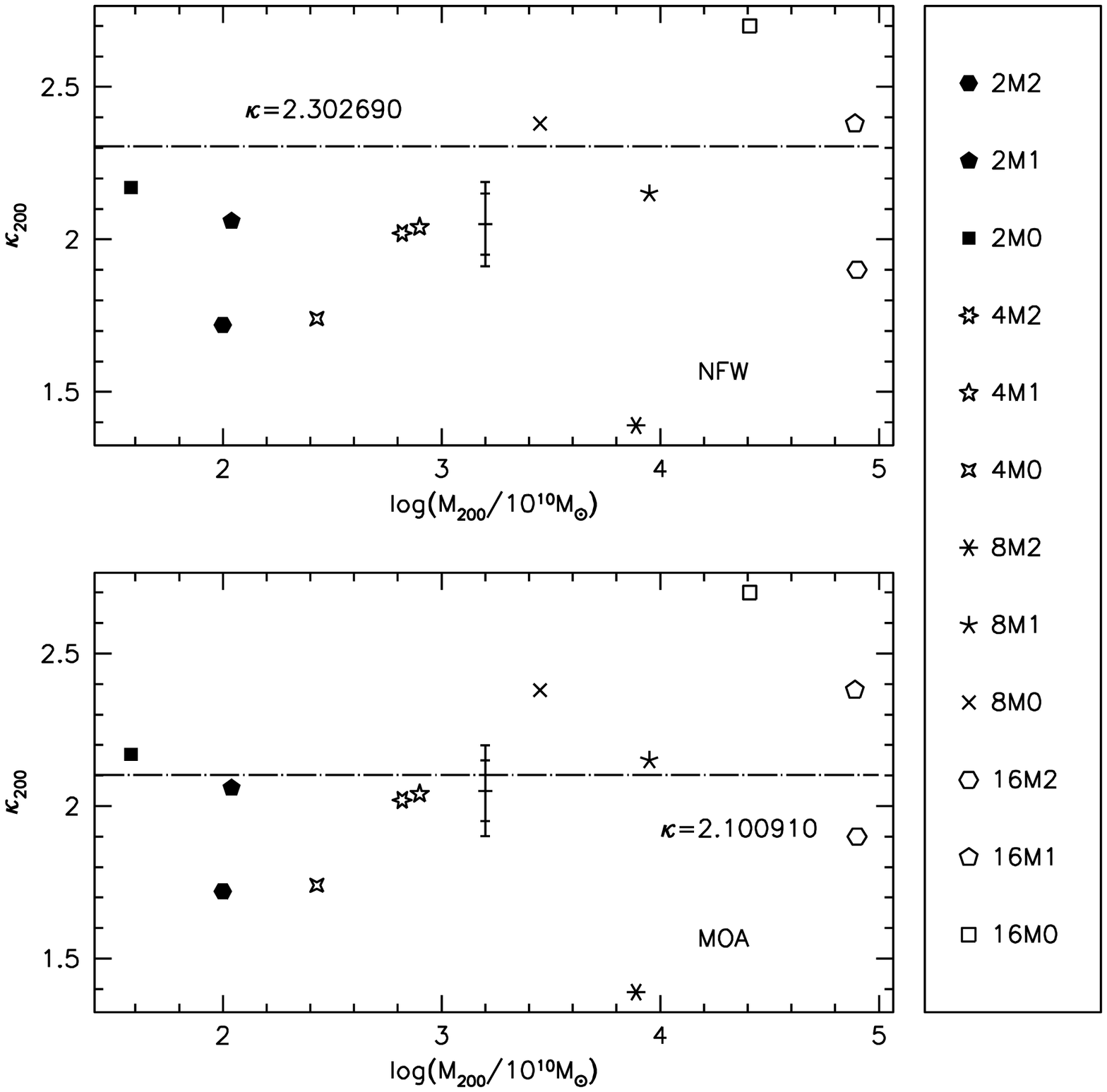,height=100mm,width=90mm}}
%\centerline{\psfig{file=kapFM.ps,height=130mm,width=120mm}}
\caption{Deviation of simulated dark matter haloes
from their fitting counterparts, with regard to the
dimensionless parameter,
$\kappa_{200}$, for both NFW (top) and MOA (bottom) 
self-similar, universal density profiles.
%The computed parameter, $\kappa_{200}$, is taken as 
%representative of its counterpart related to the 
%whole (virialized) system, $\kappa$.
Other captions
as in Fig.\,\ref{f:M200FM}.}
\label{f:C200FM}    
\end{figure}

\begin{figure}
%   \centering
%   \resizebox{\hsize}{!}{\includegraphics{masKl.eps}} 
\centerline{\psfig{file=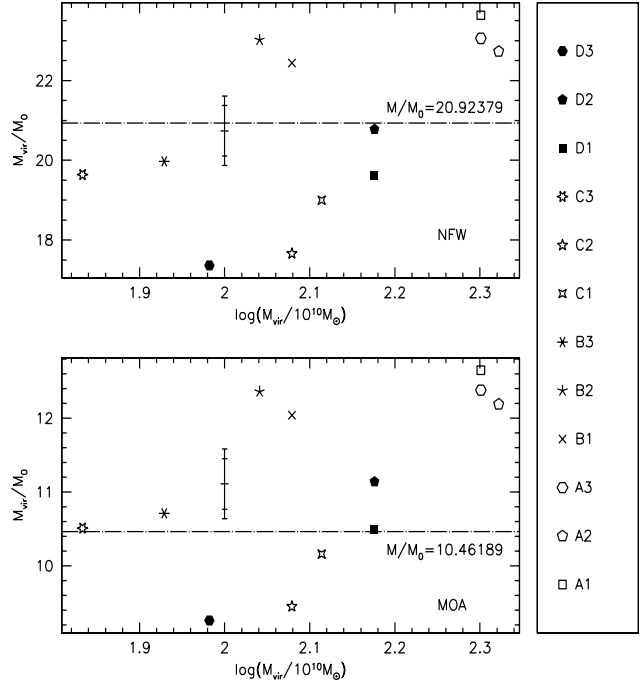,height=100mm,width=90mm}}
%\centerline{\psfig{file=masKl.ps,height=130mm,width=120mm}}
\caption{Deviation of simulated dark matter haloes 
from their fitting counterparts, with regard to the
scaled virial mass,
$M_{vir}/M_0$, for both NFW (top) and MOA (bottom) 
self-similar, universal density profiles.
%The computed mass, $M_{vir}$, is taken as representative
%of its counterpart related to the whole (virialized)
%system, $M$.
Fitting haloes lie on the horizontal
line, while simulated haloes are represented by
different symbols.   The vertical bar is centred
on the mean value of plotted data, with respect to
the ordinates (and no connection with the abscissae), 
and is limited by the standard deviation of the mean, 
without (inner boundary) and with (outer boundary) 
addition of about twice the standard
deviation from the standard deviation from the mean,
deduced from
%as in CM03I.
Eqs.\,(\ref{eq:ma}), (\ref{eq:sd}), 
and (\ref{eq:ssd}), respectively.   
Captions of symbols on the right correspond
to KLA01 runs listed in Tables\,\ref{t:SKL}, 
\ref{t:CKA}, and \ref{t:RKAN}.}
%: square-like -
%$I1$; penthagon-like - $I2$; exagon-like - 
%$I3$.   Additional captions: open - $I=A$;
%skeleton - $I=B$; star - $I=C$; filled - $I=D$.}
\label{f:MVIRKA}    
\end{figure}

\begin{figure}
%   \centering
%   \resizebox{\hsize}{!}{\includegraphics{ragKl.eps}} 
\centerline{\psfig{file=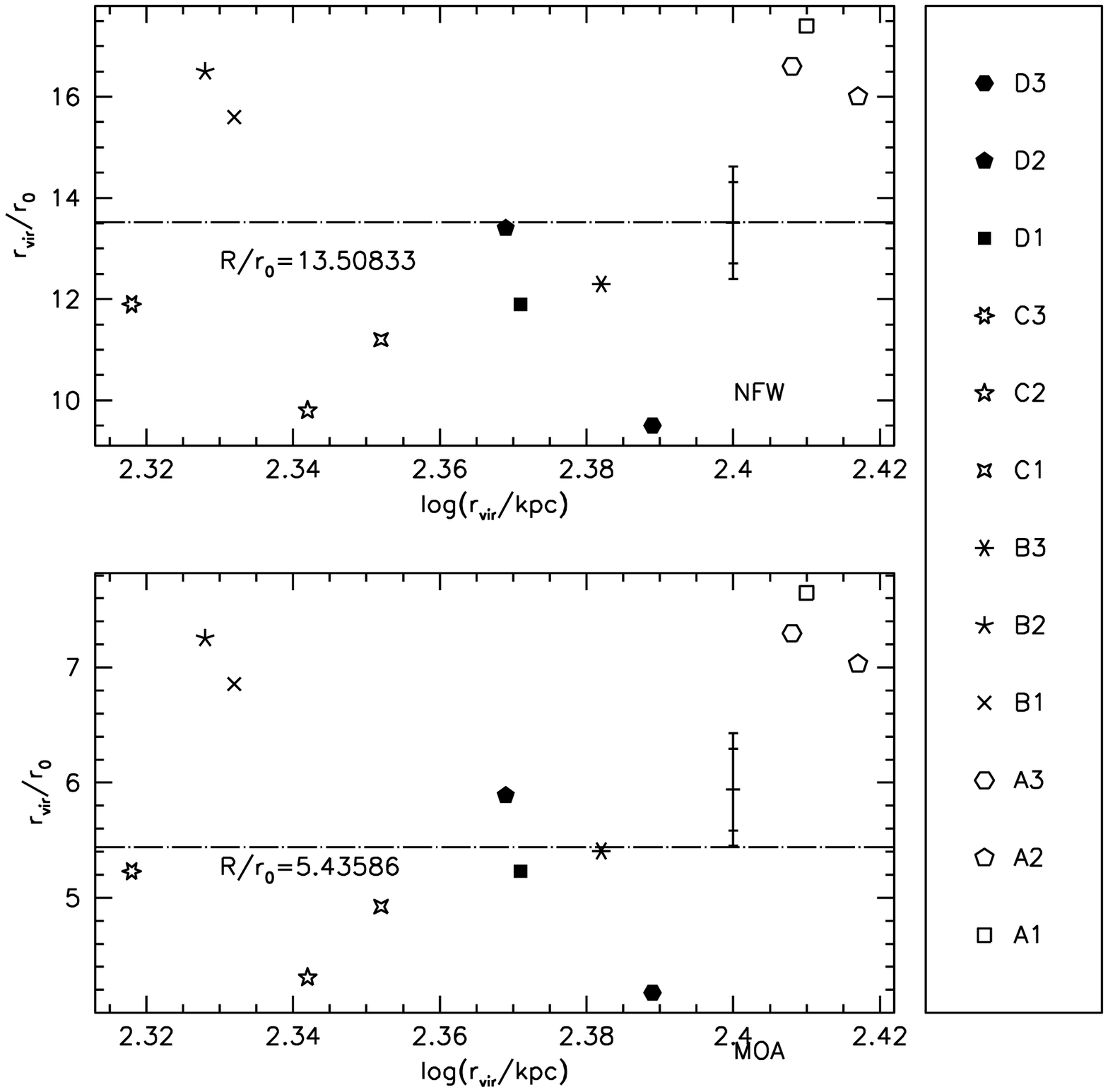,height=100mm,width=90mm}}
%\centerline{\psfig{file=ragKl.ps,height=130mm,width=120mm}}
\caption{Deviation of simulated dark matter haloes 
from their fitting counterparts, with regard to the
scaled radius, $r_{vir}
/r_0$, for both NFW (top) and MOA (bottom) 
self-similar, universal density profiles.
%The computed radius, $r_{vir}$, is taken as representative
%of its counterpart related to the whole (virialized)
%system, $R$.
Other captions as in Fig.\,\ref
{f:MVIRKA}.}
\label{f:RVIRKA}    
\end{figure}

\begin{figure}
%   \centering
%   \resizebox{\hsize}{!}{\includegraphics{densKl.eps}} 
\centerline{\psfig{file=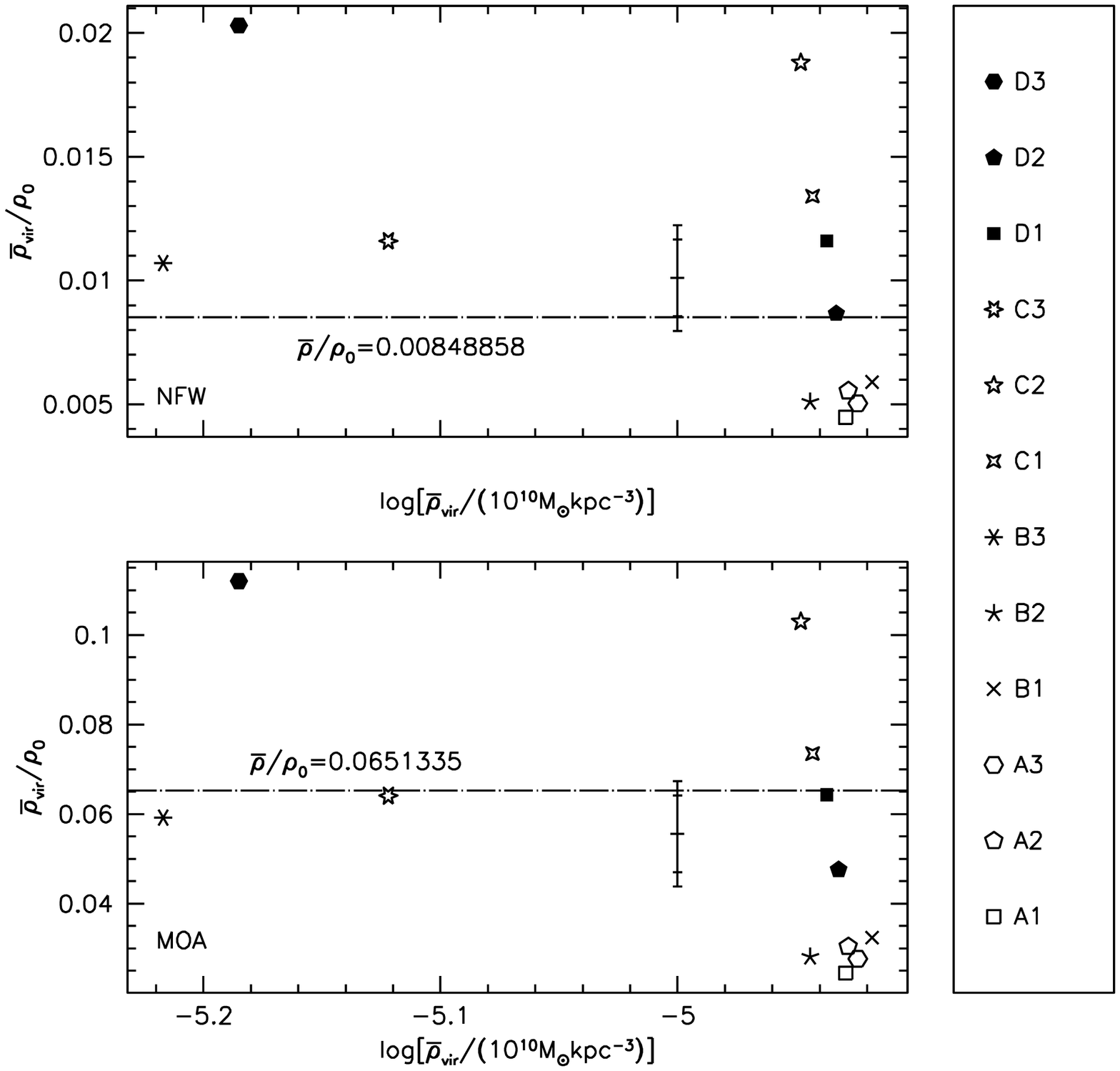,height=100mm,width=90mm}}
%\centerline{\psfig{file=densKl.ps,height=130mm,width=120mm}}
\caption{Deviation of simulated dark matter haloes 
from their fitting counterparts, with regard to the
scaled density, $\bar
{\rho}_{vir}/\rho_0$, for both NFW (top) and MOA 
(bottom) self-similar, universal density profiles.   
%The computed mean density, $\bar{\rho}_{vir}$, is 
%taken as representative of its counterpart related 
%to the whole (virialized) system, $\bar{\rho}$.   
Other captions as in Fig.\,\ref{f:MVIRKA}.}
\label{f:DVIRKA}    
\end{figure}

\begin{figure}
%   \centering
%   \resizebox{\hsize}{!}{\includegraphics{kapKl.eps}} 
\centerline{\psfig{file=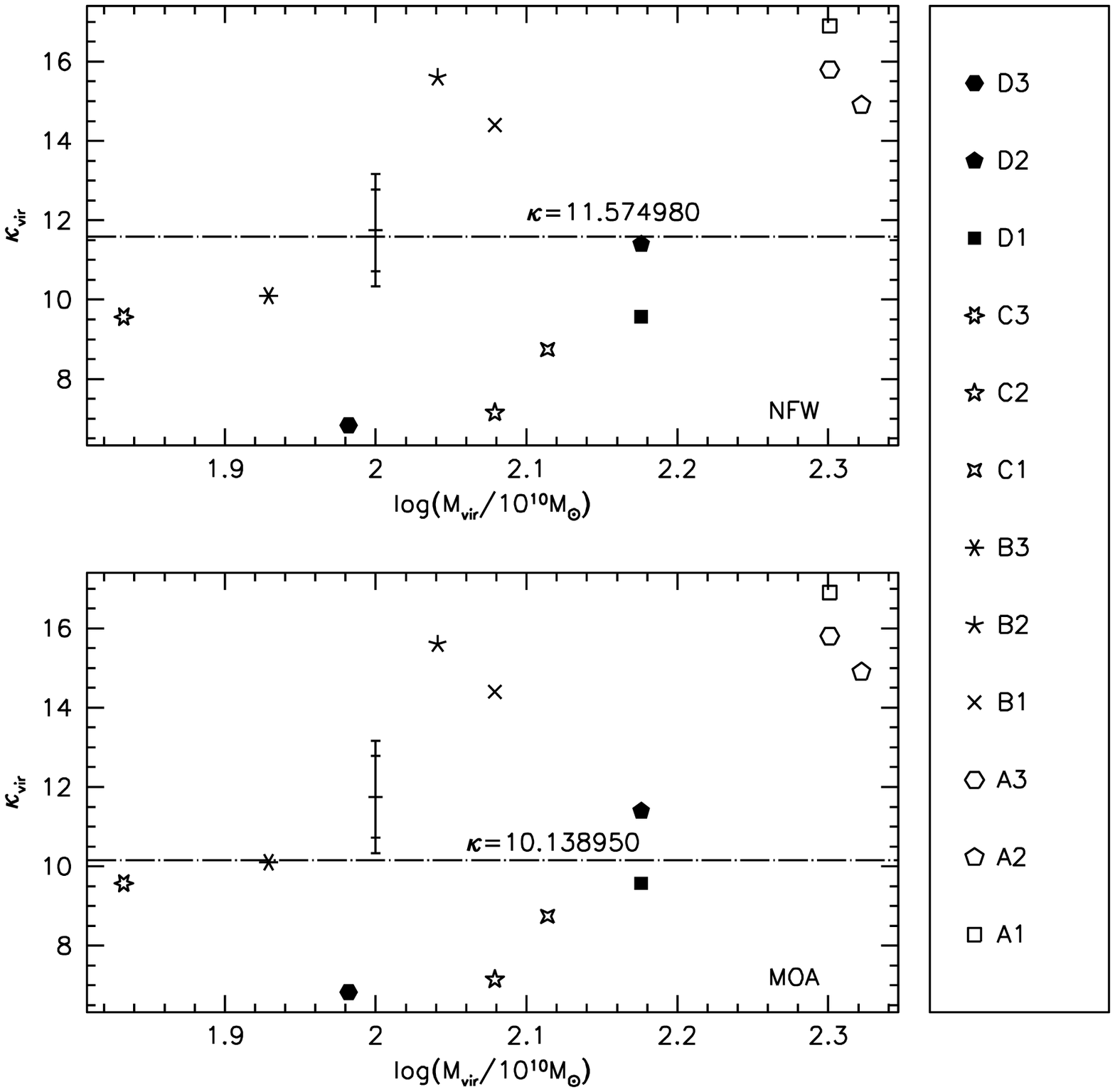,height=100mm,width=90mm}}
%\centerline{\psfig{file=kapKl.ps,height=130mm,width=120mm}}
\caption{Deviation of simulated dark matter haloes 
from their fitting counterparts, with regard to the
dimensionless parameter,
$\kappa_{vir}$, for both NFW (top) and MOA (bottom) 
self-similar, universal density profiles.
%The computed parameter, $\kappa_{vir}$, is taken as 
%representative of its counterpart related to the 
%whole (virialized) system, $\kappa$.
Other captions as in Fig.\,\ref{f:MVIRKA}.}
\label{f:CVIRKA}    
\end{figure}

The deviation of simulated dark matter haloes
from their fitting counterparts, can be seen
in Figs.\,\ref{f:M200FM}, \ref{f:R200FM},
\ref{f:D200FM}, \ref{f:C200FM}, and in
Figs.\,\ref{f:MVIRKA}, \ref{f:RVIRKA},
\ref{f:DVIRKA}, \ref{f:CVIRKA}, in connection
with FM01 and KLA01 simulations, respectively.
The following plots are represented: scaled
mass, $M_{trn}/M_0$, vs. logarithmic scaled
mass, $\log(M_{trn}/{\rm M}_{10})$; scaled
radius, $r_{trn}/r_0$, vs. logarithmic scaled
radius, $\log(r_{trn}/{{\rm kpc}})$; scaled
density, $\bar{\rho}_{trn}/{\rho_0}$, vs.
logarithmic scaled density, $\log[\bar{\rho}_
{trn}/({\rm M}_{10}{\rm kpc}^{-3})]$; and
dimensionless parameter, $\kappa_{trn}$,
vs. logarithmic scaled mass, $\log(M_{trn}/
{\rm M}_{10})$; where $tnr=200, vir$.   
The deviation of the above mentioned,
scaled parameters, from their fitting
counterparts, are clearly shown therein.

The mean value, $\bar{\eta}$, the standard
deviation from the mean value, $\sigma_{s\bar
{\eta}}$, and the standard deviation from the
standard deviation from the mean value, 
$\sigma_{s\bar{\mu}}$, which are expressed as
(e.g., Oliva \& Terrasi 1976, Chap.\,V, \S\,5.6.3):
\begin{lefteqnarray}
\label{eq:ma}
&& \bar{\eta}=\frac1n\sum_{i=1}^n\eta_i~~; \\
\label{eq:sd}
&& \sigma_{s\bar{\eta}}=\left[\frac1n\frac1{n-1}\sum_
{i=1}^n(\eta_i-\bar{\eta})^2\right]^{1/2}~~; \\
\label{eq:ssd}
&& \sigma_{s\bar{\mu}}=
%\left[\frac{2(n-1)}{n(n+1)}\right]^{1/2}
\frac{\sigma_{s\bar{\eta}}}{\sqrt{2n}}~~;\quad\bar{\mu}=
\sigma_{s\bar{\eta}}~~;
\end{lefteqnarray}
where $n=12$; $\eta=M_{trn}/M_0$, $r_{trn}/r_0$,
$\bar{\rho}_{trn}/\rho_0$, $\kappa_{trn}$; $tnr=
200, vir$; and 
$\bar{\eta}$, $\sigma_{s\bar{\eta}}$, $\sigma_
{s\bar{\mu}}$, are shown by the vertical bars on
Figs.\,\ref{f:M200FM}-\ref{f:CVIRKA}.

Numerical values of $\bar{\eta}$, $\sigma_{s\bar
{\eta}}$, $\sigma_{s\bar{\mu}}$, deduced by 
Eqs.\,(\ref{eq:ma}), (\ref{eq:sd}), and (\ref
{eq:ssd}), respectively, are listed in 
Table\,\ref{t:MFM}, together with their fitting
counterparts, $\eta^\ast$, which have been
represented as horizontal lines in Figs.\,\ref
{f:M200FM}-\ref{f:CVIRKA} and listed in 
Table\,\ref{t:pron}, in connection with NFW
and MOA density profiles.
\begin{table}
\begin{tabular}{lllllllll}
\hline
\hline
\multicolumn{1}{c}{parameter} &
\multicolumn{4}{|c|}{NFW} &
\multicolumn{4}{c}{MOA} \\
\hline
\multicolumn{1}{c}{$\eta$} &
\multicolumn{1}{c}{$\bar{\eta}$} &
\multicolumn{1}{c}{$\sigma_{s\bar{\eta}}$} &
\multicolumn{1}{c}{$\sigma_{s\bar{\mu}}$} &
\multicolumn{1}{c}{$\eta^\ast$} &
\multicolumn{1}{c}{$\bar{\eta}$} &
\multicolumn{1}{c}{$\sigma_{s\bar{\eta}}$} &
\multicolumn{1}{c}{$\sigma_{s\bar{\mu}}$} &
\multicolumn{1}{c}{$\eta^\ast$} \\
\hline
$M_{200}/M_0$ & 15.69 & 0.538 & 0.110 
& 17.05 & 8.29 & 0.309 & 0.0630 & 8.53\\
$r_{200}/r_0$ & 8.09 & 0.443 & 0.0902
& 9.21 & 3.69 & 0.209 & 0.0427 & 3.81 \\
$\bar{\rho}_{200}/\rho_0$ & 0.0346 & 0.00564 & 0.00115 
& 0.0218 & 0.195 & 0.0328 & 0.00669 & 0.154 \\
$\kappa_{200}$ & 2.05 & 0.100 & 0.0205 
& 2.30 & 2.05 & 0.100 & 0.0205 & 2.10 \\
\hline\hline
$M_{vir}/M_0$ & 20.74 & 0.632 & 0.129 
& 20.92 & 11.11 & 0.343 & 0.0701 & 10.46\\
$r_{vir}/r_0$ & 13.51 & 0.807 & 0.165 
& 13.51 & 5.94 & 0.355 & 0.0723 & 5.44 \\
$\bar{\rho}_{vir}/\rho_0$ & 0.0101 & 0.00155 & 0.000317
& 0.00849 & 0.0556 & 0.00855 & 0.00174 & 0.0651 \\
$\kappa_{vir}$ & 11.75 & 1.03 & 0.211 
& 11.57 & 11.75 & 1.03 & 0.211 & 10.14 \\
\hline\hline
\end{tabular}
\caption{The mean value, $\bar{\eta}$, the standard
deviation from the mean value, $\sigma_{s\bar
{\eta}}$, and the standard deviation from the
standard deviation from the mean value, 
$\sigma_{s\bar{\mu}}$, where $\eta=M_{trn}/M_0$, 
$r_{trn}/r_0$, $\bar{\rho}_{trn}/\rho_0$, 
$\kappa_{trn}$, $trn=200, vir$, deduced from 
simulated haloes,
data from Tables~\ref{t:SFM}, \ref{t:CFM},
\ref{t:RFMN} (FM01), and \ref{t:SKL}, \ref{t:CKA},
\ref{t:RKAN} (KLA01).   The value of the related 
fitting counterparts, $\eta^
\ast=\nu_M$, $\Xi$, $\nu_{\bar{\rho}}$, $\kappa$,
listed in Table\,\ref{t:pron}, is also reported
for comparison.   The upper and lower panel are
related to FM01 and KLA01 simulations, respectively.
The left-hand and right-hand side are related to
NFW and MOA universal density profiles, respectively.}
\label{t:MFM}
\end{table}
It is apparent that the following inequalities
hold:
\begin{equation}
\label{eq:eta}
\bar{\eta}-u\sigma_{s\bar{\eta}}-u\sigma_{s\bar{\mu}}
<\eta^\ast<\bar{\eta}+u\sigma_{s\bar{\eta}}+u\sigma_
{s\bar{\mu}}~~;
\end{equation}
where $u=2,1$ for NFW and MOA density profiles,
respectively, in connection with FM01 simulations,
and the contrary holds, in connection with KLA01
simulations.   Then the best fitting density
profile (among NFW and MOA in the case under
discussion) may be chosen as minimizing the ratio,
$\vert x_{\overline{\eta}}\vert/\sigma_{s\,\overline
{\eta}}=\vert\overline{\eta}-\eta^\ast\vert/\sigma_
{s\,\overline{\eta}}$, for the scaled parameter of
interest.

It can also be shown that a necessary condition 
for the detectability of accidental errors, 
$\Delta\eta\le\sigma_\eta$, is satisfied for
any choice of $\eta$ listed in Table\,\ref{t:MFM}.
More specifically, $\Delta\eta$ and $\sigma_\eta$
represent the sensitivity error of the simulation
and the rms error, respectively,
with regard to $\eta$.   For a formal demonstration,
see Appendix C.
%\,\ref{sene}.

It is apparent that NFW density profiles
reproduce FM01 simulations to a lesser extent
with respect to MOA density profiles, while 
the contrary holds with regard to KLA01
simulations, according to the above mentioned
criterion.   In any case, a
better fit could be obtained by use of
a different scaled density profile, i.e. different
values of the exponents, $(\alpha,\beta,
\gamma)$, appearing in Eq.\,(\ref{eq:runi}).


\subsection{Standard deviations
deduced from the propagation of the errors}
\label{prop}

Universal density profiles, expressed by
Eq.\,(\ref{eq:runi}), are
currently used in fitting simulated density
profiles (e.g., NFW; FM01; FM03).   The
choice of the exponents, $(\alpha,\beta,
\gamma)$,  
allows the scaled parameters, $\nu
_M=M/M_0$, $\nu_{\bar{\rho}}=\bar{\rho}/\rho
_0$, and $k$, depend on a single scaled
parameter i.e. the scaled radius, $\Xi$,
or concentration with regard to NFW density
profiles.   As the
concentration exhibits a lognormal
distribution (e.g., Bullock et al. 2001),
and the above mentioned parameters
may be considered as depending on the decimal
logarithm of the concentration, $\log\Xi$,
the related distribution is expected to be
(at least to a first extent) normal, via
the same procedure which leads to the
propagation of the errors.

The corresponding rms errors are:
\begin{lefteqnarray}
\label{eq:sigc}
&& \sigma_\Xi=\Xi\sigma_{\log\Xi}~~; \\
\label{eq:sigM}
&& \sigma_{\nu_M}=3f(\Xi)\Xi^3\sigma_{\log\Xi}~~; \\
\label{eq:sigd}
&& \sigma_{\nu_{\bar{\rho}}}=3\left\vert f(\Xi)-\frac1
{\Xi^3}\frac M{M_0}\right\vert\sigma_{\log\Xi}~~; \\
\label{eq:sigk}
&& \sigma_k=\frac32k\sigma_{\log\Xi}~~; 
\end{lefteqnarray}
where $f(\Xi)$ is the scaled density, expressed
by Eq.\,(\ref{eq:f}), particularized to the
scaled radius, $\xi=\Xi$.   For a formal 
demonstration, see Appendix D.
%\ref{indi}.

A comparison between rms errors, expressed
by the above relations, and standard deviations,
listed in Table \ref{t:MFM}, needs the following 
steps. 
\begin{description}%                          Beccari,  p. 26
\item[\rm{(a)}] Assume a rms error of the
lognormal distribution of the concentration
as in Bullock el al. (2001), $\sigma_{\log
\Xi}=0.18$.   
\item[\rm{(b)}] Assume a scaled radius related
to the expected value of the lognormal 
distribution of the concentration, $\Xi^\ast=
\exp_{10}(\log\Xi)^\ast$, as listed in Table 
\ref{t:MFM} for each universal density profile
and each set of simulations.
\item[\rm{(c)}] Divide the rms errors, $\sigma_
\eta$, $\eta=\Xi,~\nu_M,~\nu_{\bar{\rho}},~k,$
by the square root of the number of measures
that have been averaged in calculating the
standard deviations from the mean, which is
equal to $\sqrt{12}$.
\item[\rm{(d)}] For a selected, universal density 
profile and a selected set of simulations, use
the mean values, $\overline{\eta}$, and their
fitting counterparts, $\eta^\ast$,
listed in Table \ref{t:MFM}.
\end{description}

The rms errors, $\sigma_{\bar{\eta}}$, and
the ratios of rms error to the expected value
of the related distribution, $\sigma_{\bar
{\eta}}/\eta^\ast$, the absolute errors,
$\vert x_{\bar{\eta}}\vert=\vert\bar{\eta}-
\eta^\ast\vert$,
and the ratios of absolute error to
rms error, $\vert x_{\bar{\eta}}\vert/\sigma_
{\bar{\eta}}$, are listed in Table \ref{t:indi}
\begin{table}
\begin{tabular}{lllllllll}
\hline
\hline
\multicolumn{1}{c}{parameter} &
\multicolumn{4}{|c|}{NFW} &
\multicolumn{4}{c}{MOA} \\
\hline
\multicolumn{1}{c}{$\eta$} &
\multicolumn{1}{c}{$\sigma_{\bar{\eta}}$} &
\multicolumn{1}{c}{$\sigma_{\bar{\eta}}/\eta^\ast$} &
\multicolumn{1}{c}{$\vert x_{\bar{\eta}}\vert$} &
\multicolumn{1}{c}{$\vert x_{\bar{\eta}}\vert/\sigma_{\bar{\eta}}$} &
\multicolumn{1}{c}{$\sigma_{\bar{\eta}}$} &
\multicolumn{1}{c}{$\sigma_{\bar{\eta}}/\eta^\ast$} &
\multicolumn{1}{c}{$\vert x_{\bar{\eta}}\vert$} &
\multicolumn{1}{c}{$\vert x_{\bar{\eta}}\vert/\sigma_{\bar{\eta}}$} \\
\hline
$M_{200}/M_0$ & 1.17 & 0.0685 & 1.36
& 1.16 & 0.633 & 0.0742 & 0.24 & 0.379\\
$r_{200}/r_0$ & 1.10 & 0.120 & 1.12 
& 1.02 & 0.455 & 0.120 & 0.12 & 0.264 \\
$\bar{\rho}_{200}/\rho_0$ & 0.0635 & 0.290 & 0.0128 
& 2.02 & 0.0440 & 0.285 & 0.041 & 0.932 \\
$\kappa_{200}$ & 0.413 & 0.179 & 0.25 
& 0.605 & 0.377 & 0.179 & 0.05 & 0.133 \\
\hline\hline
$M_{vir}/M_0$ & 1.24 & 0.0595 & 0.18 
& 0.145 & 0.665 & 0.0636 & 0.65 & 0.977 \\
$r_{vir}/r_0$ & 1.62 & 0.120 & 0 
& 0 & 0.650 & 0.120 & 0.5 & 0.769 \\
$\bar{\rho}_{vir}/\rho_0$ & 0.00254 & 0.299 & 0.00161
& 0.634 & 0.0192 & 0.295 & 0.0095 & 0.495 \\
$\kappa_{vir}$ & 2.30 & 0.179 & 0.18 
& 0.0783 & 1.82 & 0.179 & 1.61 & 0.885 \\
\hline\hline
\end{tabular}
\caption{The rms errors, $\sigma_{\bar{\eta}}$,
the ratios of rms error to the expected value
of the related distribution, $\sigma_{\bar
{\eta}}/\eta^\ast$, the absolute errors,
$\vert x_{\bar{\eta}}\vert=\vert\bar{\eta}-
\eta^\ast\vert$,
and the ratios of absolute error to rms
error, $\vert x_{\bar{\eta}}\vert/\sigma_{\bar
{\eta}}$, where $\eta=M_{trn}/M_0$, 
$r_{trn}/r_0$, $\bar{\rho}_{trn}/\rho_0$, 
$\kappa_{trn}$, $trn=200, vir$, deduced from 
simulated haloes,
data from Tables~\ref{t:SFM}, \ref{t:CFM},
\ref{t:RFMN} (FM01), and \ref{t:SKL}, \ref{t:CKA},
\ref{t:RKAN} (KLA01).   Values of parameters
related to their fitting counterparts, $\eta^
\ast=\nu_M$, $\Xi$, $\nu_{\bar{\rho}}$, $\kappa$,
are taken from Table\,\ref{t:pron}.   The upper
and lower panel are related to FM01 and KLA01
simulations, respectively.   The left-hand and
right-hand side are related to NFW and MOA density
profiles, respectively.}
\label{t:indi}
\end{table}
with regard to FM01 (upper panel) and KLA01 
(lower panel) simulations, related to NFW 
(left-hand side) and MOA (right-hand side) 
universal density profiles.

The rms errors appear to be systematically
larger than the related standard deviations
listed in Table \ref{t:MFM}, essentially for
three orders of reasons.   First, the 
statistical significance of the samples 
considered is low, owing to the small
number of objects $(N=12)$.   In addition,
KLA01 runs make in fact three
sets of simulations of only four dark
matter haloes with resolution varied in
each set.   For this reason, they cannot
be treated as twelve independent runs,
and a lower standard deviation is expected.

Second, the rms errors calculated by use
of Eqs.\,(\ref{eq:sigc})-(\ref{eq:sigk})
need a normal distribution for the related
random variables, which holds to a good
extent only if the fluctuations are
sufficiently small, to neglect higher-order
terms, with respect to the first, in the 
related series developments.   For further
details, see Appendix D.
%\ref{indi}.

Third, the comparison between rms errors 
and standard deviations should be valid
only in connection with the cosmological
model, and the universal density profile,
used by Bullock et al. (2001) in building
up the statistical sample of simulated
dark matter haloes, from which a lognormal
distribution of the concentration has been
deduced, with rms error $\sigma_{\log\Xi_
{trn}}=0.18$.  It needs $\Lambda_0=0.7$,
$\Omega_0=0.3$, $h=0.7$, $\sigma_8=1.0$,
and a NFW density profile.   In fact,
larger discrepancies occur for FM01
simulations, where $\Lambda_0=0$, $\Omega
_0=1$, $h=0.5$, and $\sigma_8=0.7$.   On
the other hand, KLA01 simulations were 
performed with the same choice of
cosmological parameters as in Bullock et 
al. (2001), with the exception of $\sigma
_8=0.9$.   Then we expect that a lognormal 
distribution of concentration for an 
assigned mass bin occurs for any plausible 
choice of cosmological parameters, and its
expected value and rms error do not change
dramatically for any plausible variation
of cosmological parameters.

The existence of a lognormal distribution
is a necessary, but not sufficient condition,
for the validity of the central limit theorem.
In this view, the concentration is related 
to the final properties of a simulated halo,
which are connected with the initial conditions,
$\alpha_1$, $\alpha_2$, ..., $\alpha_n$, by a
transformation, $\Xi=\alpha_1\alpha_2...\alpha_
n$, as in dealing with the process of
star formation, where the stellar mass follows
a lognormal distribution (Adams \& Fatuzzo
1996; Padoan et al. 1997).    With regard to
dark matter haloes within a fixed mass bin,
an interpretation of the lognormal distribution,
depending on the concentration,
in terms of the central limit theorem, is
outlined in Appendix E.
%\ref{lice}.

It is worth recalling that standard deviations
of scaled parameters from their mean values,
have been deduced directly from the results of
simulations, with regard to a selected, 
fitting density profile.   On the other
hand, rms errors of scaled parameters have
been deduced from their dependence on the
(decimal) logarithm of the concentration,
$\log\Xi$, and the lognormal distribution
of the concentration.

An inspection to Table \ref{t:indi} shows
that the ratio of absolute error to
rms error, is closer to zero for MOA 
density profiles with regard to FM01 
simulations, and for NFW density profiles
with regard to KLA01 simulations.   This
agrees with the results found using
standard deviations from the mean, 
represented in Figs.\,\ref
{f:M200FM}-\ref{f:CVIRKA}, which have
been deduced directly from simulations,
with regard to a selected, fitting
density profile.   Then a valid criterion
for the choice of a fitting density
profile, in connection with a given set 
of simulated, dark matter haloes, appears
to be the following.

\begin{trivlist}
\item[\hspace\labelsep{\bf Statistic razor for fitting
density profiles to simulated, dark matter haloes}] \sl
Given two or more fitting density 
profiles and a set of simulated
haloes, the best fit is related to the minimum value
of the ratio of the absolute error to the corresponding
standard deviation from the mean, $\vert x_{\overline
{\eta}}\vert/\sigma_{s\,\overline{\eta}}=\vert
\overline{\eta}-\eta^\ast\vert/\sigma_{s\,\overline
{\eta}}$.
\end{trivlist}

\subsection{Interpretation in terms of the
spherical top-hat model}
\label{cth}

The spherical, top-hat model makes a valid
reference for simulated haloes (e.g., Cole
\& Lacey 1996; KLA01), and for this reason
we think an interpretation of the
dimensionless parameter, $\delta$,
in terms of the spherical top-hat model,
may be of some utility.   With regard to
fitting, dark matter haloes, the combination
of Eqs.\,(\ref{eq:r0a}), (\ref{eq:k1}), and 
(\ref{eq:k2}) yields:
\begin{equation}
\label{eq:delMrh}
\delta=\frac{4\pi}3C_r^3\Xi^3\frac{\bar
{\rho}}{{\rm M}_{10}{\rm kpc}^{-3}}\frac
M{{\rm M}_{10}}~~;
\end{equation}
which shows the dependence of $\delta$
on the product, $\bar{\rho}M$.

With regard to FM01 simulations, an
inspection of Table \ref{t:SFM} shows
that while the mean density, $\bar
{\rho}_{200}$, is inversely proportional
to the (fiducial) total mass, $M_{200}$,
the dimensionless parameter, $\delta$, 
is directly proportional to $M_{200}$.
It can also be seen that the product,
$\bar{\rho}_{200}M_{200}$, also increases
as $M_{200}$ does and vice versa, in
agreement with Eq.\,(\ref{eq:delMrh}).
The parameter, $\delta$, appears to
depend on the mass, $M$, and a second 
independent parameter, which may be
chosen as the mass excess at the start 
of simulation, $(\delta M/M)_{start}$,
or the related mean peak height, $\bar{\nu}
_{start}=(\delta M/M)_{start}/<(\delta 
M/M)_{start}^2>^{1/2}$.

In the framework of the spherical, 
top-hat model, the total mass is
conserved and the following relations 
hold:
\begin{equation}
\label{eq:rrhR3}
\frac{\bar{\rho}_{vir}}{\bar{\rho}_{max}}=
\frac{R_{max}^3}{R_{vir}^3}~~;\qquad
\frac{\bar{\rho}_{max}}{\bar{\rho}_{rec}}=
\frac{R_{rec}^3}{R_{max}^3}~~;
\end{equation}
where the indices, $rec$, $max$, $vir$, 
denote recombination, maximum expansion,
virialization, respectively, and the
ratio, $(R_{rec}/R_{max})^{-1}$, is a 
solution of the third-degree equation:
\begin{lefteqnarray}
\label{eq:gr3}
&& 
%\left(\frac{a_{rec}}{a_{max}}\right)^{3(1+w)}\Psi
\Lambda_{rec}x^3+\left\{1-\Omega_{rec}
\left[1+\left(\frac{\delta M^\ast}M\right)_
{rec}\right]-\Lambda_{rec}\right\}x+\Omega
_{rec}\left[1+\left(\frac{\delta M^\ast}M
\right)_{rec}\right]=0~~; 
\nonumber \\
&& \phantom{10}
\end{lefteqnarray}
where 
%$a$ is the scale factor, 
$\Omega$ and 
$\Lambda$ represent the density parameter 
related to matter and cosmological constant,
respectively,
%
%quintessence,
%respectively, $w$ is the quintessence index
%($w<-1/3$ for an accelerated expansion and $w
%=-1$ in the special case of the cosmological
%constant), 
%
and $\delta M^\ast/M$ is the mean mass excess 
within a spherical volume where the mass of 
the material Hubble flow equals $M$
%
%, and $z$ is the redshift
%
(e.g., Lokas \& Hoffman 2001a,b).

The solution to Eq.\,(\ref{eq:gr3})
which is of interest here, in absence
of cosmological constant i.e. $\Lambda
\to0$, has 
%quintessence i.e. $\Psi\to0$, has
to attain the limiting expression 
(e.g., Peebles 1980, Chap.\,II, \S 19a):
\begin{lefteqnarray}
\label{eq:rRrm}
&& \frac{R_{rec}}{(R_{max})_{\Lambda=0}}
=\frac{1-\Omega_
{rec}^{-1}+(\delta M^\ast/M)_{rec}}{1+
(\delta M^\ast/M)_{rec}}\approx\left(\frac
{\delta M^\ast}M\right)_{rec}~~; \nonumber \\
%\label{eq:rRrm}
&& \vert1-\Omega_{rec}^{-1}\vert\ll\left(
\frac{\delta M^\ast}M\right)_{rec}\ll1~~;
\end{lefteqnarray}
%\label{seq:rRrm}
where the mass excess, at recombination
epoch, has a substantial contribution
from both the growing and the decreasing 
mode of the density perturbation.   For 
further details see e.g., Caimmi et al. 
(1990).

The combination of Eqs.\,(\ref{eq:rrhR3})
and (\ref{eq:rRrm}) yields:
\begin{equation}
\label{eq:rrhmr}
\frac{\bar{\rho}_{max}}{\bar{\rho}_{rec}}=
\zeta_{max}\left[\left(\frac{\delta M^\ast}
M\right)_{rec}\right]^
3;~\zeta_{max}=\left[\frac{(R_{max})_
{\Lambda=0}}{R_{max}}\right]^3;
~\vert1-\Omega_{rec}^{-1}\vert\ll
\left(\frac{\delta M^\ast}M\right)_{rec}\ll1;
\end{equation}
and, in addition:
\begin{equation}
\label{eq:zita}
\frac{\bar{\rho}_{vir}}{\bar{\rho}_{max}}=
\zeta_{vir}~~;\qquad1\le\zeta_{vir}\le8~~;
\end{equation}
where the upper limit corresponds to null
kinetic energy at maximum expansion,
and the lower limit is related to a necessary 
condition for maximum expansion, $R_{vir}\le
R_{max}$.   For further details see e.g., 
Caimmi et al. (1990).   Finally, the combination
of Eqs.\,(\ref{eq:rrhR3}), (\ref{eq:rrhmr}),
and (\ref{eq:zita}), produces:
\begin{equation}
\label{eq:rhv}
\bar{\rho}_{vir}=\zeta_{max}\zeta_{vir}(\rho_h)
_{rec}\left[\left(\frac
{\delta M^\ast}M\right)_{rec}\right]^3~~;\qquad\vert
1-\Omega_{rec}^{-1}\vert\ll
\left(\frac{\delta M^\ast}M\right)_{rec}\ll1~~;
\end{equation}
where $(\rho_h)_{rec}=\bar{\rho}_{rec}/[1+(\delta 
M^\ast/M)_{rec}]$ is the density of the material Hubble
flow at recombination epoch.   The above relation
may be cast under the equivalent form:
\begin{equation}
\label{eq:rhvo}
\bar{\rho}_{vir}=\frac{125}{72\pi}\zeta_{max}
\zeta_{vir}\frac
{H_0^2\Omega_0}G\left[\left(\frac{\delta M}M\right)_0
\right]^3
~~;\qquad\vert1-\Omega_{rec}^{-1}\vert\ll
\left(\frac{\delta M^\ast}M\right)_{rec}\ll1~~;
\end{equation}
where $(\delta M/M)_0=(3/5)(\delta M^\ast/M)
_{rec}(1+z_{rec})$ represents the present-day
mass excess of the growing mode predicted by the
top-hat model, in a flat cosmology with a
vanishing quintessence or, in particular,
cosmological constant.   For a formal
derivation, see Appendix A.
%\,\ref{rhov}.

The combination of Eqs.\,(\ref{eq:delMrh}) and 
(\ref{eq:rhvo}) yields:
\begin{leftsubeqnarray} 
\slabel{eq:delna}
&& \delta=C_\delta\zeta_{max}\left[\left(\frac{\delta M}M
\right)_0\right]^3\frac M{{\rm M}_{10}}~~; \\
\slabel{eq:delnb}
&& C_\delta=\frac{125}{54}\zeta_{vir}C_r^3\Xi^3
\frac{H_0^2\Omega_0}G\frac{{\rm kpc}^3}{{\rm 
M}_{10}}~~;\qquad\vert1-\Omega_{rec}^{-1}\vert
\ll1~~;
\label{seq:deln}
\end{leftsubeqnarray}
where the coefficient, $C_\delta$, may be
evaluated numerically by use of Eqs.\,(\ref
{eq:r0b}), (\ref{eq:Crrh}), and (\ref
{eq:zita}), taking the value of the scaled
radius, $\Xi$, from Table \ref{t:pron},
and keeping in mind that
$H_0=50$ and $70~{\rm km~s}^{-1}{\rm~Mpc}^
{-1}$ have been assumed in FM01 and KLA01
simulations, respectively.   The result is:
\begin{leftsubeqnarray}
\slabel{eq:Cda}
&& 7.12786~10^{-7}\le (C_\delta)_{NFW,FM}\le5.70229~10^{-6}
~~; \\
\slabel{eq:Cdb}
&& 5.93342~10^{-7}\le (C_\delta)_{MOA,FM}\le4.74673~10^{-6}
~~; \\ 
\slabel{eq:Cdc}
&& 1.05903~10^{-5}\le (C_\delta)_{NFW,KLA}\le8.47224~10^{-5}
~~; \\
\slabel{eq:Cdd}
&& 8.12553~10^{-6}\le (C_\delta)_{MOA,KLA}\le6.50043~10^{-5}
~~;
\label{seq:Cd}
\end{leftsubeqnarray}
in connection with both NFW and MOA density profiles,
related to both FM01 and KLA01 simulations.

\begin{figure}
%   \centering
%   \resizebox{\hsize}{!}{\includegraphics{ultimo.eps}} 
\centerline{\psfig{file=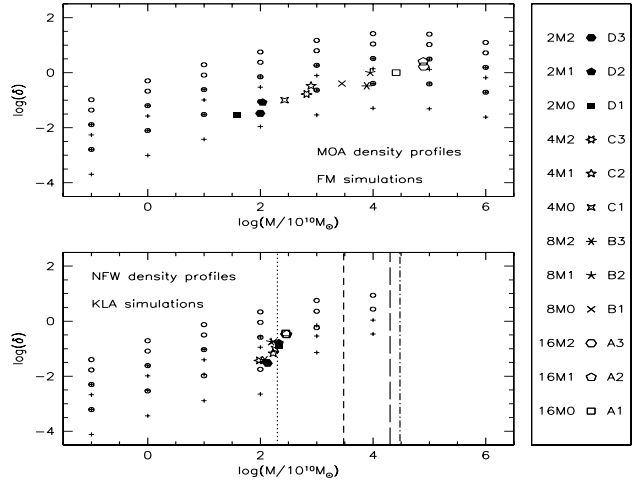,height=70mm,width=90mm}}
%\centerline{\psfig{file=ultimo.ps,height=100mm,width=120mm}}
\caption{The dimensionless parameter, $\delta$, vs.
the fitting mass, $M$, for top-hat, spherical 
perturbations, with no (open circles) and maximum
allowed (crosses) acquisition of angular momentum,
according to Eqs.\,(\ref{eq:Cdb}) and (\ref
{eq:Cdc}), in connection with MOA (top panel) and
NFW (bottom panel) density profiles.   For each
mass and class of symbols, values from down to up
are related to mean peak heights at recombination
epoch, $\bar{\nu}_{rec}=1,2,3,4,$ respectively.
The mass of the critical density perturbation,
for which the turnaround radius is infinite, 
%and $\delta=0$, owing to quintessence,
is marked by a vertical line,
from the left to the right for increasing peak 
height (bottom panel). 
Values related to FM01 (top panel) and KLA01
(bottom panel) simulations are also represented,
with same symbol captions as in 
Figs.\ref{f:M200FM}-\ref{f:C200FM} and 
\ref{f:MVIRKA}-\ref{f:CVIRKA}, respectively.
Passing from NFW to MOA density profile for
each set of simulations, would lower values
related to top-hat, spherical perturbations,
by about one dex, and vice versa.}
\label{f:del1}    
\end{figure}

With regard to FM01 ($\sigma_8=0.7$) and KLA01
($\sigma_8=0.9$) simulations, the mass enclosed
within a spherical region of radius $R_8=8h^{-1}
$~Mpc, is:
\begin{leftsubeqnarray}
\slabel{eq:M08na}
&& (M_8)_{FM}=2.021868~10^5 {\rm M}_{10}~~; \\
\slabel{eq:M08nb}
&& (M_8)_{KLA}=4.842291~10^4 {\rm M}_{10}~~;
\label{seq:M08n}
\end{leftsubeqnarray}
for a formal derivation, see Appendix A.
%\ref{rhov}.
The normalization of top-hat,
spherical perturbations with same spectrum 
as in Gunn (1987), to the above values,
demands a multiplication of the rms mass
excess plotted in Gunn (1987) by a factor
of 14/13 and 3/10, respectively.   The 
related values,
together with some other parameters which
are characteristic of top-hat, spherical
perturbations, are listed in Table\,\ref
{t:del}.
\begin{table}
\begin{tabular}{rllllllll}
\hline
\hline
\multicolumn{1}{c}{mass} &
\multicolumn{4}{|c|}{FM01} &
\multicolumn{4}{c}{KLA01} \\
\hline
\multicolumn{1}{c}{${\log\frac M{{\rm M}_{10}}}$} &
\multicolumn{1}{c}{$\sigma_0$} &
\multicolumn{1}{c}{$b_1$} &
\multicolumn{1}{c}{$b_2$} &
\multicolumn{1}{c}{$\log b_2$} &
\multicolumn{1}{c}{$\sigma_0$} &
\multicolumn{1}{c}{$b_1$} &
\multicolumn{1}{c}{$b_2$} &
\multicolumn{1}{c}{$\log b_2$} \\
\hline
-1$\quad$ & 15.08 & $1.00~10^1$ & $3.43~10^2$ & 2.535 & 
4.200 & $1.72~10^1$ & $7.41~10^0$ & 0.870 \\
0$\quad$ & 11.85 & $2.75~10^1$ & $1.66~10^3$ & 3.221 & 
3.300 & $4.73~10^1$ & $3.59~10^1$ & 1.555 \\
1$\quad$ &  $\phantom{1}$8.61 & $8.14~10^1$ & $6.38~10^3$ & 3.805 & 
2.400 & $1.40~10^2$ & $1.38~10^2$ & 2.141 \\
2$\quad$ &  $\phantom{1}$5.71 & $2.64~10^2$ & $1.86~10^4$ & 4.270 & 
1.590 & $4.55~10^2$ & $4.02~10^2$ & 2.604 \\
3$\quad$ &  $\phantom{1}$3.66 & $8.89~10^2$ & $4.90~10^4$ & 4.690 & 
1.020 & $1.53~10^3$ & $1.06~10^3$ & 3.026 \\
4$\quad$ &  $\phantom{1}$2.05 & $3.42~10^3$ & $8.61~10^4$ & 4.935 & 
0.570 & $5.90~10^3$ & $1.85~10^3$ & 3.268 \\
5$\quad$ & $\phantom{1}$0.937 & $1.61~10^4$ & $8.23~10^4$ & 4.915 & 
0.261 & $2.77~10^4$ & $1.77~10^3$ & 3.249 \\
6$\quad$ & $\phantom{1}$0.345 & $9.43~10^4$ & $4.11~10^4$ & 4.613 & 
0.096 & $1.62~10^5$ & $8.84~10^2$ & 2.947 \\
\hline\hline
\end{tabular}
\caption{The mass (in decimal logarithm and unit
${\rm M}_{10}=10^{10}{\rm M}_\odot$), the
present-day rms mass excess, $\sigma_0$=$<(\delta
M/M)_0^2>^{1/2}$, and the parameters,
$b_1=\bar{\nu}_{rec}\zeta_{max}
R_{max}/{{\rm kpc}}$, $b_2=\delta/(\bar{\nu}_
{rec}^3C_\delta\zeta_{max})$,
related to top-hat, spherical
perturbations with same mass spectrum as in 
Gunn (1987), but normalized to cosmological
models assumed in FM01 and KLA01 simulations,
respectively.   The plot of the mass spectrum 
(Gunn 1987) has been assumed to reproduce only
the growing mode.}
\label{t:del}
\end{table}
The parameter, $\delta$, expressed by
Eqs.\,(\ref{seq:deln}), is plotted as a
function of the mass, for lower and
upper values of the factor, $C_\delta$,
expressed by Eq.\,(\ref{seq:Cd}), and
mean peak heights $\bar{\nu}_{rec}=1,
2,3,4,$ in Fig.\,\ref{f:del1}, where
the values deduced from FM01 (top panel)
and KLA01 (bottom panel) simulations, in
connection with MOA (top panel) and NFW
(bottom panel) density profiles, 
respectively, are also represented.   
In both cases it is apparent that, for
increasing masses, simulated haloes 
correspond to fitting, top-hat haloes
with increasing (mean) peak height.

The occurrence of cosmological constant changes
the value of the treshold, beyond which
density perturbations are expanding 
forever (e.g., Lokas 
\& Hoffman 2001a,b).   The mass 
of the critical density perturbation,
related to an infinite turnaround 
radius,
%and $\delta=0$, 
is marked by a vertical line in Fig.\,\ref
{f:del1}, from the left
to the right for increasing peak
height.

Keeping in mind that, for a fixed
peak, upper curves are related to
no acquisition of angular momentum
$(\zeta_{vir}=8)$ and lower curves are
related to maximal, allowed acquisition
of angular momentum $(\zeta_{vir}=1)$, the
correspondence between simulated and
fitting, spherical, top-hat haloes 
implies an efficient acquisition of 
angular momentum, especially for low 
masses.   On the other hand, with
regard to FM01 runs, simulated 
haloes - especially at low masses - 
appear to be consistent with fitting, 
spherical, top-hat haloes with mean 
peak heights within the range
$1\appleq\bar{\nu}_{rec}\appleq2$, 
contrary to what would be expected,
$\bar{\nu}_{rec}\appgeq2$; with
regard to KLA01 runs, the above
mentioned discrepancy disappears.

\section{Concluding remarks}\label{core}

Simulated dark matter haloes have been fitted by
self-similar, universal density profiles, where
the scaled parameters depend only on a scaled
(truncation) radius, $\Xi=R/r_0$, which, in turn,
has been supposed to be independent of the mass
and the formation redshift.
The further assumption of a lognormal distribution
(for a selected mass bin) of the scaled 
radius, or concentration, in agreement with the
data from a large statistical sample of simulated 
haloes (Bullock et al., 2001), has allowed (at
least to a first extent) a normal or lognormal
distribution for other scaled parameters, via 
the same procedure which leads to the propagation
of the errors.

A criterion for the choice of the best fitting density
profile has been proposed, with regard to a set of
high-resolution simulations, where some averaging
procedure on scaled density profiles has been
performed, in connection with a number of fitting
density profiles.   To this aim, a minimum value of
the ratio, $\vert x_{\overline{\eta}}\vert/\sigma_{s\,
\overline{\eta}}=
\vert\overline{\eta}-\eta^\ast\vert/\sigma_{s\,
\overline{\eta}}$, has been required to
yield the best fit, where $\overline{\eta}$ is the
arithmetic mean over the whole set; $\eta^\ast$ is
its counterpart related to the fitting density profile;
$\sigma_{s\,\overline{\eta}}$ is the standard deviation
from the mean; and $\eta$ is a selected, scaled i.e.
dimensionless parameter.

The above criterion has been
applied to a pair of sets each made of a dozen of
high-resolution simulations, FM01 (Fukushige \& Makino
2001) and KLA01 (Klypin et al. 2001), in connection
with two currently used density profiles,  NFW (e.g.,
Navarro et al. 1997) and MOA (e.g., Moore et al. 1999),
where the dependence of the scaled radius on the mass
and the formation redshift, may be neglected to a first
extent.
%and in any case for a sufficiently narrow mass range.
With regard to FM01
and KLA01 samples, the best fit has revealed to be 
MOA and NFW, respectively.   In addition, the above results
have been found to hold also in dealing with rms errors
derived via the propagation of the errors, with
regard to the distributions of scaled parameters.
The sensitivity error of simulations has also been estimated
and shown to be less than the related, standard deviation,
that is a necessary condition for detectability of
accidental errors.

Some features of the early
evolution of dark matter haloes represented by fitting
density profiles, have been discussed in the limit of the
spherical top-hat model.   Though the related matter
distributions have appeared to be poorly representative of simulated
haloes, unless the (mean) peak height is an increasing function
of the mass, the results have been shown to be consistent,
provided considerable acquisition
of angular momentum took place during the expansion phase. 

\section{acknowledgements}
We thank A. Gardini, L. Moscardini, E. Rasia, and 
G. Tormen for helpful 
discussions, and E. Lokas
for enlightening explanations on Lokas \& Hoffman papers  
referenced below.   We are also grateful to S.H. Hansen
for pointing our attention to a paper of his, referenced
below, and related comments.
%\end{acknowledgements}

\section{Appendix}

\subsection{A. Some properties of spherical, top-hat
density perturbations}\label{rhov}

With regard to a special class of QCDM cosmological 
models, where quintessence obeys the equation of
state $p_\Psi=w\rho_\Psi$ relating its density,
$\rho_\Psi$, to its pressure, $p_\Psi$, via a
time-independent parameter or quintessence index,
$w$, where $-1\le w<0$, the following relations hold 
(e.g., Lokas 2002; Lokas \& Hoffman 2002):
\begin{lefteqnarray}
\label{eq:a}
&& \frac{\diff a}{\diff t}=\frac{H_0}{\phi(a)}~~;
\qquad a=\frac R{R_0}=\frac1{1+z}~~; \\
\label{eq:phia}
&& \phi(a)=\left[1+\Omega_0\left(a^{-1}-1\right)
+\Psi_0\left(a^{-1-3w}-1\right)\right]^{-1/2}~~; \\
\label{eq:Omega}
&& \Omega(z)=\Omega_0(1+z)^3\frac{H_0^2}{H^2(z)}
~~; \\
\label{eq:Psi}
&& \Psi(z)=\Psi_0(1+z)^{3(1+w)}\frac{H_0^2}{H^2(z)}
~~; \\
\label{eq:H}
&& \frac{H^2(z)}{H_0^2}=(1+z)^2(1+\Omega_0z)+\Psi_0
(1+z)^{3(1+w)}\left[1-(1+z)^{-(1+3w)}\right]~~;
\end{lefteqnarray}
where $a$ is the scale factor normalized to unity
at present, $z$ is the redshift, $\Omega$ and 
$\Psi$ the density parameter of matter and
quintessence, respectively, $H$ the Hubble 
parameter, and the index 0 denotes the current 
time.   In the limit of a vanishing quintessence,
Eqs.\,(\ref{eq:a}), (\ref{eq:Omega}), and (\ref
{eq:H}) reduce to their counterparts in cosmological
models with sole matter and radiation (e.g., 
Zeldovich \& Novikov 1982, Chap.\,III, \S 4).
The special case, $w=-1$, corresponds to the
cosmological constant.

An equivalent expression of Eqs.\,(\ref{eq:Omega})
and (\ref{eq:Psi}), using (\ref{eq:H}), is:
\begin{lefteqnarray}
\label{eq:O-1}
&& \Omega^{-1}(z)=1+\frac1{1+z}\frac{1-\Omega_0-
\Psi_0}{\Omega_0}+(1+z)^{3w}\frac{\Psi_0}{\Omega
_0}~~; \\
\label{eq:P-1}
&& \Psi^{-1}(z)=1+\frac{\Omega_0}{\Psi_0}(1+z)^
{-3w}\left[1+\frac1{1+z}\frac{1-\Omega_0-\Psi_0}
{\Omega_0}\right]~~;
\end{lefteqnarray}
and Eq.\,(\ref{eq:O-1}) allows the validity of
the inequality:
\begin{equation}
\label{eq:uli}
1+\frac1{1+z}\frac{1-\Omega_0-\Psi_0}{\Omega_0+
\Psi_0}\le\Omega^{-1}(z)\le1+\frac1{1+z}\frac
{1-\Omega_0}{\Omega_0}~~;\qquad-1\le w\le-
\frac13~~;
\end{equation}
which shows that the evolution of the matter
density parameter in a cosmological model
defined by assigned values of $(\Omega_0, 
\Psi_0)$, is comprised between its counterparts 
related to cosmological models defined by 
$(\Omega_0, 0)$ and $(\Omega_0+\Psi_0, 0)$,
respectively.

With regard to the function, $\phi(a)$,
defined by Eq.\,(\ref{eq:phia}), a detailed
analysis involving the solution of a third-degree
equation yields the final result:
\begin{equation}
\label{eq:phad}
\left[1+\Omega_0\left(a^{-1}-1\right)\right]^
{-1/2}\le\phi(a)<\left[1+\frac{\Omega_0}2\left(
a^{-1}-1\right)\right]^{-1/2}~~;\qquad\Omega_0
\ge0.25~~;
\end{equation}
which shows that the evolution of the scale 
factor in a cosmological model
defined by assigned values of $(\Omega_0, 
\Psi_0)$, is comprised between its counterparts 
related to cosmological models defined by 
$(\Omega_0/2, 0)$ and $(\Omega_0, 0)$,
respectively, provided $\Omega_0\ge0.25$.

With regard to the Hubble parameter, $H(t)$,
defined by Eq.\,(\ref{eq:H}), a detailed
analysis involving the solution of a transcendental
equation yields the final result:
\begin{equation}
\label{eq:Hd}
\frac12(1+z)^2(1+\Omega_0z)<\frac{H^2(z)}{H_0^2}
\le(1+z)^2(1+\Omega_0z)~~;\qquad\Omega_0\ge0.25~~;
\end{equation}
which shows that the evolution of the Hubble
parameter in a cosmological model
defined by assigned values of $(\Omega_0, \Psi_0,
H_0)$, is comprised between its counterparts 
related to cosmological models defined by 
$(\Omega_0, 0, H_0)$ and $(\Omega_0, 0, H_0/
\sqrt{2})$, respectively, provided $\Omega_0
\ge0.25$.

Cosmological models with sole matter and radiation 
evolve at the same rate in the limit $\vert1-
\Omega^{-1}\vert\ll1$ i.e. at early times (e.g., 
Zeldovich \& Novikov 1982, Chap.\,III, \S 4)
and, owing to inequalities (\ref{eq:uli}),
(\ref{eq:phad}), and (\ref{eq:Hd}), the same
holds for cosmological models with
quintessence.   In the above mentioned limit,
the matter density of the Hubble flow, $\rho_
h$, reads (e.g., Peebles 1993, Chap.\,II, \S 
13):
\begin{equation}
\label{eq:rhhL}
\rho_h=\frac3{8\pi}\frac{H_0^2\Omega_0}G(1+z)^3~~;
%\qquad\vert1-(\Omega+\Psi)^{-1}\vert\ll1~~;
\end{equation}
where $G$ is the constant of gravitation.

The combination of Eqs.\,(\ref{eq:rhv}) and (\ref
{eq:rhhL}) yields:
\begin{lefteqnarray}
\label{eq:rhvr}
&& \bar{\rho}_{vir}=\frac3{8\pi}\zeta_{max}\zeta_{vir}
\frac{H_0^2\Omega_0}G(1+z_{rec})^3\left[\left(\frac
{\delta M^\ast}M\right)_{rec}\right]^3~~; \nonumber \\
&& \vert1-\Omega_{rec}
%+\Psi_{rec})
^{-1}
\vert\ll\left(\frac{\delta M^\ast}M\right)_{rec}
\ll1~~;
\end{lefteqnarray}
in the framework of the top-hat, spherical model,
related to a flat universe with vanishing quintessence,
the growing mode of the density perturbation 
attains the present-day value (e.g., Peebles 1980, 
Chap.\,II, \S 15):
\begin{equation}
\label{eq:pero}
\left(\frac{\delta M}M\right)_0=\frac35\left(
\frac{\delta M^\ast}M\right)_{rec}(1+z_{rec})
%=\frac35\left(\frac{\delta M^\ast}M\right)_0
~~;
\end{equation}
and the combination of Eqs.\,(\ref{eq:rhvr}) and 
(\ref{eq:pero}) yields Eq.\,(\ref{eq:rhvo}).

In addition, the combination of Eqs.\,(\ref
{eq:rRrm}), (\ref{eq:rrhmr}), (\ref{eq:rhhL}), 
and (\ref{eq:pero}) produces:
\begin{equation}
\label{eq:nuRm}
\bar{\nu}_{rec}\zeta_{max}R_{max}=\frac35\left(\frac{H_0^2
\Omega_0}{2G}\right)^{-1/3}M^{1/3}\left\langle\left[\left(
\frac{\delta M}M\right)_0\right]^2\right\rangle^{-1/2};~
\vert1-\Omega_{rec}
%+\Psi_{rec})
^{-1}\vert\ll1;
\end{equation}
where the mass excess has been expressed as
the product of the present-day rms mass excess
and the peak height at recombination epoch,
averaged over the whole volume, that is:
\begin{equation}
\label{eq:rmsml}
\left(\frac{\delta M}M\right)_{rec}=\bar{\nu}_
{rec}\left\langle\left[\left(\frac{\delta M}M
\right)_{rec}\right]^2\right\rangle^{1/2}~~;
\end{equation}
for more details see e.g., Caimmi et al. 
(1990).   Of course, Eq.\,(\ref{eq:nuRm})
makes a lower limit to the product, $\bar
{\nu}_{rec}R_{max}$, as Eq.\,(\ref{eq:rRrm})
holds in the limit of a vanishing quintessence.

On the other hand, in open or flat universes,
density perturbations below a treshold are
destined to expand forever.   In the special
case of cosmological constant, the critical
value is (e.g., Lokas \& Hoffman 2001a,b): 
\begin{leftsubeqnarray}
\slabel{eq:Minfa}
&& \left[\left(\frac{\delta M}M\right)_{rec}
\right]_\infty=\frac{U(\Lambda_{rec})}
{\Omega_{rec}}-1~~; \\
%\slabel{eq:Minfb}
%&& U(\Lambda)=1-\Psi~~;\qquad w>-1~~; \\
\slabel{eq:Minfc}
&& U(\Lambda)=1+\frac54\Lambda+
\frac34\frac{\Lambda(8+\Lambda)}{T(\Lambda)}
+\frac34T(\Lambda)~~;
%\qquad w=-1~~; 
\\
\slabel{eq:Minfd}
&& T(\Lambda)=\Lambda^{1/3}\left[8-\Lambda^
2+20\Lambda+8(1-\Lambda)^{3/2}\right]^{1/3}
~~;
\label{seq:Minf}
\end{leftsubeqnarray}
where the turnaround occurs at an infinite
radius.
%, which tends to be infinite in
%the case under discussion, and $\Gamma\le1$ 
%for $w\ge-1.
%It is apparent that $U[\Psi_{rec},(\Gamma_
%{rec})_\infty]\le0$ for $w>-1$ and $U[\Psi_
%{rec},(\Gamma_{rec})_\infty]\ge0$ for $w=-1$.
%Accordingly, in universes with cosmological
%constant $(w=-1)$ density perturbations
%have to be larger in order to collapse than
%in universes with same density parameter,
%$\Omega_{rec}$, but no cosmological constant,
%where, in turn, density perturbations have to 
%be larger in order to collapse than in
%universes with same density parameter,
%$\Omega_{rec}$, but with quintessence $(w>-1)$.

With regard to FM01 and KLA01 simulations,
the present-day rms mass excess in a spherical
region of radius $R_8=8h^{-1}~$Mpc, takes the
value:
\begin{leftsubeqnarray}
\slabel{eq:sig08a}
&& (\sigma_8)_{FM}=\left\{\left\langle\left[\left(\frac
{\delta M}{M_8}\right)_0\right]^2\right\rangle^{1/2}
\right\}_{FM}=0.7~~; \\
\slabel{eq:sig08b}
&& (\sigma_8)_{KLA}=\left\{\left\langle\left[\left(\frac
{\delta M}{M_8}\right)_0\right]^2\right\rangle^{1/2}
\right\}_{KLA}=0.9~~;
\label{seq:sig08}
\end{leftsubeqnarray}
owing to the general definition of mass excess:
\begin{equation}
\label{eq:delM}
\frac{\delta M}M=\frac{\bar{\rho}-\rho_h}{\rho_
h}~~;
\end{equation}
the mass within the region under consideration
may be obtained using Eqs.\,(\ref{eq:rhhL}), 
(\ref{seq:sig08}), and (\ref{eq:delM}).    The
result is:
\begin{equation}
\label{eq:M08}
M_8=2^8~10^9~(1+\sigma_8)\frac{H_0^2\Omega_0}
{h^3G}~~;
%\qquad\vert1-(\Omega+\Psi)^{-1}\vert\ll1~~;
\end{equation}
which may be specified for any flat cosmological
model with assigned values of $H_0$, $\Omega_0$,
%$\Lambda$, 
and $\sigma_8$.

\subsection{B. Determination of fitting haloes to
FM01 and KLA01 simulated haloes, with regard
to NFW and MOA density profiles}\label{k1}

Given a selected, FM01 simulated halo among
the runs listed in Table\,\ref{t:SFM} i.e.
for which the values of the parameters,
$M_{200}$, $r_{200}$, and $\delta$, are
known, we aim to derive the values of the
scaling density, $\rho_0$, and the scaling
radius, $r_0$, related to the corresponding,
fitting halo, and then the remaining
parameters, in connection with both NFW and
MOA density profiles.   The total mass of
the simulated halo, $M$, appears in FM01
prescriptions, expressed by Eqs.\,(\ref
{seq:rho0}) and (\ref{seq:r0}), but the
related values are not reported therein.
%
%it is approximated by the virial mass,
%$M_{200}$, enclosed within the region
%bounded by an isopycnic surface, enclosing
%a mean density, $\bar{\rho}_{200}=200\rho_
%{crit}$, which is reported for each run in 
%FM01.
%
Then the key parameter is the scaled
radius, $\xi_{200}$.

Owing to the general definition of scaled 
radius, Eq.\,(\ref
{eq:r0a}) may be written under the
equivalent form:
\begin{equation}
\label{eq:csi2N}
\xi_{200}=\frac{r_{200}}{r_0}=C_r^{-1}\delta^
{1/3}\frac{r_{200}}{{\rm kpc}}\left(\frac M{{\rm M}_{10}}
\right)^{-2/3}~~;
\end{equation}
on the other hand, the particularization of
the general expression of the mass enclosed
within a generic, isopycnic surface,
to the case under discussion,
via Eqs.\,(\ref{eq:f}) and (\ref{eq:csi}),
reads: 
\begin{equation}
\label{eq:M200}
M_{200}=3M_0\int_0^{\xi_{200}}f(\xi)\xi^2\diff
\xi~~;
\end{equation}
and the combination of Eqs.\,(\ref{eq:csi2N})
and (\ref{eq:M200}) produces a transcendental 
equation in $M$, which can be solved in
connection with an assigned (NFW or MOA)
density profile, provided the values of the
parameters, $C_r$, $\delta$, $r_{200}$, and
$M_{200}$, are specified using the data
listed in Table\,\ref{t:SFM}.

The knowledge of the total mass of the
simulated halo, assumed to coincide with the
mass of the fitting halo, $M$, allows
the calculation of the scaling density, $\rho
_0$, and the scaling radius, $r_0$, via
Eqs.\,(\ref{seq:rho0}) and (\ref{seq:r0}),
and then the radius along a fixed direction,
$R=\Xi r_0$, the scaling mass, $M_0$,
appearing in Eq.\,(\ref{eq:MM0}),
and the dimensionless parameter, 
$\kappa$, expressed by Eq.\,(\ref{eq:k2}).

With regard to NFW density profiles, the
particularization of Eq.\,(\ref{eq:M200})
to the case under discussion, 
yields:
\begin{equation}
\label{eq:M200N}
M_{200}=12M_0\left[\ln(1+\xi_{200})-\frac
{\xi_{200}}{1+\xi_{200}}\right]~~;
\end{equation}
where, owing to Eqs.\,(\ref{eq:MM0}) and
(\ref{eq:snuMN}):
\begin{equation}
\label{eq:nuM2N}
\frac1{12}\frac{M_{200}}{M_0}=\frac{125}{28\pi}
\frac{M_{200}}M~~;
\end{equation}
and the combination of Eqs.\,(\ref{eq:M200N})
and (\ref{eq:nuM2N}) yields:
\begin{equation}
\label{eq:traN}
\frac{M_{200}}M=\frac{28\pi}{125}\left[\frac1
{1+\xi_{200}}-\ln\frac1{1+\xi_{200}}-1\right]~~;
\end{equation}
finally, the combination of Eqs.\,(\ref
{eq:r0b}), (\ref{eq:csi2N}), and (\ref
{eq:traN}) produces the 
ultimate transcendental equation in $M$.

With regard to MOA density profiles, the
particularization of Eq.\,(\ref{eq:M200})
to the case under discussion, 
yields:
\begin{equation}
\label{eq:M200M}
M_{200}=4M_0\ln\left(1+\xi_{200}^{3/2}\right)~~;
\end{equation}
where, owing to Eqs.\,(\ref{eq:MM0}) and
(\ref{eq:snuMM}):
\begin{equation}
\label{eq:nuM2M}
\frac14\frac{M_{200}}{M_0}=\frac{375}{56\pi}
\frac{M_{200}}M~~;
\end{equation}
and the combination of Eqs.\,(\ref{eq:M200M})
and (\ref{eq:nuM2M}) yields:
\begin{equation}
\label{eq:traM}
\frac{M_{200}}M=\frac{56\pi}{375}\ln\left(1+\xi_{200}
^{3/2}\right)~~;
\end{equation}
finally, the combination of Eqs.\,(\ref
{eq:r0b}), (\ref{eq:csi2N}), and (\ref
{eq:traM}), produces the ultimate 
transcendental equation in $M$.

The above procedure, via Eqs.\,(\ref
{eq:Crrh})-(\ref{eq:nuMN}), also holds
for a selected, KLA01 simulated halo
among the runs listed in Table \ref
{t:SKL}, for which the values of the
parameters, $M_{vir}$, $r_{vir}$, and
$\delta$, are known.

With regard to both NFW and MOA density
profiles, some parameters related to
fitting haloes, in connection with
simulated haloes from FM01 and KLA01, 
are listed in Tables \ref{t:RFMN} and
\ref{t:RKAN}, respectively.

\subsection{C. Sensitivity errors of 
dark matter halo simulations}\label{sene}

Bearing in mind the general results listed
in Tab.\,\ref{t:proMN}, together with
Eq.\,(\ref{eq:k2}), the scaled parameters,
$M/M_0$, $R/r_0$, $\bar{\rho}/{\rho_0}$,
and $\kappa$, depend only on the fitting density
profile.
%, which is assumed to be self-similar and universal.
For a selected choice of
exponents $(\alpha,\beta,\gamma)$, fitting 
haloes depend on two parameters, $(r_0,\rho_0)$, 
or $(M,\delta)$, via Eqs.\,(\ref
{seq:rho0}) and (\ref{seq:r0}).

Given a computer run with $N$ identical
particles of mass $m$, the sensitivity
error with respect to the mass is clearly
expressed as $\Delta M=m$.   It follows
that:
\begin{equation}
\label{eq:esrM}
\Delta\frac{M_{200}}{M_0}=\left(1+\frac
{M_{200}}{M_0}\right)\frac{M_{200}}{M_0}
\frac m{M_{200}}~~;\quad\Delta M_0=\Delta
M_{200}=m~~;
\end{equation}
the second parameter, $\delta$, is 
proportional to the present-day 
mass excess of the growing mode
predicted by the top-hat model,
$(\delta M/M)_0$, as:
\begin{leftsubeqnarray} 
\slabel{eq:delna2}
&& \delta=C_\delta\left[\left(\frac{\delta M}M
\right)_0\right]^3\frac M{{\rm M}_{10}}~~; \\
\slabel{eq:delnb2}
&& C_\delta=\frac{125}{54}\zeta_{vir} C_r^3\Xi^3
\frac{H_0^2\Omega_0}G\frac{{\rm kpc}^3}{{\rm 
M}_{10}}~~;
\label{seq:deln2}
\end{leftsubeqnarray}
where $\Omega_0$ is the present-day, matter 
density parameter ($\Omega_0+\Lambda_0
=1$), and $\zeta_{vir}$ depends on the evolution
of the density perturbation during the expansion 
phase, and lies within the range $1\le\zeta_{vir}
\le8$.
For a formal demonstration, see Appendix A.
%\ref{rhov}.
%,according to Eq.\,(\ref{eq:delna}).  

The repetition of the above procedure
yields:
\begin{leftsubeqnarray}
\slabel{eq:esda}
&& \Delta\delta=\delta\frac{3M_{200}+2(\delta
M)_0}{M_{200}(\delta M)_0}m=\delta\left[2+3
C_\delta^{1/3}\delta^{-1/3}\left(\frac{M_
{200}}{{\rm M}_{10}}\right)^{1/3}\right]
\frac m{M_{200}}~~; \\
\slabel{eq:esdb}
&& \Delta(\delta M)_0=\Delta M_{200}=m~~;
\label{seq:esd}
\end{leftsubeqnarray}
and the sensitivity error with respect to
$\rho_0$ and $r_0$, by use of Eqs.\,(\ref
{seq:rho0}) and (\ref{eq:MM0}), is:
\begin{lefteqnarray}
\label{eq:esrh}
&& \Delta\rho_0=3\rho_0\frac{M_{200}+(\delta M)_0}
{M_{200}(\delta M)_0}m=3\rho_0\left[1+C_\delta^
{1/3}\delta^{-1/3}\left(\frac{M_{200}}{{\rm 
M}_{10}}\right)^{1/3}\right]\frac m{M_{200}}~~; \\
\label{eq:esr}
&& \Delta r_0=\frac13r_0\left[4+3C_\delta^
{1/3}\delta^{-1/3}\left(\frac{M_{200}}{{\rm 
M}_{10}}\right)^{1/3}\right]\frac m{M_{200}}~~;
\end{lefteqnarray}
finally, the further assumption $\Delta r_{200}
=\Delta r_0$ allows the following results:
\begin{lefteqnarray}
\label{eq:esrR}
&& \Delta\frac{r_{200}}{r_0}=\frac13\left(1+\frac
{r_{200}}{r_0}\right)\left[4+3C_\delta^{1/3}\delta
^{-1/3}\left(\frac{M_{200}}{{\rm M}_{10}}\right)^
{1/3}\right]\frac m{M_{200}}~~; \\
\label{eq:esrrh}
&& \Delta\frac{\bar{\rho}_{200}}{\rho_0}=\Delta
\left(\frac{M_{200}}{M_0}\frac{r_0^3}{r_{200}^3}
\right) \nonumber \\
&& \phantom{\Delta\frac{\bar{\rho}_{200}}{\rho_0}}
=\frac{\bar{\rho}_{200}}{\rho_0}\left[5+
\frac{M_{200}}{M_0}+3C_\delta^{1/3}\delta
^{-1/3}\left(\frac{M_{200}}{{\rm M}_{10}}\right)^
{1/3}\right]\left(1+\frac{r_0}{r_{200}}\right)
\frac m{M_{200}}~~; \\
\label{eq:esc}
&& \Delta\kappa_{200}=\kappa_{200}\left(1+
\frac{r_0}{r_{200}}\right)\left[2+\frac32C_\delta
^{1/3}\delta^{-1/3}\left(\frac{M_{200}}{{\rm M}_{10}}
\right)^{1/3}\right]\frac m{M_{200}}~~;
\end{lefteqnarray}
where Eqs.\,(\ref{seq:rho0}) and (\ref
{eq:k1}) have been used.

The sensitivity error of FM01 computer
runs in dark matter halo simulations,
expressed by Eqs.\,(\ref{eq:esrM}), 
(\ref{eq:esrR}), (\ref{eq:esrrh}) and
(\ref{eq:esc}), may be calculated and
compared with their rms counterparts,
to see if a necessary condition for
the detectability of accidental errors
is satisfied, namely:
\begin{equation}
\label{eq:teor}
\Delta\eta\le\sigma_\eta~~;
\end{equation}
where $\eta=M_{200}/M_0$, $r_{200}/r_0$, 
$\bar{\rho}_{200}/\rho_0$, $\kappa_{200}$.
To this aim, an inspection of Tabs.\,\ref
{t:SFM}, \ref{t:CFM}, and \ref{t:RFMN}, 
shows that the following inequalities hold:
\begin{leftsubeqnarray}
\slabel{eq:ine1a}
&& \frac m{M_{200}}<10^{-6}~~;\quad\frac{M_{200}}
{M_0}<20~~;\quad\frac{r_{200}}{r_0}<20~~;\quad
\frac{r_0}{r_{200}}<\frac12~~; \\
\slabel{eq:ine1b}
&& \frac{\bar{\rho}_{200}}{\rho_0}<1~~;
\quad\kappa_{200}<2.8~~;\quad C_\delta^{1/3}
\delta^{-1/3}\left(\frac{M_{200}}{{\rm M}_{10}}
\right)^{1/3}<0.2~~;
\label{seq:ine1}
\end{leftsubeqnarray}
where upper values of $C_\delta$ have been used,
in connection with the range $1\le\zeta_{vir}\le8$,
according to Eq.\,(\ref{eq:delnb}).
%according to Eqs.\,(\ref{eq:Cda}) and (\ref{eq:Cdb}).
The combination of Eqs.\,(\ref{eq:esrM}), 
(\ref{eq:esrR}), (\ref{eq:esrrh}), (\ref
{eq:esc}), and (\ref{seq:ine1}) yields:
\begin{equation}
\label{eq:ine2}
\Delta\frac{M_{200}}{M_0}<10^{-3}~~;\quad
\Delta\frac{r_{200}}{r_0}<10^{-4}~~;\quad
\Delta\frac{\bar{\rho}_{200}}{\rho_0}<10^{-4}~~;
\quad\Delta\kappa_{200}<10^{-5}~~.
\end{equation}

The sensitivity errors of KLA01 computer runs
in dark matter halo simulations, are expressed
in the same way, provided
$u_{200}$ is replaced by $u_{vir}$
therein, where $u=M$, $r$, $\bar{\rho}$,
$\kappa$.
An inspection of Tabs.\,\ref
{t:SKL}, \ref{t:CKA}, \ref{t:RKAN}, shows that the following
inequalities hold:
\begin{lefteqnarray}
\label{eq:ine3a}
&& \frac m{M_{vir}}<6~10^{-4}~~;\quad\frac{M_{vir}}
{M_0}<30~~;\quad\frac{r_{vir}}{r_0}<20~~;\quad
\frac{r_0}{r_{vir}}<0.2~~; \\
\label{eq:ine3b}
&& \frac{\bar{\rho}_{vir}}{\rho_0}<0.02~~;
\quad\kappa_{vir}<20~~;\quad C_\delta^{1/3}
\delta^{-1/3}\left(\frac{M_{vir}}{{\rm M}_{10}}
\right)^{1/3}<1.2~~;
\end{lefteqnarray}
where upper values of $C_\delta$ have been used,
in connection with the range $1\le\zeta_{vir}\le8$,
according to Eq.\,(\ref{eq:delnb}).
%Eqs.\,(\ref{eq:Cdc}) and (\ref{eq:Cdd}).
Following the same procedure used for FM01, 
yields the final result:
\begin{equation}
\label{eq:ine4}
\Delta\frac{M_{vir}}{M_0}<1~~;\quad
\Delta\frac{r_{vir}}{r_0}<4~10^{-2}~~;\quad
\Delta\frac{\bar{\rho}_{vir}}{\rho_0}<5~10^{-4}~~;
\quad\Delta\kappa_{vir}<6~10^{-2}~~;
\end{equation}

The comparison between the sensitivity
errors, expressed by Eqs.\,(\ref{eq:ine2})
and (\ref{eq:ine4}), and their rms 
counterparts, deduced from 
Table\,\ref{t:MFM}, shows that a necessary
condition for the detectability of accidental
errors, expressed by Eq.\,(\ref{eq:teor}), is
satisfied for both FM01 and KLA01 simulations.   
To this aim, it is worth remembering
that $\sigma_{\bar{\eta}}=\sigma_{\eta}/
\sqrt{N}$,
where $N=12$ in the case under consideration,
according to the theory of the errors.

\subsection{D. rms errors of distributions
depending on scaled parameters}\label{indi}

Let dark matter haloes be fitted by 
universal density profiles, expressed by
Eq.\,(\ref{eq:f}), and let the distribution
depending on the scaled radius, $\Xi$ (or
concentration with regard to NFW density
profiles), be lognormal.    The 
scaled mass enclosed within the generic
scaled distance, $\xi$, and the related
scaled mean density, are:
\begin{lefteqnarray}
\label{eq:Mcsi}
&& \frac{M(\xi)}{M_0}=3\int_0^\xi f(\xi)\xi^2
\diff\xi~~; \\
\label{eq:romc}
&& \frac{\bar{\rho}(\xi)}{\rho_0}=\frac3\xi
\int_0^\xi f(\xi)\xi^2\diff\xi~~;
\end{lefteqnarray}
and the generalization of the dimensionless
parameter, $k$, defined by Eq.\,(\ref{eq:k2}),
to the generic scaled radius, $\xi$, reads:
\begin{equation}
\label{eq:kcsi}
k(\xi)=C_r^{3/2}\xi^{3/2}~~;
\end{equation}
where the constant, $C_r$, is determined by
averaging on the results of simulations 
(FM01), and for the cases of interest it
is expressed by Eqs.\,(\ref{eq:r0b}) and 
(\ref{eq:Crrh}).

The first derivatives of the functions on
the left-hand side of Eqs.\,(\ref{eq:Mcsi}), 
(\ref{eq:romc}), and (\ref{eq:kcsi}), are:
\begin{lefteqnarray}
\label{eq:dMc}
&& \frac{\diff(M/M_0)}{\diff\xi}=3f(\xi)
\xi^2~~; \\
\label{eq:droc}
&& \frac{\diff(\bar{\rho}/\rho_0)}{\diff\xi}=
\frac3\xi\left[f(\xi)-\frac1{\xi^3}\frac{M(\xi)}
{M_0}\right]~~; \\
\label{eq:dkc}
&& \frac{\diff k}{\diff\xi}=\frac32\frac{k(\xi)}
\xi~~;
\end{lefteqnarray}
and, in addition:
\begin{equation}
\label{eq:dlgc}
\frac{\diff\log\xi}{\diff\xi}=\frac1{\ln10}\frac1\xi~~;
\end{equation}
following the standard rules of derivation.

Let us suppose that (i) the scaled parameters,
$\eta$, $\eta=M_{trn}/{M_0}$, $\Xi_{trn}$, 
$\bar{\rho}_{trn}/\rho_0$, and $k_{trn}$, 
$trn=200,~vir$, as functions of $\log\Xi_{trn}$,
are expressible as Taylor series where the
starting point coincides with the expected 
value of the lognormal distribution of the
concentration, $\Xi=\exp_{10}(\log\Xi_{trn})
^\ast$; (ii) the convergence radius of the
series under discussion exceeds at least
three times the rms error of the lognormal
distribution, $\sigma_{\log\Xi_{trn}}$, or
in other words the convergence occurs at
least within the range, $\log\Xi\mp3\sigma_
{\log\Xi_{trn}}$; (iii) the series under
discussion can safely be approximated by
neglecting all the terms of higher order
with respect to the first.   It is worth
remembering that the propagation of the
errors, and the related formulae currently 
used in literature, are grounded on the 
above mentioned assumptions.

Let $\phi(\xi)$ be a generic, derivable
function of an independent variable, $\xi$.
The application of the theorem of the
derivative of a function of a function,
where the second function is $\log\xi$,
yields:
\begin{equation}
\label{eq:dffl}
\frac{\diff\phi}{\diff\log\xi}=\frac{\diff\phi}
{\diff\xi}\frac{\diff\xi}{\diff\log\xi}=\ln10~\xi
\frac{\diff\phi}{\diff\xi}~~;
\end{equation}
and the particularization to the scaled 
parameters considered here, reads:
\begin{lefteqnarray}
\label{eq:dMl}
&& \frac{\diff(M/M_0)}{\diff\log\xi}=3\ln10~f(\xi)
\xi^3~~; \\
\label{eq:drol}
&& \frac{\diff(\bar{\rho}/\rho_0)}{\diff\log\xi}=
3\ln10\left[f(\xi)-\frac1{\xi^3}\frac{M(\xi)}
{M_0}\right]~~; \\
\label{eq:dkl}
&& \frac{\diff k}{\diff\log\xi}=\frac32\ln10~k(\xi)~~;
\end{lefteqnarray}
owing to Eqs.\,(\ref{eq:dMc}), (\ref{eq:droc}), 
and (\ref{eq:dkc}).

On the other hand, the validity of the above
assumptions implies that the relations:
\begin{leftsubeqnarray}
\slabel{eq:setaa}
&& \eta(\log\Xi_{trn})=\eta(\log\Xi)+\left(
\frac{\diff\eta}{\diff\log\Xi_{trn}}\right)_
{\log\Xi}(\log\Xi_{trn}-\log\Xi)~~; \\
\slabel{eq:setab}
&& \eta=\frac M{M_0},~\Xi,~\frac{\bar{\rho}}
{\rho_0},~k~~;\qquad trn=200,vir~~; 
\label{seq:seta}
\end{leftsubeqnarray}
hold to a good extent.   The combination
of Eqs.\,(\ref{eq:dlgc}), (\ref{eq:dMl}), 
(\ref{eq:drol}), (\ref{eq:dkl}), and 
(\ref{seq:seta}), yields:
\begin{lefteqnarray}
\label{eq:sM}
&& \frac{M(\log\Xi_{trn})}{M_0}=\frac M{M_0}
+3\ln10f(\Xi)\Xi^3(\log\Xi_{trn}-\log\Xi)~~; \\
\label{eq:scs}
&& \Xi_{trn}(\log\Xi_{trn})=\Xi+\ln10\Xi(\log
\Xi_{trn}-\log\Xi)~~; \\
\label{eq:sro}
&& \frac{\bar{\rho}(\log(\Xi_{trn})}{\rho_0}=
\frac{\bar{\rho}}{\rho_0}+3\ln10\left[f(\Xi)-
\frac1{\Xi^3}\frac M{M_0}\right](\log\Xi_{trn}
-\log\Xi)~~; \\
\label{eq:sk}
&& k(\log\Xi_{trn})=k+\frac32\ln10(\log
\Xi_{trn}-\log\Xi)~~; \\
\end{lefteqnarray}
where $M=M(\Xi)$, $\bar{\rho}=\bar
{\rho}(\Xi)$, and $k=k(\Xi)$, for sake of
brevity.

The scaled parameter, $\log\Xi_{trn}$,
%
%s on the left-hand
%side of Eqs.\,(\ref{eq:sM}), (\ref{eq:scs}), 
%(\ref{eq:sro}), and (\ref{eq:sk}),
%
may be
considered as a physical quantity to be
measured directly.   Accordingly, the
distribution depending on $\log\Xi_{trn}$
has necessarily to be normal.   A theorem
related to the theory of errors
ensures that the scaled parameters, $\eta$,
defined by Eq.\,(\ref{seq:seta}), also
follow normal distributions, whose expected 
values and rms errors, via Eqs.\,(\ref
{eq:sM}), (\ref{eq:scs}), (\ref{eq:sro}), 
and (\ref{eq:sk}), are expressed as:
\begin{lefteqnarray}
\label{eq:etasp}
&& \eta^\ast=\frac M{M_0},~\Xi,~\frac{\bar
{\rho}}{\rho_0},~k~~; \\
\label{eq:etava}
&& \sigma_\eta=\left\vert\left(\frac{\diff
\eta}{\diff\log\Xi_{trn}}\right)_{\log\Xi}
\right\vert\sigma_{\log\Xi}~~; 
\end{lefteqnarray}
and the last relation, owing to Eqs.\,(\ref
{eq:dlgc}), (\ref{eq:dMl}), (\ref{eq:drol}), 
and (\ref{eq:dkl}), takes the explicit form:
\begin{lefteqnarray}
\label{eq:varM}
&& \sigma_{M/M_0}=3\ln10~\Xi^3f(\Xi)~\sigma_
{\log\Xi}~~; \\
\label{eq:varc}
&& \sigma_\Xi=\ln10~\Xi~\sigma_{\log\Xi}~~; \\
\label{eq:varr}
&& \sigma_{\bar{\rho}/\rho_0}=3\ln10\left
\vert f(\Xi)-\frac1{\Xi^3}\frac M{M_0}\right
\vert\sigma_{\log\Xi}~~; \\
\label{eq:vark}
&& \sigma_k=\frac32\ln10~k~\sigma_{\log\Xi}~~;
\end{lefteqnarray}
for the scaled parameters under consideration.

%The particularization of Eq.\,(\ref{eq:dffl})
%to the decimal logarithm of the scaled
%distance, $\eta$, via Eq.\,(\ref{eq:dlgc})
%yields:
%\begin{equation}
%\label{eq:dll}
%\frac{\diff\eta}{\diff\log\Xi_{trn}}=\frac1
%{\ln10}\frac1\eta\frac{\diff\eta}{\diff\Xi_
%{trn}}~~;
%\end{equation}
%and the combination of
%
Starting from Eq.\,(\ref{eq:etava}),
after replacing $\eta$ with $\log\eta$,
and using Eq.\,(\ref{eq:dlgc}),
after replacing $\xi$ with $\eta$, yields
%
%with $\log\eta$, (\ref{eq:dll}), allows
%
an expression of the
rms errors of lognormal distributions, in
terms of their counterparts related to
normal distributions.   The result is:
\begin{equation}
\label{eq:letav}
\sigma_{\log\eta}=\frac1{\ln10}\frac1\eta
\left\vert\left(\frac{\diff\eta}{\diff\log
\Xi_{trn}}\right)_{\log\Xi}\right\vert
\sigma_{\log\Xi}~~;
\end{equation}
or, after comparison with Eq.\,(\ref
{eq:etava}):
\begin{equation}
\label{eq:letet}
\sigma_{\log\eta}=\frac1{\ln10}\frac1\eta~
\sigma_\eta~~;
\end{equation}
which, owing to Eqs.\,(\ref{eq:varM}), 
(\ref{eq:varc}), (\ref{eq:varr}), and 
(\ref{eq:vark}), take the explicit form:
\begin{lefteqnarray}
\label{eq:valM}
&& \sigma_{\log(M/M_0)}=3\Xi^3f(\Xi)\left(
\frac M{M_0}\right)^{-1}\sigma_{\log\Xi}~~; \\
\label{eq:valc}
&& \sigma_{\log\Xi}=\sigma_{\log\Xi}~~; \\
\label{eq:valr}
&& \sigma_{\log(\bar{\rho}/\rho_0)}=3\left
\vert f(\Xi)-\frac1{\Xi^3}\frac M{M_0}\right
\vert\left(\frac{\bar{\rho}}{\rho_0}\right)^
{-1}\sigma_{\log\Xi}~~; \\
\label{eq:valk}
&& \sigma_{\log k}=\frac32~\sigma_{\log\Xi}~~;
\end{lefteqnarray}
where the identity has been written for
sake of completeness.

\subsection{E. Random model: the concentration
distribution as a result of the central
limit theorem}\label{lice}

Dark matter halo and star formation take
place in a similar fashion, namely a
transition from an undifferentiated fluid
to substructures.   Though a molecular
cloud is neither expanding nor subjected
to the Copernican principle, contrary to
the Hubble flow, still the above mentioned
processes are expected to exhibit some
common features.

The initial mass function in a star 
generation may safely be fitted by a
lognormal distribution (e.g., Adams \&
Fatuzzo 1996; Padoan et al. 1997), which,
in turn, can be interpreted in terms of
the central limit theorem (Adams \&
Fatuzzo 1996).   On the other hand, 
data from a statistical sample of about
five thousands of simulated dark matter
haloes (Bullock et al. 2001), show - to
a good extent - a lognormal distribution 
of the concentration within mass bins of
(0.5-1.0)$\times10^nh^{-1}{\rm M}_\odot$, where
$11\le n\le14$ and $n$ is an integer.

Here we adopt a statistical approach to
the calculation of the lognormal distribution
of the concentration, following the same 
procedure used by Adams \& Fatuzzo (1996),
in dealing with the initial mass function of
a star generation.   To this aim, first let
us suppose that a transformation exists 
between initial conditions and the final
properties of the dark matter halo with
assigned mass.

Given a cosmological model and a perturbation 
spectrum, the initial conditions of a simulation
are defined by a generation of complex numbers 
with a phase randomly distributed in the range 
$0\le\phi<2\pi$ and with amplitude normally 
distributed, where the variance is provided 
by the selected spectrum, in the simplest case of a
Gaussian distributed random field.   For further
details see e.g., Moscardini (1993); Tormen et al.
(1997).   

Given a universal density profile, assumed to fit
to dark matter haloes under consideration,
the final properties are related to the scaled 
radius, or the concentration with regard to NFW
density profiles, via the results
listed in Table \ref{t:proMN}.

Second, let us suppose that the transformation
under consideration is expressible as a product:
\begin{equation}
\label{eq:alc}
\Xi_{trn}=A\prod_{j=1}^n\beta_j^{\gamma_j}=
\prod_{j=1}^n\alpha_j~~;
\end{equation}
where the constant, $A$, and the exponents,
$\gamma_j$, are fixed, and the variables,
$\beta_j$ or $\alpha_j$, are conceived as 
random variables.   Though Eq.\,(\ref{eq:alc})
cannot be motivated by the existence of a
semiempirical relation, as in the case of
star formation (Adams \& Fatuzzo 1996),
neverthless it cannot be excluded unless
further knowledge about the genesis of
dark matter haloes will be available.

The central limit theorem holds provided
the random variables, $\alpha_j$, appearing
in Eq.\,(\ref{eq:alc}), are completely
independent, and their number, $n$, tends
to infinite.   For the more realistic case
of a finite number of not completely
independent random variables, the resulting
distribution is expected to be different
from a (log)normal distribution.

Taking the decimal logarithm on both sides
of Eq.\,(\ref{eq:alc}) yields:
\begin{equation}
\label{eq:lalc}
\log\Xi_{trn}=\sum_{j=1}^n\log\alpha_j~~;
\end{equation}
and the expected value of the distribution,
$f_j(\log\alpha_j)\diff\log\alpha_j$,
depending on the random variable, $\log
\alpha_j$, is:
\begin{equation}
\label{eq:spalj}
(\log\alpha_j)^\ast=\int_{-\infty}^{+\infty}
\log\alpha_jf_j(\log\alpha_j)\diff\log\alpha_j~~;
\quad1\le j\le n~~;
\end{equation}
accordingly, the error, $x_{\log\alpha_j}$, is:
\begin{leftsubeqnarray}
\slabel{eq:eralja}
&& x_{\log\alpha_j}=\log\alpha_j-(\log\alpha_j)^
\ast=\log\frac{\alpha_j}{\alpha_j^\ast}~~; \\
\slabel{eq:eraljb}
&& \alpha_j^\ast=\exp_{10}(\log\alpha_j)^\ast~~;
\qquad1\le j\le n~~;
\label{seq:eralj}
\end{leftsubeqnarray}
where, of course, $\alpha_j^\ast$ is different
from the expected value of the distribution,
$f_j(\alpha_j)\diff\alpha_j$, depending on
the random variable, $\alpha_j$.

The variance of the distribution, $f_j(x_
{\log\alpha_j})\diff x_{\log\alpha_j}$, is:
\begin{equation}
\label{eq:sixlj}
\sigma_{\log\alpha_j}^2=\int_{-\infty}^{+\infty}
x_{\log\alpha_j}^2f_j(x_{\log\alpha_j})\diff 
x_{\log\alpha_j}~~;\qquad1\le j\le n~~;
\end{equation}
and the related, expected value, equals zero.

Let us define the random variable:
\begin{equation}
\label{eq:zitx}
\zeta=\sum_{j=1}^n x_{\log\alpha_j}=\sum_{j=1}^
n\log\frac{\alpha_j}{\alpha_j^\ast}=\sum_{j=1}^
n\log\alpha_j-\sum_{j=1}^n\log\alpha_j^\ast~~;
\end{equation}
and combine Eqs.\,(\ref{eq:lalc}) and (\ref
{eq:zitx}), to obtain:
\begin{equation}
\label{eq:lalz}
\log\Xi_{trn}=\zeta+\sum_{j=1}^n\log\alpha_j^\ast~~;
\end{equation}
which is equivalent to:
\begin{leftsubeqnarray}
\slabel{eq:csiza}
&& \Xi_{trn}=\Xi^\ast\exp_{10}(\zeta)~~; \\
\slabel{eq:csizb}
&& \Xi^\ast=\prod_{j=1}^n\alpha_j^\ast~~;
\label{seq:csiz}
\end{leftsubeqnarray}
where, of course, $\Xi^\ast$ is different
from the expected value of the distribution,
$f(\Xi_{trn})\diff\Xi_{trn}$, depending on
the random variable, $\Xi_{trn}$.

The application of the central limit theorem
to the distribution, $f(\zeta)\diff\zeta$,
depending on the random variable, $\zeta$,
yields (e.g., Adams \& Fatuzzo 1996):
\begin{equation}
\label{eq:sigz}
\sigma_\zeta^2=\sum_{j=1}^n\sigma_{\log\alpha_
j}^2~~;
\end{equation}
and aiming to find a distribution, $f(\tilde
{\zeta})\diff\tilde{\zeta}$, characterized by
unit variance, $\sigma_{\tilde{\zeta}}^2=1$,
and null expected value, $\tilde{\zeta}^\ast=0$,
let us define the random variable:
\begin{equation}
\label{eq:zitt}
\tilde{\zeta}=\frac\zeta{\sigma_\zeta^2}~~;
\end{equation}
where, owing to the central limit theorem,
the distribution is normal:
\begin{equation}
\label{eq:fzitt}
f(\tilde{\zeta})\diff\tilde{\zeta}=\frac1
{\sqrt{2\pi}}\exp\left(-\frac{\tilde{\zeta}^2}
2\right)\diff\tilde{\zeta}~~;
\end{equation}
independent of the initial distributions,
$f_j(x_{\log\alpha_j})\diff x_{\log\alpha_j}$. 

Taking the decimal logarithm of both sides
of Eq.\,(\ref{eq:csiza}), and using Eq.\,(\ref
{eq:zitt}), yields:
\begin{equation}
\label{eq:lcsiz}
\log\Xi_{trn}=\log\Xi^\ast+\zeta=\log\Xi^\ast+
\sigma_\zeta^2\tilde{\zeta}~~;
\end{equation}
and the distribution, $f(\Xi_{trn})\diff\Xi_
{trn}$, depending on the random variable, 
$\Xi_{trn}$, is lognormal.    The related
expected value and variance are:
\begin{lefteqnarray}
\label{eq:ewa}
&& (\log\Xi_{trn})^\ast=\log\Xi^\ast~~; \\
\label{eq:ewb}
&& \sigma_{\log\Xi_{trn}}^2=\sigma_\zeta^2~~; 
\end{lefteqnarray}
accordingly, the distribution reads:
\begin{lefteqnarray}
\label{eq:flis}
&& f(\log\Xi_{trn})\diff\log\Xi_{trn}=\frac1
{\sqrt{2\pi}\sigma_\zeta}\exp\left[-\frac{(\log
\Xi_{trn}-\log\Xi^\ast)^2}{2\sigma_\zeta^2}
\right]\diff\log\Xi_{trn}~~; 
\end{lefteqnarray}
and the decimal logarithm of the probability
density, $f(\log\Xi_{trn})$, may be written as:
\begin{lefteqnarray}
\label{eq:lfli}
&& \log[f(\log\Xi_{trn})]=-\frac12\log(2\pi)-
\log\sigma_\zeta-\frac1{\ln10}\frac1
{2\sigma_\zeta^2}
\left(\log\frac{\Xi_{trn}}{\Xi^\ast}\right)^2~~;
\end{lefteqnarray}
where the first term on the right-hand side 
member may be conceived as a normalization 
constant.

The values of the expected value and the
variance, expressed by Eqs.\,(\ref{eq:ewa})
and (\ref{eq:ewb}), related to the lognormal
distribution, defined by Eq.\,(\ref{eq:flis}),
may be deduced from the results of simulations
(e.g., Bullock et al. 2001).   For further
details on the procedure outlined above, see
Adams \& Fatuzzo (1996).

By analogy with the theory of measure, a
computer run may be 
considered as an execution of measure operations, the related
computer code as a measure instrument, the dark halo as a 
measure subject, and the sequences of random numbers used in
the definition of initial conditions as contributors to the
accidental error.   Then the computer output may be thought
about as a measure of the corresponding scaled parameter,
which may be conceived as fluctuating around its fitting
counterpart.

In addition, it is worth of note that the application of
a least-squares or least-distances method in fitting
simulated with universal density profiles (e.g., 
Dubinski \& Carlberg 1991; KLA01; FM03) implies a
(fiducial) normal distribution of the simulated density
(in decimal logarithm) around the expected value deduced
from the fitting density profile, at any fixed distance
(in decimal logarithm).   It is the particularization, to
the case of interest, of a well known result of the theory
of errors (e.g., Taylor 2000, Chap.\,8, \S\,8.2).

\end{document}